CLARKSON UNIVERSITY

# Hyperbolic Metamaterial Filter for Angle-independent TM-Transmission in Imaging Applications

A Dissertation by

**Golsa Mirbagheri**

Department of Electrical and Computer Engineering

Submitted in partial fulfillment of the requirements for the degree of

Doctor of Philosophy

**Electrical and Computer Engineering**

Date

**11/30/2020**

The undersigned have examined the thesis entitled 'Hyperbolic Metamaterial Filter for Angle-independent TM-Transmission in Imaging Applications' presented by Golsa Mirbagheri, a candidate for the degree of Doctor of Philosophy – Computer and Electrical Engineering and hereby certify that it is worthy of acceptance.

12/7/20     _[signature]_     **David Crouse**
Date                                                 Advisors name

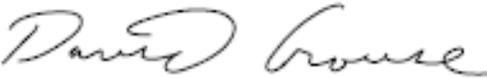

Jan Scrimgeour
Date                                                 committee member name

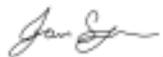

Chee-Keong Tan
Date                                                 committee member name

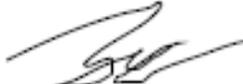

Zijie Yan
Date                                                 committee member name

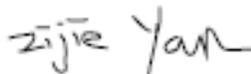

12/10/2020     Francesco Monticone     **Francesco Monticone**
Date                                                 committee member name

William Jemison     Digitally signed by William Jemison Date: 2020.12.11 09:34:37 -05'00'
Date                                                 committee member name

# ABSTRACT

With conventional narrowband filters, the center-wavelength of the narrow transmission band undergoes large shifts as the angle of incident light changes. In this project, we designed and experimentally verified a type of hyperbolic metamaterial Bragg stack that eliminates, or vastly reduces, this angle-of-incidence dependence (i.e., dispersion) of the transmission band for TM polarized beam. The filter developed in this project is composed of array of subwavelength sized metal wires vertically penetrating in the dielectric layers of the Bragg stack. We first discuss the physical reasons why such a structure is expected to minimize dispersion, then we performed detailed electromagnetic modeling, fabrication of proof-of-concept structures, and optical testing of the structures. The filter is fabricated by CMOS fabrication techniques at Cornell University's Nanoscale Science and Technology Center, and optical characterized using Fourier-transform infrared spectroscopy. Both simulated and experimental data show that narrow band transmission filters can be designed such that the center-wavelength of the transmission peak for TM polarized incident light does not change as the angle of incidence of an incoming beam changes. These types of narrowband notch filters have been used in many sensing and imaging applications, including remote sensing and hyperspectral imaging.

# ACKNOWLEDGMENTS

I would like to thank my advisor, Dr. David Crouse for guiding and supporting me over the years. I would especially like to thank CNF staff at Cornell University for the support and constant encouragement I have gotten last three years. I'd like to thank my thesis committee members for all their guidance through this process and my fellow graduate students and undergraduates who contributed to this research. I am very grateful to all of you.

Finally, I would like to thank for my family for the love and support, I'd like to dedicate this thesis to my mother.

# TABLE OF CONTENTS

# LIST OF TABLES

**Table** **Page**

# LIST OF FIGURES

# CHAPTER I: Introduction

**INTRODUCTION**

Narrowband notch filters have been used in many sensing and imaging applications, including remote sensing and hyperspectral imaging. However, angle shift to short wavelength districts the performance of these operational applications. Hyperbolic metamaterial (HMM) has demonstrated their great performance in infrared regime. The homogeneous HMM has a unique anisotropic dispersion called hyperbolic, with the unit cell much smaller than the working wavelength. The experimental data show that by implementing nanowires perpendicular to the alternative dielectric layers of Bragg stack, the effect of angle shift is eliminated, makes the TM center wavelength constant. The related electronic vibrations and electric field add up together, lead to an enormous collective response in HMM filter. This design is implemented in many systems that require high angle stability. Hyperspectral imaging system is one of the most interesting and effective areas of application of HMMs [1].

In this work, a hyperbolic metamaterial structure is designed by two concepts; conventional Bragg stack with a hyperbolic metamaterial, (i.e., a metal wire mesh) in order to reduce the dependency of the angle of incidence (AoI) to the center wavelength ($\lambda_c$) of the narrow band of TM transmission light. Section 1 explains the hyperspectral imaging, Section 2 talks about the related works and background, section 3 presents the theory and the results of the simulations performed in the design of the structure and the fabrication procedures, and Section 4 concludes the thesis.

**Hyperbolic Metamaterial**

When light propagates through the hyperbolic metamaterials, they act like metal in one axis and behaves like a dielectric in another direction of the crystal. Although metals are not great optical materials, they have promising performance in nano-photonics. Metals offer nano in the photonic devices that barely can scale down their feature, which are as big as 1μm, the wavelength of light. The wavelength of light propagates and then absorbs along the surface of a metal and can't penetrate through the hyperbolic materials. Therefore, metals and dielectric are used together to form the hyperbolic metamaterials with nano dimensions. In addition to nano characters, they introduce extraordinary properties, including negative refraction. All optical materials are defined by their dielectric function, which describes how the field changes inside of the material during light propagation. Dielectric function is denoted by epsilon and guided light. In metal, the field of light incidence is canceled by rearrangement of free electrons. The wave propagation of optical devices and materials is characterized by the dielectric function, which is denoted by dispersion relation. In the dispersion relation, the wavelength of light is proportional to the reciprocal of the frequency, as Eq.1 [2]:

$$\lambda \sim \frac{1}{\sqrt{\varepsilon}\,\omega} \sim \frac{2\pi}{k} \quad \text{Eq. 1}$$

Where k is the wave vector. Since the epsilon of metal is negative, the Eq.1 can't be satisfied for metals and that's why the light is not transmitted through the metals and wave can't be propagated. In some crystals, the lattice constant in one side is different than the other side, makes the crystal anisotropic and depending on the direction of light propagation, the dielectric function is different in each side. Each dielectric function is defined by the separate wave vector related to direction of light propagation and shows a three-dimension surface with allowed propagation directions for a given frequency, as shown in Eq. 2.

$$k_x \sim \sqrt{\varepsilon_z}\, \omega\ ,\ k_y \sim \sqrt{\varepsilon_y}\, \omega,\ k_z \sim \sqrt{\varepsilon_x}\, \omega$$

$$\frac{K_x^2}{\varepsilon_z} + \frac{K_y^2}{\varepsilon_y} + \frac{K_z^2}{\varepsilon_x} \sim \omega^2 \qquad\qquad \text{Eq. 2}$$

This surface is usually in the form of an ellipsoid and sphere. However, when the metal is part of the surface, the dielectric function of the surface in that specific direction is negative and call the surface hyperbolic mathematically. On the other hand, metamaterials composed of metal can create the hyperbolic dispersion in a wide range of wavelength and transfer the energy with no loss.

The intensity of surface plasmon can be adjusted by changing the geometry of unit cell of hyperbolic metamaterials. In contrast with ellipse, hyperbola extends out to infinity, indicates that every point inside of this hyperbola shows an acceptable wave vector and therefore, very large wave vector of hyperbolic metamaterial accepts the corresponding small (nano) wavelengths. Optical absorbers and resonators are build based on these small allowed wavelengths. One of the promising applications using these unusual properties hyperbolic metamaterials are imaging applications in nano-photonics. In next section, we describe our proposed filter implemented in the hyperspectral imaging applications [2].

**Hyperspectral Imaging System**

Hyperspectral imaging, as part of remote sensing, combines imaging and spectrometry, is used in a wide variety of applications, including remote sensing systems. The imaging system takes a picture of the reflected electromagnetic radiation of the scene infrared band. In addition, spectrometry captures information about the composition of the materials by measuring the variation of wavelength of light. The hyperspectral imaging sensors can capture spatial and spectral resolution at the same time and form a hypercube. The hypercube is a 3-dimension data set composed of layers of images; each pixel of hypercube shows the information related to the material composition. The spectrum measurement includes information about the light wavelength and chemical material composition that shape the pixel spectrum. These pixels create the multispectral imagery to get information from the scene, however, a continuous spectrum creates a greater amount of information (hyperspectral sensors) in comparison with the more discrete spectral bands (multispectral sensors). In order to capture the spectral characteristics of scene materials, a perfect set of measurements is needed to enable separation between sensor noise and interface of the scene. The term hyper can be associated with capturing spectral measurement dimensions to support this separation in the applications. The hyperspectral remote sensing includes material spectroscopy, radiative transfer, imaging spectrometry, and hyperspectral data processing, as seen in Figure 1. The material spectroscopy defines the interaction between electromagnetic radiation and material, considering the wavelength dependence features of substance [3].

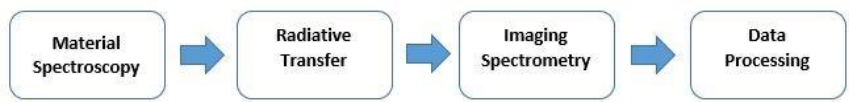

Figure 1. Fundamental elements of hyperspectral remote sensing [3]

The bonding resonance between the molecules of materials corresponds to the electromagnetic radiation of different wavelength, and shows the intrinsic spectral properties of material, like $SiO_2$ molecules interaction with the frequencies in the LWIR spectral region. The intrinsic spectral properties show the interaction of the material with the different wavelengths of electromagnetic radiation and come from the real and imaginary part of the refractive index. Although these properties are unique for each material, but the complex part of the index of refraction changes from different spectral reflectance or transmittance measurement. Therefore, we must use apparent spectral properties that are captured by measurement as a function of wavelength. Radiative transfer is the science of transferring the electromagnetic radiation from the source to final senor, while the radiation interacts with the environment. On the other hand, there is a relationship between the apparent characterization of the material and measured spectral radiance at sensor location defines by the radiative transfer [3].

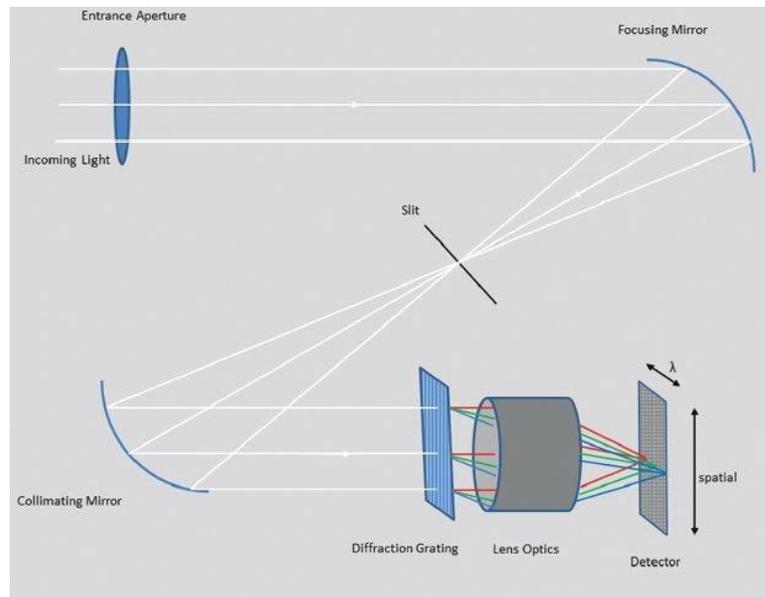

Figure 2. Hyperspectral imaging systems schematic [4]

In imaging spectrometry, a sensor is developed to capture the hyperspectral imagery in the form of prism, grating spectrometer or dielectric filters. Dispersive imaging spectrometer uses a 2D detector array to map the spectral information. The detector first collects a single spatial slice of the scene at different light wavelengths, then scans more slices to find the other spatial dimension. The information in hyperspectral imagery cannot be readable by human visualization and need the computer processing to interpret it.

Therefore, algorithmic techniques in hyperspectral data processing are designed to capture the atmospheric transmission and scattering radiations [3].

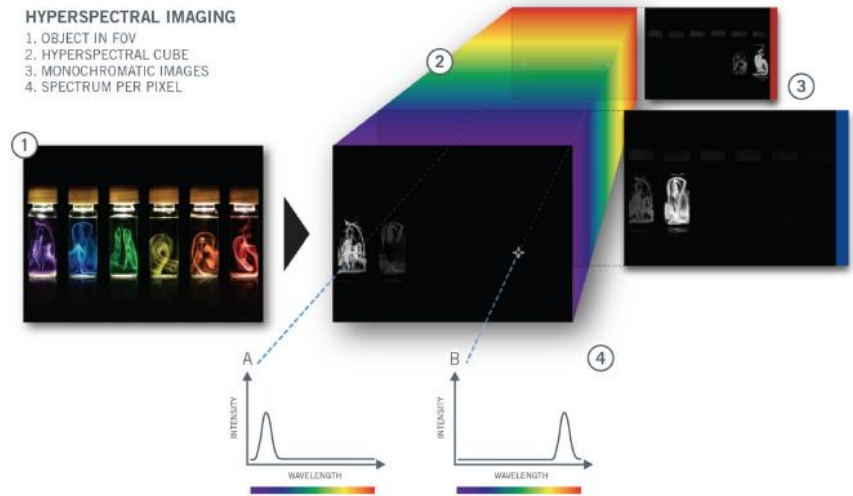

Figure 3. Hyperspectral Data Cube [5]

The Distributed Bragg Reflector (DBR) can be widely used in hyperspectral imaging systems to identify the materials or detect the processes. A hyperspectral camera needs the intensity of light for many continuous spectral bands for each pixel of an image. Therefore, each pixel in the image has a continuous spectrum and can specify objects in the scene with the wavelength details.

The camera takes image from the scene line by line at a time, using a technique called pushbroom scanning method, as shown in Figure 2. Each narrow light only passes from a narrow line of slit and collimates the light into a dispersive element. The slit blocks all light to shape a rectangular area, with dimension of one-pixel spatial in x,y axis (equivalent to detector element size) and many pixels spectral in z axis (equivalent to detector array width), called the hyperspectral cube. The dispersive element (in our project a HMBS) splits the light to different spectral wavelength. Then, the light is focused on the sensor array. The light contains a slice of a hyperspectral image, including two dimensions, one with spectral information and the other one with spatial information. The hyperspectral camera gathers slices of all adjacent lines and creates a hyperspectral image cube, illustrated in Figure 3, which has one spectral (many wavelengths) dimension and two spatial dimensions. The resolution of the final image is directly dependent on the lattice constant. [3, 6, 4, 1]

Current hyperspectral imaging systems typically use pushbroom systems for the spatial scanning, need to the relative motion between the scene and the system; which leads to increase the size, weight and cost. Therefore, the hyperspectral imaging can't be used in industry in wide scale. For this purpose, there is a necessary need to minimize spectrometer.

Spectral measuring instruments implement either multiband filter radiometer or interferometer sensor to gather the discrete or continuous spectral bands. In order to gather continuous optical channels, the huge high-resolution data is required to gathered over a spectral range that is not easy due to the time consuming of data analyzing and bandwidth limitations. The proposed arrayed metamaterial filter in this project is nanofabricated in a small size and overcomes these limitations.

Using spectrometer technology, the sub-nanometer passband filter supports a wide spectral and light cone without additional mirrors or dispersive elements. The spectrometers are bigger and more complex than the proposed filter, since the incident light should be collimated before reaching to the dispersive element. However, the angle independent filter uses the focused light, without dispersive tool to make spectral channels, as depicted in Figure 4.

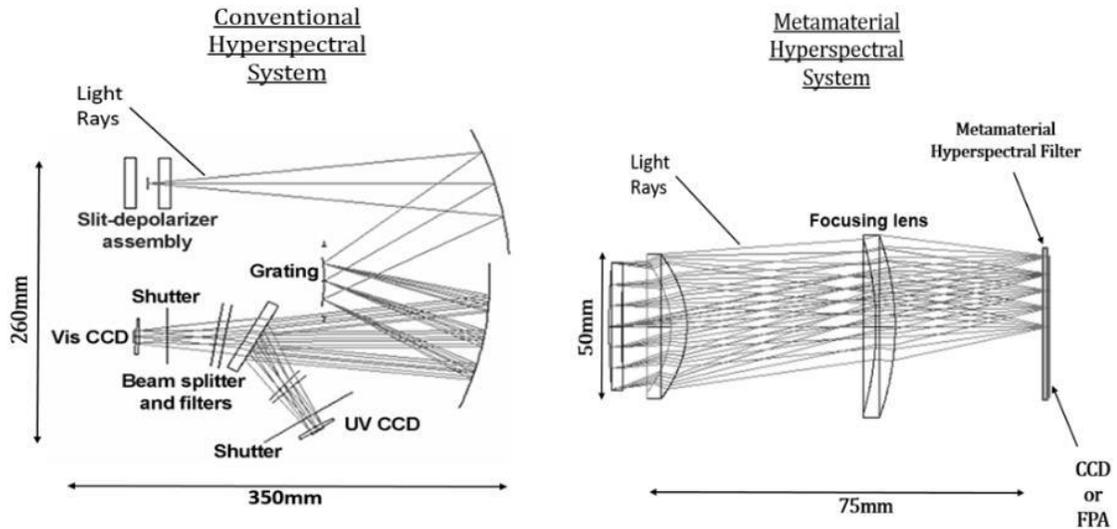

Figure 4. Conventional Hyperspectral System vs Metamaterial Hyperspectral System [7]

The angle-of-incidence independence (AoI) of proposed wafer comes from change we made in the middle resonant layer. Unlike traditional Bragg filter that are based on Fabry-Perot optical mode [8], the proposed fast F/number (more sensitive in wider range) optic method offers unchanged center wavelength for different AoI, thanks to the perpendicular arrays of metal wires in the middle layer of metamaterial filter. By changing the geometry properties of unit cell of the filter, different center-wavelength of the transmission band and therefore, wide spectral channels are supported.

# CHAPTER II: Background and Literature Review

## Related Work

Hyperbolic metamaterial Bragg stack has gained serious attentions recently and implemented in considerable number of imaging applications and opto-electronics, thanks to surface plasmon. In this section, we start to review some related articles which are close to the proposed research work.

In [9], the dispersion of SP is engineered by tuning the electromagnetic fields of the metal-like multilayers structures, that is possible by implementing multiple dielectric layers or adding doped semiconductors to control the permittivity. The permittivity of the metal is independent of their position, makes it hard to shape the SP. The applications of SP need the structures that the modes of electromagnetic waves have group velocities that changes the from positive to negative (or negative to positive), as the SP wave vector increases. The $k_x$ and $k_y$ are calculated as Eq 3.

$$k_{sp} = k_x = k_0 \left(\frac{\varepsilon_m \epsilon_d}{\varepsilon_m + \epsilon_d}\right) \quad \text{and} \quad k_y = \sqrt{\varepsilon k_0^2 - k_x^2} = \frac{i}{l} \quad \text{Eq. 3 [9]}$$

The electromagnetic field is dependent on the SP momentum $k_x$ and the decay length $l$. As the SP momentum $k_x$ increases, the $l$ is decreasing related to $k_y$ and the fields are concentrated in the area closer to the dielectric metal interface (thinner layer). Otherwise, the SP field spread out for length far from $l$, effect more layers (thicker layers for small $k_x$). As shown in Figure 5, for higher SP momentum $k_x$ the thinner dielectric layer affects the frequencies of even modes that have less energies in comparison to odd modes. The dielectric constant ($\epsilon_d < \epsilon_{Superstrate}$) and thinner films help the dispersion curves to decrease energy as $k_x$ goes up or turns over.

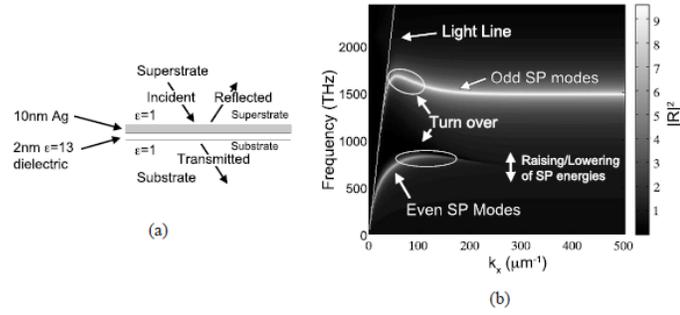

Figure 5. Thin metal-dielectric films (a) Corresponding odd and even SP modes (b) [9]

In addition to using multiple dielectric layers, the multiple conductive layers can be implemented. However, large $\varepsilon_m$ makes the field confined in one thin layer and not the other metal layers. This problem can be adjusted by using doped semiconductors that have smaller negative primitivities [9].

In [10], subwavelength apertures with extraordinary transmission is introduced as one of the SP structures. The bull's eye with periodic concentric circles is a subwavelength structure acts as an antenna that couples the incident light to the SP modes, as shown in Figure 6. This leads to the high field above the hole aperture and makes the high transmission.

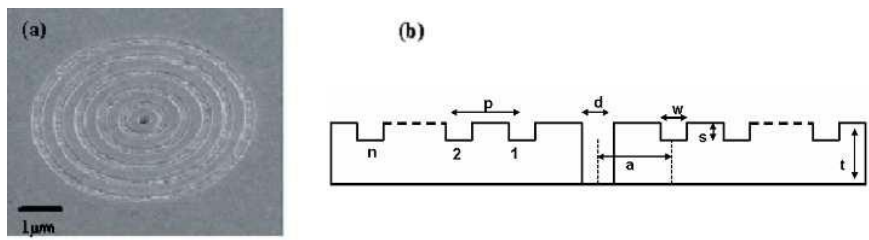

Figure 6. Periodic concentric circles of the bull's eye (a) Subwavelength structure(b) [10].

In [11], the all angle negative refraction created by dielectric matrix composed of metallic nanowires is discussed. The structure with features smaller than incident electromagnetic wavelength of light is called indefinite media. With the different signs of parallel permittivity (positive) and perpendicular permittivity (negative), the dispersion relation of indefinite media is hyperbolic for TM polarization, with the help of fakir bed of nanowires in a wide range of light. In metamaterial with negative index (NIMs), in opposite with snell's law, light is bended in opposite side. However, by engineering the dispersion relation in photonic crystals (PCs), negative refractive index can be achieved as well, without NIMs. The other significant different between Metamaterials and NIM is that NIMs have more losses in visible range due to their resonance-based nature. The metamaterials with negative index are azimuthal angle independent, working in wavelength far from resonant and have less loss. In [11], the silver fakir bed is used which has the lowest loss in visible light regime. As shown in Figure 7, $\varepsilon_\perp$ has the strong dispersion due to the resonant of SPs and p is the filling ratio. The condition of $\varepsilon_\perp \cdot \varepsilon_\parallel < 0$ for negative index of refraction, is satisfied in most of the wavelength range in Figure 7.

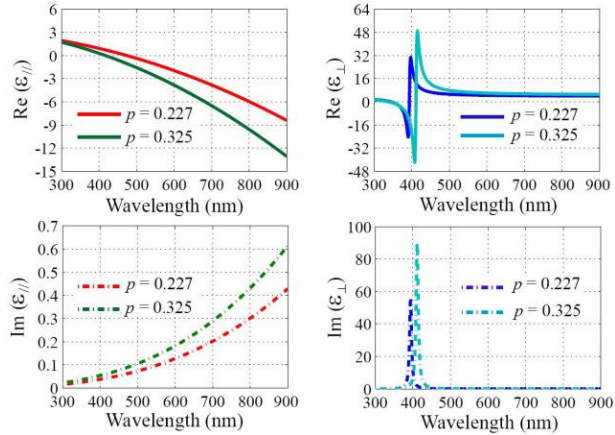

Figure 7. Real and imaginary parts of the permittivity for nanowires embedded in the alumina matrix [11].

The theta between the wave vector and pointing vector was calculated for the proposed structure, which is the opposite of NIM with anti-parallel direction.

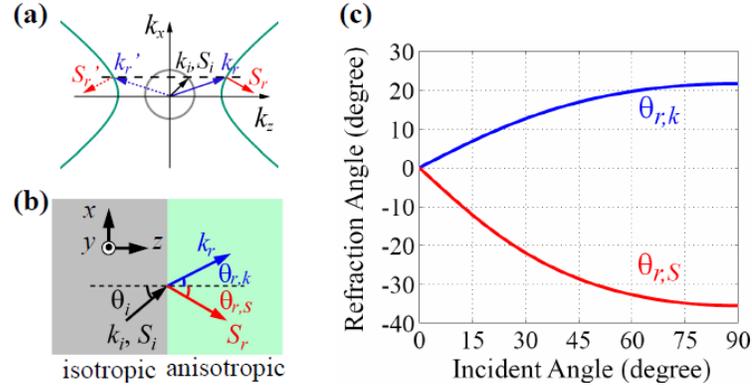

Figure 8. The hyperbola (green) and isotropic media (gray) in (a). negative refraction of anisotropic hyperbola in (b) Negative refraction of pointing vector for all angle of incident light in (c) [11].

In Figure 8, the green hyperbola in (a) has the equifrequency contour, also the wave vector (blue) and pointing vector are illustrated. In (b) the negative refraction for the proposed indefinite material is shown. The pointing vector shows the negative refraction in (c) [11].

In [12], the design of optical properties of surfaces is explained. Surface plasmon (SP) is the resonant oscillation of coherent conduction electrons with a TM polarization at the boundary of metallic and dielectric media which are stimulated by incident light. The surface charges are intensively oscillated back and forth by the electric field. The field of the surface plasmon is confined close to the surface and evanesces from the material exponentially. Furthermore, the oscillation of charges changes the electromagnetic energy into heat, leads to a big absorption at the resonance frequency. The wave vectors of the surface plasmon are bigger than the free space wave vector, according to the Eq 4 indicates that SP wave vectors don't couple with the modes of EM waves in air ($k_{sp} > k_0$). In Eq 4, the wave vector of plasmon is defined by the wave vector in vacuum ($k_0=\omega/c_0$) multiplied by a component composed of permittivity of dielectric ($\varepsilon_1$) and permittivity of metal ($\varepsilon_2$) [12].

$$k_{sp} = k_0 \sqrt{\frac{\varepsilon_1 \varepsilon_2}{\varepsilon_1 + \varepsilon_2}} \qquad \text{Eq 4 [12]}$$

It should be mentioned that the Eq 4 is defined to excite SPs for the non-magnet materials (TM Polarization), for TE polarization magnetic materials with negative permeability are required. To excite the SPs, the electromagnetic wave vectors should have tangential elements equal to $k_{sp}$. One way to coupling the electromagnetic waves with incident light is using the high index prisms to make incident light wave vector bigger. The other way is using diffracted light of the periodic nanostructures [12].

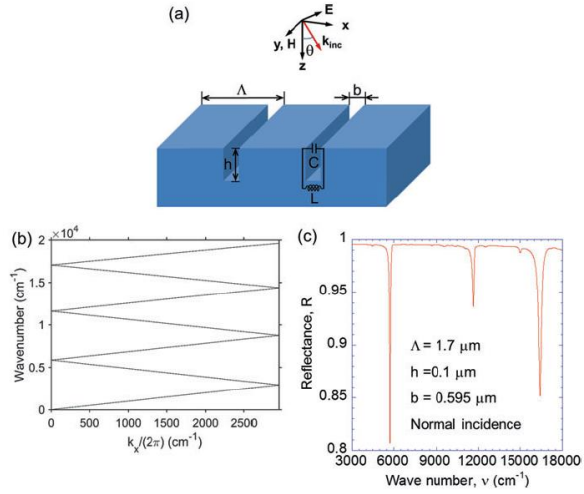

Figure 9. The grating structure in (a). The folded dispersion relation in (b) and corresponding normal reflection peaks for TM polarization in (c) [12].

In Figure 9, the magnetic field oscillating in y-direction, have the incident wave in the xz plane for TM polarization. The structure is one-dimension Ag grating with $\Lambda = 1.7$ μm, h = 0.1 μm, and b = 0.595 μm. In Figure 9.b, the dispersion relation of Bloch-Floquet $k_{x,m} = k_{x,inc} + \frac{2\pi m}{\Lambda}$ is folded in the region $k_x \leq \frac{\pi}{\Lambda}$ for Ag grating, with $\Lambda$ as period of grating, indicates that surface plasmon can be excited on a grating surface [13]. Also, folded dispersions have the intersection with the axis, indicate the location of excited surface plasmons for normal incidence light, shown in Figure 9.c. In Eq 4, the permittivity of metal ($\varepsilon_2$) is a large negative real number in comparison to $\varepsilon_1$, makes the dispersion of SPP very close to the light line $k_0 = \frac{\omega}{c_0}$ and therefore, light can couple with surface plasmon [12].

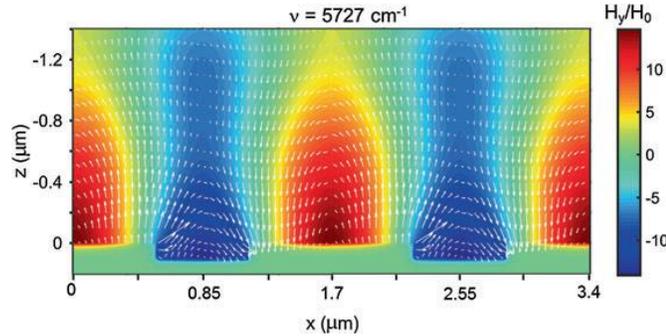

Figure 10. For the normal incidence, the field distribution represents at the resonant wave number[12]

In Figure 10, the distribution of instantaneous field represented with the arrows, illustrated at $\nu = 5{,}727$ cm$^{-1}$ corresponded to the first narrow dip in Figure 9.c. The normal incident light can be excited and coupled with two surface plasmons in $k_{sp} = \frac{2\,m\pi}{\Lambda}$, one relates to m=+1 going to right, the other one related to m=-1 going to left, forming a standing wave. This is different for the oblique incident light, where incident photon only can couple with one surface plasmon [12].

The pixelated guided-mode resonant filters (GMRFs) for measuring the absorption band of $CO_2$ and $H_2O$ in MWIR is described in [14]. For each gas, a narrowband hyperspectral filter is designed for register the pixels of the filter to the pixels of the sensor. The focal plane array (FPA) sensor is an optical hyperspectral arrayed device [14]. FPA is device with a pixelated light-sensing array at the focal plane of a lens. The resonant subwavelength grating (RSG) technology is used in this filter. The term subwavelength is used to express the structures which one or more dimensions of the structure are smaller than the wavelength of incident light. The all-dielectric RSG filter reflects the desired wavelength, while different narrowband wavebands can be reached by scaling the nanostructure of the filter. In the both transmissive filters, the RSG structure is used.

In [14] the absorption spectra data in LWIR (8-12μm) have low resolution and integrated with the atmosphere. The proposed technology shows the local information about the climate changing effect, specially the water and $CO_2$. Since the temperature and altitude of water vapor in the greenhouse effect changes, it is not easy to detect and understand them. Therefore, the local measurement, shows a better depiction of water and $CO_2$ distribution of column height of air between upper atmosphere and sea level. The spectral region used in proposed method, cover the desire MWID (4-8μm) for water and $CO_2$ absorption band and their altitude in column can be easily measured. This wavelength range has high absorption, which is a feature for this local monitoring the gases in short distance. The current filter technologies for water vapor and $CO_2$ detection are matched filter and hyperspectral filter. In matched filter method, the measured absorption spectrum of target gas and atmosphere compared to each other and the match parts show the large narrow band peak. However, water has complex narrowband peaks and it's difficult to find the match peaks among the long wave range. The other technology, called hyperspectral filter, measures separate wavelength in a wide absorption band. The narrowband thin-film filters, in the shape of the array, are used in this technology. However, each filter pixel should be isolated from the neighbor, which leads to misalignment in nanofabrication process. Therefore, RSG technology is being used in multilayer thin-film filter to avoid the fabrication problems. The materials used in the both transmissive and reflective filters are ZnS and Ge which have low absorption in MWIR. Also, the grating periods are easy to fabricate, because of the long wavelength of MWIR.

The rigorous coupled-wave analysis (RCWA) semi-analytical method is used to compute electromagnetics [14].

The reflective filter showed in [14] has two layers with shallow etch. The $CO_2$ and $H_2O$ bands have different wavelength which is achievable by using different period in each pixel of the filter. If the thickness of layers is chosen wisely as described in [14], the sideband across reflection are reduced and the narrow bands have FWHM close to zero in period range. The pixelated FPA detector detects the incident light through 1m of atmosphere which passed by the reflective filter. The detector recognizes the integral of incident light limited to the filter element bandwidth. The transmissive filter, which has more challenging fabrication, are based on Fabry-Petrot micro-cavity.

In Fabry-Perot technique, the incidence light radiates through a cavity surrounded by two reflective mirrors (surfaces). The light is broken to multiple beams, transmitted and reflected. If the total path of the reflected lights is the multiple of the wavelength light, then the reflected beams will arise constructively. The difference between two beams is written as Eq. 5.

$$\Delta = 2nd\, cos(\theta) \qquad \text{Eq. 5 [15]}$$

And the phase difference is written as Eq. 6.

$$\delta = \frac{2\pi}{\lambda}\Delta \qquad \text{Eq. 6 [15]}$$

The schematic of optical path difference is illustrated calculated as Figure 11 [15].

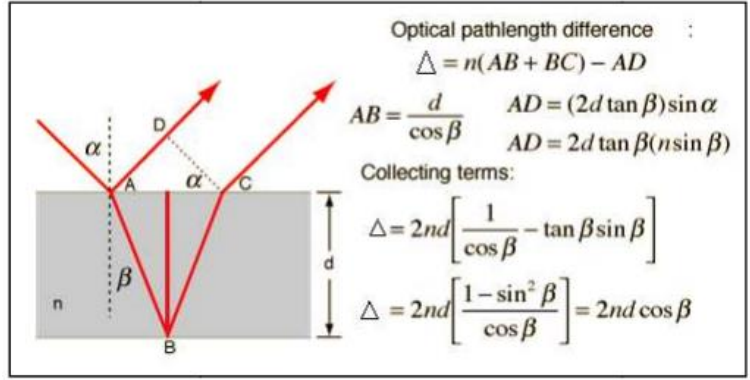
Figure 11. Optical Path Difference [16]

Regardless of any center wavelength, different spectral resolution can be achieved with the designed transmissive filter in [14]. The very narrow band filters in an optical filter array have wavelength filtering function control, possible by the RSG, to scale the wide range of wavelength. The transmission peak of the transmissive filter varies when the period of filter changes, but in the reflective filter the peaks are the same as the periods are varied (as shown in Figure 12). However, by changing the design of RSG and mirror stack, the transmission peak of transmissive filter can have uniform responses.

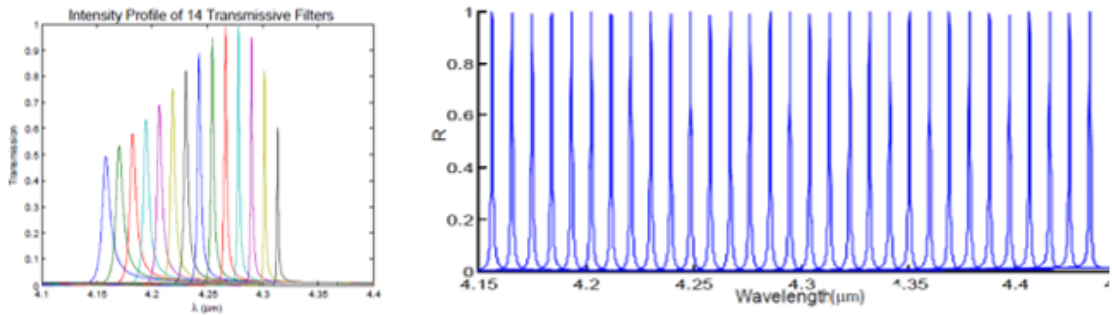
Figure 12. Simulated transmission for 14 filters (left), for 31 gratings with different periods using RCWA (right) [14]

Fabry-Perot in spherical and flat mirror is discussed in [17]. The plane mirror resonator has two parallel plane mirrors with reflectivity between 0 and 1. The plane wave $E_e$ goes through the mirror and then, we have infinite coherent superpositions of all $E_i$, as depicted in Figure 13.

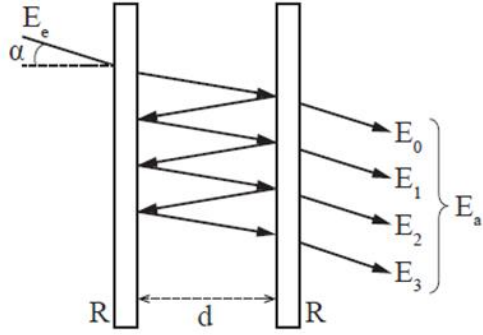

Figure 13. Interference of the multi-beams at parallel mirrors of resonator [17]

The amplitude of total transmitted beam is shown in Eq. 7.

$$E_a = \sum_{n=0}^{\infty} E_n = TE_e \sum_{n=0}^{\infty} R^n e^{-in\Delta\phi} = \frac{TE_e}{1 - Re^{-i\Delta\phi}}$$

Eq. 7 [17]

And the transmitted intensity is calculated as Eq 8.

$$I_T = I_e \frac{(1-R)^2}{(1-R)^2 + 4R\sin^2(\Delta\phi/2)}$$

Eq. 8 [17]

In [18] the angle-independent color filtering is described. The proposed resonant cavity included an amorphous silicon (a-Si) layer which deposited in a specific temperature to reduce the angle dependency of the filter. Using high refractive index material as dielectric layer can be used in Fabry-Petrot cavity-based to have less sensitivity to the light of incidence. The a-Si layer contacted with the metal layer in the proposed filter has a high reflectivity and low absorption loss in visible wavelength range, leads to a big reflection phase shift and therefore, the constructive light interference, that increase the tolerance of the filter to the angle of the incident light.

The proposed filter has the a-Si layer (phase matching layer), which is on top the silver layer and under the chrome layer. Also, the titanium oxide (TiO$_2$) layer on top of the filter acts as an anti-reflective layer. The high absorption of silver and silicon is adjusted by deposition in the suitable temperature. The temperature should be such that the dielectric layer (here a-Si) has suitable index of refraction and balanced multilayer's surface roughness. After deposition of Ag on fused silica substrate, the temperature of the filter remained on 150C and the rest of the layers, including a-Si, Cr and $TiO_2$, by e-beam evaporation. It showed that the real part (n) and imaginary part of complex index of a-Si increases as temperature increases and leads to angular tolerance. The phase shift of propagation in a-Si layer is σ = 2*nd*cos(θ) / λ, which shows that reflection changes with the phase shift based on the angle of incident (θ).On the other hand, if the phase not change, the reflection remain the same by change the angle of light, which is verified in [18]. The big difference between index of refraction of a-Si layer and air makes the critical angle in a-Si layer very small, which leads to limited angle of propagation in critical a-Si layer. Therefore, very little phase shift changes and higher index of refraction of a-Si makes the filter insensitive to the angle of incident light. The higher index of refraction is reached in higher packing density by implying higher deposition temperature.

In [19] the planar array of bandpass filter composed of two distributed Bragg reflectors separated by a dielectric low loss metasurface layer. The filter acts as an Fabry-Perot resonator which the middle transmittive metasurface layer controls the center wavelength as phasing shift function. The center wavelength is changed by the geometry parameter of the middle layer independently. The metasurfaces are subwavelength arrays that can control the phase shift and polarization. The high transmission of metasurface is related to the high index material sandwiched between low index materials and phase shifting feature is because of the geometry design of the nanoposts.

The two-dimensional transmittive metasurface between two FP reflectors controls the roundtrip phase inside the filter and therefore, the resonance passband wavelength can be manipulated without changing the reflector mirrors distance. The metasurface layer includes of an a-Si nanoposts surrounded by low-index SU-8 polymer to change transmission phase by varying the nano-posts width, as depicted in Figure 14.

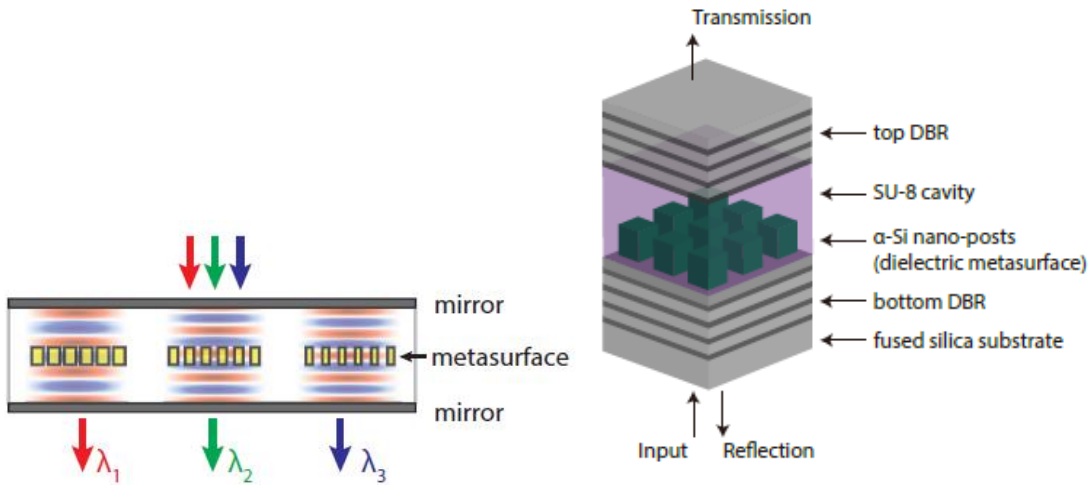

Figure 14. Schematic of two DBR micro-cavities around the metasurface layer (left) and proposed filter(right) [19]

The resonance condition happens when the thickness of cavity is equal to the half of wavelength divided by the index of refraction of cavity. Then, the metasurface layer (phase shifting layer) in the polymer cavity layer shifts the phase (resonance wavelengths) of FP resonators. Latter happens without changing the distance of the mirrors of resonators. By calculating the complex coefficients of transmission/reflection of transfer matrix for the metasurface layer with RCWA simulation, the transmission of the whole filter was calculated.

The angle independent near-infrared bandpass filter, proposed in [20], is based on the etalon resonator (Fabry–Pérot interferometer) is made of a high index $TiO_2$ cavity layer between the two Ag/Ge metallic mirrors. The effect of angle on the resonant wavelength, transmission and polarization dependent in the high index $TiO_2$ in proposed filter shows smaller variation theoretically and experimentally, as compared to the filter composed of low index cavity layer, such as SiO2.

An Oxide layer on top of the etalon is for protecting the oxidation of Ag layer, as depicted in Figure 15. The Ge film in metallic mirror is used to reduce the roughness of Ag layer which leads to optical loss, also, Ge acts as adhesion layer for Ag to TiO2 layer.

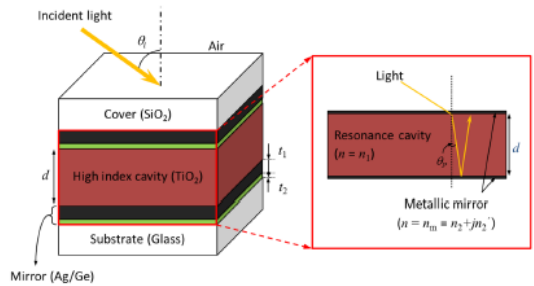

Figure 15. The angular tolerance spectral filter [20]

The propagation angle θ$_p$ is defined by the index of recreation of the resonance cavity and angle of incident light. This angle has a direct relation to the resonant wavelength. In order to decrease angle dependency of the filter, the filter should have less sensitivity to θ$_i$ by decreasing the propagation angle.

$$\frac{(\partial \lambda_c / \partial \theta_i)}{\lambda_c} = \frac{\sin \theta_i \cos \theta_i}{n_1^2 - \sin^2 \theta_i}$$
Eq. 9 [20]

As it can be seen in Eq 9, by increasing the refractive index of the resonance cavity for a constant incident angle, the relative wavelength shift is reduced. The relative center wavelength shift for SiO2 and TiO2 is shown in Figure 16 for unpolarized light.

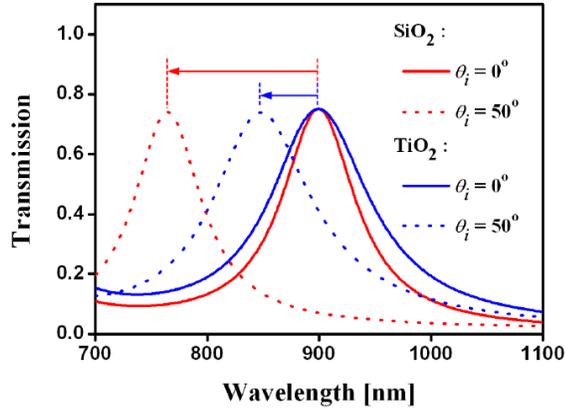

Figure 16. Spectral response of the etalon for cavity materials of SiO2 and TiO2 for $\theta_i = 0°$ and $50°$ [20]

## Different Methods for Modeling of Subwavelength Structures

There are different methods for mathematical modeling of anti-reflective subwavelength structures, including Effective Medium Theory (EMT), Finite-Difference Time-Domain (FDTD), Transfer Matrix Method (TMM), Finite Element Method (FEM) and Fourier Modal Method (FMM)/Rigorous Coupled-Wave Analysis (RCWA). In this project FMM/RCWA and TMM are described. This chapter is inspired by the notes of the "Electromagnetic Structure" course with the help of Dr. David Crouse in ECE Dept. at Clarkson University.

### Rigorous Coupled-Wave Analysis

Rigorous coupled-wave analysis (RCWA) is a semi-analytical method that is used to calculate the scattering from anisotropic periodic dielectric layer stacks by solving the electromagnetic, with the help of the Fourier transform calculation. The layers of the structure are uniformed in the homogenous direction (longitudinal z direction) and electromagnetic modes of each layer are calculated by staircase approximation. In order to solve the electromagnetism modes defined by the wave vectors, the boundary conditions and Maxwell's equations are expanded by discrete Fourier transform. This means that fields and materials are represented as a s set of plane waves, leads to infinite number of equations for periodic dielectric structure. These infinite algebraic equations can be finite by the higher order of Floquet functions. Each layer of multilayer structures can be picometer to megameters without changing the efficient speed. The boundary conditions are implemented at the interfaces between the layers [21].

**Constructing RSGW Algorithm**
In order to solve the optical properties of the layered Bragg Stack, first we need to define the RSGW algorithm. This approach will be used to model our complex structure. A Bragg stack composed of several layers is illustrated as Figure 17.

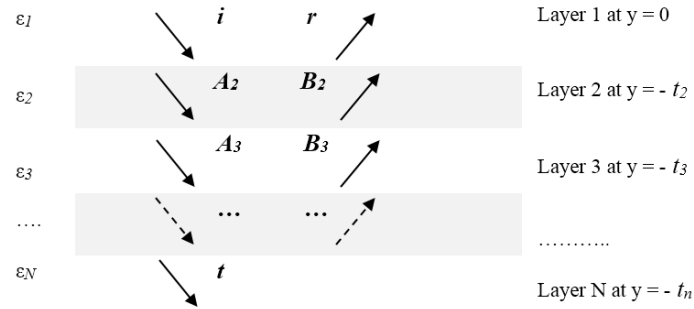

Figure 17. Coupled propagated waves of a layer stack

The propagation wave equations for each layer of BS are written as Eq 10, 11 and 12.

$$f_1 = e^{i(k_x x + \beta_1 y)} + r\, e^{i(k_x x + \beta_1 y)} \quad \text{Eq 10}$$

$$f_n = A_n\, e^{i(k_x x + \beta_n (y+d_n))} + B_n\, e^{i(k_x x - \beta_n (y+d_n))} \quad \text{Eq. 11}$$

$$f_N = t\, e^{i(k_x x + \beta_n (y+d_n))} \quad \text{Eq. 12}$$

with f is $H_z$ for TM polarization. we have Eq 13 based on Figure 18 :

$$k_x = \sqrt{\varepsilon_0}\, k_0 \sin(\theta_i), \quad k_0 = \frac{\omega}{c} = \frac{2\pi}{\lambda}, \quad \beta_n = \sqrt{\varepsilon_n k_0^2 - k_x^2} \quad \text{Eq. 13}$$

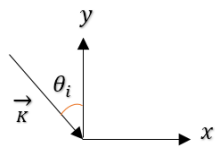

Figure 18. k vector

with a time dependence of $e^{-i\omega t}$ and $d_n = \sum_{s=2}^{n} t_s$ with n=2, 3, …, N-1

For TM polarization, electric and magnetic fields are in the form of Eq 14:

$$\vec{H} = H_z \hat{z} \quad \text{and} \quad \vec{E} = E_x \hat{x} + E_y \hat{y} \quad \text{Eq 14}$$

Faraday's law in Eq 15 and Ampere's law Eq 16 show the Maxwell's equations (for $\mu = \mu_0 = 1$) are shown.

$$\Delta \times \vec{E} = \frac{-1}{c} \frac{\delta \vec{B}}{\delta t} = i \frac{\omega}{c} \vec{H} = i k_0 \vec{H} \quad \text{Eq 15}$$

$$\Delta \times \vec{H} = \frac{1}{c} \frac{\delta \vec{D}}{\delta t} = -i \frac{\omega}{c} \varepsilon \vec{E} = -i \varepsilon k_0 \vec{E} \quad \text{Eq 16}$$

Eq 16 can be written in other form of Eq 17:

$$\begin{vmatrix} i & j & k \\ \frac{\delta}{\delta x} & \frac{\delta}{\delta y} & 0 \\ 0 & 0 & H_z \end{vmatrix} = \frac{\delta H_z}{\delta y} \hat{x} - \frac{\delta H_z}{\delta x} \hat{y} = -i \varepsilon k_0 (E_x \hat{x} + E_y \hat{y})$$

$$E_x = \frac{i}{\varepsilon k_0} \frac{\delta H_z}{\delta y} \quad \text{and} \quad E_y = -\frac{i}{\varepsilon k_0} \frac{\delta H_z}{\delta x} \quad \text{Eq 17}$$

Considering the tangent component of fields $E_x$ and $H_z$, the boundary condition at layer y=0 for $H_z$ is applied as Eq 18:

$$1 + r = A_2 e^{i\beta_2 t_2} + B_2 e^{i\beta_2 t_2} \quad \text{Eq 18}$$

If we consider $e^{i\beta_n t_n}$ as $\phi_n$, the equation Eq 18 can be written as Eq 19:

$$1 + r = A_2 \phi_2 + B_2 \phi_2^{-1} \quad \text{E 19}$$

By using the boundary condition at y=0 and using Eq 17, $E_x$ is written as Eq 20:

$$-\frac{\beta_1}{\varepsilon_1 k_0} + \frac{\beta_1}{\varepsilon_1 k_0} r = -\frac{\beta_2}{\varepsilon_2 k_0} A_2 \phi_2 + \frac{\beta_2}{\varepsilon_2 k_0} B_2 \phi_2^{-1} \quad \text{Eq 20}$$

By using the boundary condition at y=- $t_2$, $H_z$ is written as Eq 21.

$$A_2 + B_2 = A_3 \phi_3 + B_3 \phi_3^{-1} \quad \text{Eq 21}$$

And $E_x$ is represented as Eq 22:

$$-\frac{\beta_2}{\varepsilon_2 k_0} A_2 + \frac{\beta_2}{\varepsilon_2 k_0} B_2 = -\frac{\beta_3}{\varepsilon_3 k_0} A_3 \phi_3 + \frac{\beta_3}{\varepsilon_3 k_0} B_3 \phi_3^{-1} \qquad \text{Eq 22}$$

The rest of interfaces except the last interface, have the boundary conditions for $H_z$, shown in Eq 23:

$$A_n + B_n = A_{n+1} \phi_{n+1} + B_{n+1} \phi_{n+1}^{-1} \qquad \text{Eq 23}$$

And the $E_x$ is represented as Eq 24.

$$-\frac{\beta_n}{\varepsilon_n k_0} A_n + \frac{\beta_n}{\varepsilon_n k_0} B_n = -\frac{\beta_{n+1}}{\varepsilon_{n+1} k_0} A_{n+1} \phi_{n+1} + \frac{\beta_{n+1}}{\varepsilon_{n+1} k_0} B_{n+1} \phi_{n+1}^{-1} \qquad \text{Eq 24}$$

The N-1/N interface has the boundary condition for $H_z$, shown as Eq 25:

$$A_{N-1} + B_{N-1} = t \qquad \text{Eq 25}$$

And $E_x$ is represented as Eq 26:

$$-\frac{\beta_{N-1}}{\varepsilon_{N-1} k_0} A_{N-1} + \frac{\beta_{N-1}}{\varepsilon_{N-1} k_0} B_{N-1} = -\frac{\beta_N}{\varepsilon_N k_0} t \qquad \text{Eq 26}$$

Based on the separate equation for each interface, we can construct a matrix Equation. The column of the matrix of 2N-2 unknown variables is shown as Eq 27:

$$\Phi = \begin{pmatrix} r \\ A_2 \\ B_2 \\ A_3 \\ B_3 \\ \ldots \\ A_{N-1} \\ B_{N-1} \\ t \end{pmatrix} \qquad \text{Eq 27}$$

The matrix for 2N-2 initial conditions or knowns is shown as Eq 28:

$$L = \begin{pmatrix} -1 \\ \frac{\beta_1}{\varepsilon_1 k_0} \\ 0 \\ 0 \\ 0 \\ \vdots \\ 0 \\ 0 \\ 0 \end{pmatrix} \quad \text{Eq 28}$$

Then, we can easily construct the matrix of equation, as $(M)(\phi) = L$ q 29:

$$(M)(\phi) = (L) \quad \text{Eq 29}$$

On the other hand, the relation of coupling matrix M, ϕ and L is illustrated as represented in Eq 30:

$$\begin{pmatrix} 1 & -\phi_2 & -\phi_2^{-1} & 0 & 0 & \cdots & 0 & 0 & 0 \\ \frac{\beta_1}{\varepsilon_1 k_0} & \frac{\beta_2 \phi_2}{\varepsilon_2 k_0} & \frac{-\beta_2 \phi_2^{-1}}{\varepsilon_2 k_0} & 0 & 0 & \cdots & 0 & 0 & 0 \\ 0 & 1 & 1 & -\phi_3 & -\phi_3^{-1} & \cdots & 0 & 0 & 0 \\ 0 & \frac{-\beta_2}{\varepsilon_2 k_0} & \frac{\beta_2}{\varepsilon_2 k_0} & \frac{\beta_3 \phi_3}{\varepsilon_3 k_0} & \frac{-\beta_3 \phi_3^{-1}}{\varepsilon_3 k_0} & \cdots & 0 & 0 & 0 \\ \vdots & \vdots & \vdots & \vdots & \vdots & \vdots & \vdots & \vdots & \vdots \\ \vdots & \vdots & \vdots & \vdots & \vdots & \cdots & 1 & 1 & -1 \\ 0 & 0 & 0 & 0 & 0 & \cdots & \frac{-\beta_{N-1}}{\varepsilon_{N-1} k_0} & \frac{\beta_{N-1}}{\varepsilon_{N-1} k_0} & \frac{\beta_N}{\varepsilon_N k_0} \end{pmatrix} \begin{pmatrix} r \\ A_2 \\ B_2 \\ A_3 \\ B_3 \\ \vdots \\ \vdots \\ A_{N-1} \\ B_{N-1} \\ t \end{pmatrix} = \begin{pmatrix} -1 \\ \frac{\beta_1}{\varepsilon_1 k_0} \\ 0 \\ 0 \\ 0 \\ \vdots \\ \vdots \\ 0 \\ 0 \\ 0 \end{pmatrix} \quad \text{Eq 30}$$

The ϕ is represented as Eq 31:

$$(\phi) = (M)^{-1}(L) \quad \text{Eq 31}$$

**The Diffraction Equation**

Scattering of waves from periodic materials makes different optical phenomena. In this section the Optical grating is explained. Considering just two isotropic scatters (Equal in all direction), $d_i$ (extra distance beam travel to get to scatterer2) and $d_s$ (extra distance scattered before catching up to scattered beam 2) are calculated as Eq 32, based on what is represented in Figure 19:

$$d_i = \Lambda \sin(\theta_i) \quad \& \quad d_s = \Lambda \sin(\theta_s) \quad \text{Eq 32}$$

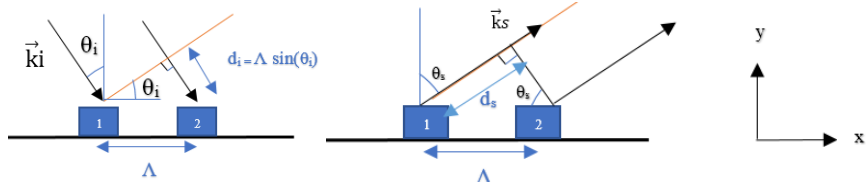

Figure 19. Scattering of the waves

And the total amplitude of the scattered wave is shown as Eq 33:

$$\xi = A\, e^{i\vec{k}_s \cdot \vec{r}} + B\, e^{i(\vec{k}_s \cdot \vec{r} + \Delta\Phi)} \quad \text{Eq 33}$$

Since each scatter is identical, both coefficients A and B are equal. Due to the conservation of energy, $|\vec{k}_s| = |\vec{k}_i| = \frac{2\pi n}{\lambda}$, where n is the index of refraction. In order to have constructive interfering, $\Delta\Phi$ (phase different of two scattered wave) needs to be an integer multiple of $2\pi$, as shown in Eq 34:

$$\Delta\Phi = 2\pi m = k_i\, d_i - k_s\, d_s = \frac{2\pi n}{\lambda} \Lambda\, (\sin(\theta_i) - \sin(\theta_s)) \quad \text{Eq 34}$$

And for normal incidence ($\theta_i = 0$), the diffraction equation is concluded as Eq 35:

$$\Lambda \sin(\theta_s) = \frac{\lambda}{n}\, m \quad \text{Eq 35}$$

The generalized scattering equation can be written in different way base on Figure 20. The constructive interference condition can be written as Eq 36:

$$\Delta\Phi_{12} = |\vec{k}_i||\vec{R}_{12}| \cos(\Phi_i) - |\vec{k}_s||\vec{R}_{12}| \cos(\Phi_s) = (\vec{k}_i - \vec{k}_s) \cdot \vec{R}_{12} = \vec{K} \cdot \vec{R}_{12} \quad \text{Eq 36}$$

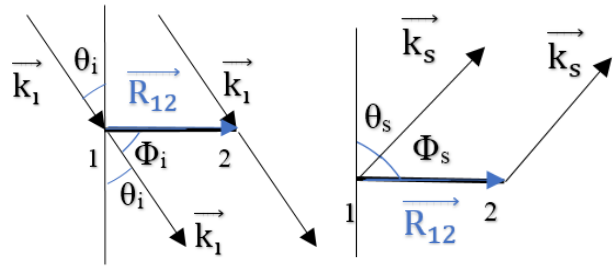

Figure 20. Incident and scattered waves schema

The infinite array of scatters is given by Eq 37:

$$\xi = \sum_n A_n e^{i(\vec{k}_s \cdot \vec{r} + \Delta \Phi_{1n})} \quad \text{Eq 37}$$

where $A_n$ is A for all n for the all identical scatterers nature and $\Delta\Phi_{1n}$ is $\vec{K} \cdot \vec{R_{1n}}$ where $\vec{R}$ is a translation vector between two scatters.

$$\xi = \left( \sum_n e^{i\vec{K} \cdot \vec{R}_{1n}} \right) A e^{i(\vec{k}_s \cdot \vec{r})} \quad \text{Eq. 38}$$

In Eq 38, $\sum_n e^{i\vec{K} \cdot \vec{R}_{1n}}$ is the multiplicative scattering factor and $A e^{i(\vec{k}_s \cdot \vec{r})}$ is the particular plane wave with amplitude A. Furthermore, $e^{i\vec{K} \cdot \vec{R}}$ can take on any value in the complex plane with unit amplitude, based on what value vector $\vec{K}$ and $\vec{R}_{1n}$ (n=2,3,…,∞) can take. Therefore, the sum is zero and there will be no scattered wave, unless $\vec{K} \cdot \vec{R}_{1n} = 2\pi m$ for all $\vec{R}_{1n}$.

Based on Eq 38, the scattered wave's wavevector is written as Eq 39:

$$\vec{k_s} = \vec{K} + \vec{k_\iota} \quad \text{Eq 39}$$

Using Floquet's theory, the sum of plane waves (Floquet modes) shows the waves in a crystal, as Eq 40 is used to construct RCWA modeling.

$$\xi = \sum_{\vec{K}} A_K e^{i((\vec{k} + \vec{K}) \cdot \vec{r})} \quad \text{Eq 40}$$

**Implement RCWA in Dielectric Grating**

The one-dimensionally periodic all-dielectric grating is represented in Figure 21, composed of three layers: first layer is superstrate; second layer is grating layer and third layer is substrate.

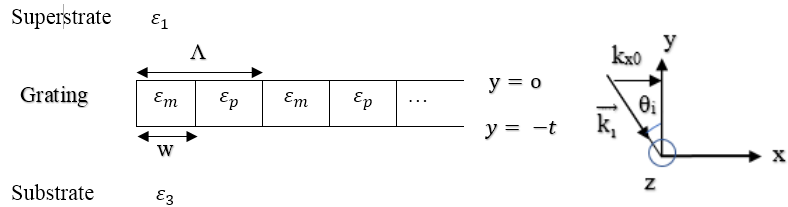

Figure 21. Three-layer stack composed of middle grating layer schematic

The fields in whole layers not only have plane wave $k_{x0}$, but also have K for the periodicity of each layer, as written in Eq 41

$$k_x = k_{x0} + nK \qquad \text{Eq 41}$$

where $K = \frac{2\pi}{\Lambda}$, n = (-N, -N+1, ..., -1, 0, 1, ..., N-1, N) and N is integer and $k_{x0} = n_1 k_0 \sin(\theta)$.

**Electric and Magnetic Fields in Three Layers for TM Polarization**

The $H_z$ and $E_x$ in superstrate layer are represented as Eq 42 and 43

$$H_z = e^{i(k_{x0}x + \beta_0 y)} + \sum_n r_n e^{i(k_{xn}x - \beta_n y)} \qquad \text{Eq 42}$$

$$E_x = \frac{i}{k_0 \varepsilon} \frac{\delta H_z}{\delta y} = \frac{-\beta_0}{k_0 \varepsilon} e^{i(k_{x0}x + \beta_0 y)} + \frac{1}{k_0 \varepsilon} \sum_n \beta_n r_n e^{i(k_{xn}x - \beta_n y)} \qquad \text{Eq 43}$$

Where $\beta_n = \sqrt{\varepsilon_1 k_0^2 - k_{xn}^2}$ and $k_{xn} = \sqrt{\varepsilon_1} k_0 \sin(\theta) + mK$, $K = \frac{2\pi}{\lambda}$.

In substrate layer, $H_z$ and $E_x$ are represented as Eq 44 and 45.

$$H_z = \sum_n t_n e^{i(k_{xn}x + \tilde{\beta}_n(y+t))} \qquad \text{Eq 44}$$

$$E_x = \frac{i}{k_0 \varepsilon} \frac{\delta H_z}{\delta y} = -\frac{1}{k_0 \varepsilon} \sum_n \tilde{\beta}_n t_n e^{i(k_{xn}x + \tilde{\beta}_n(y+t))} \qquad \text{Eq 45}$$

For finding the fields in the patterned layer, the Maxwell equations should be reminded at first, as written in Eq 46 and 47

$$\vec{\Delta} \times \vec{E} = -\frac{1}{c} \frac{\delta \vec{B}}{\delta t} = i k_0 \vec{H} \qquad \text{Eq 46}$$

$$\vec{\Delta} \times \vec{H} = \frac{1}{c} \frac{\delta \vec{D}}{\delta t} = -i \varepsilon k_0 \vec{E} \qquad \text{Eq 47}$$

The Eq 46 can be written as Eq 48.

$$\begin{vmatrix} i & j & k \\ \frac{\delta}{\delta x} & \frac{\delta}{\delta y} & 0 \\ E_x & E_y & 0 \end{vmatrix} = \left( \frac{\delta E_y}{\delta x} - \frac{\delta E_x}{\delta y} \right) \hat{z} = i k_0 H_z \hat{z}$$

$$\frac{\delta(-E_x)}{\delta y} = i k_0 H_z - \frac{\delta E_y}{\delta x} \qquad \text{Eq 48}$$

The Eq 47 can be written in the form of Eq 49.

$$\begin{vmatrix} i & j & k \\ \frac{\delta}{\delta x} & \frac{\delta}{\delta y} & 0 \\ 0 & 0 & H_z \end{vmatrix} = \frac{\delta H_z}{\delta y}\hat{x} - \frac{\delta H_z}{\delta x}\hat{y} = -i\,\varepsilon\,k_0\,(E_x\,\hat{x} + E_y\,\hat{y}) \qquad \text{Eq 49}$$

Considering $\frac{\delta H_z}{\delta y} = -i\,\varepsilon\,k_0\,E_x$ and $E_y = -\frac{i}{\varepsilon\,k_0}\frac{\delta H_z}{\delta x}$ resulted from Eq 47 and 48, the Eq 50 and 51 are written.

$$\frac{\delta H_z}{\delta y} = i\,\varepsilon\,k_0(-E_x) \qquad \text{Eq 50}$$

$$\frac{\delta(-E_x)}{\delta y} = \left(\frac{i}{k_0}\frac{\delta}{\delta x}\frac{1}{\varepsilon}\frac{\delta}{\delta x} + i\,k_0\right) H_z \qquad \text{Eq 51}$$

Adding y-dependence of $e^{i\tau y}$ to $H_z$ and $-E_x$, the new formula is written in the form of Eq 52 and 53.

$$H_z(x,y) = H_z(x)\,e^{i\tau y} \qquad \text{Eq 52}$$

$$-E_x(x,y) = -\zeta_x(x) = e^{i\tau y} \qquad \text{Eq 53}$$

If we name $\Psi_{TM}(x) = \begin{pmatrix} H_z(x) \\ -\zeta_x(x) \end{pmatrix} e^{i\tau y}$ where $\begin{pmatrix} H_z(x) \\ -\zeta_x(x) \end{pmatrix} = \Phi_{TM(x)}$, the Eq 54 is written in the form of :

$$\frac{\delta}{\delta y}\Psi_{TM}(x) = \begin{pmatrix} 0 & i\,\varepsilon\,k_0 \\ \frac{i}{k_0}\frac{\delta}{\delta x}\frac{1}{\varepsilon}\frac{\delta}{\delta x} + i\,k_0 & 0 \end{pmatrix} \begin{pmatrix} H_z(x) \\ -\zeta_x(x) \end{pmatrix} e^{i\tau y} \qquad \text{Eq 54}$$

Assuming $\begin{pmatrix} 0 & i\,\varepsilon\,k_0 \\ \frac{i}{k_0}\frac{\delta}{\delta x}\frac{1}{\varepsilon}\frac{\delta}{\delta x} + I\,k_0 & 0 \end{pmatrix} = L_{TM}(x)$, the Eq 55 is shown as:

$$i\,\tau\,\Phi_{TM}(x) = L_{TM}(x)\,\Phi_{TM}(x) \qquad \text{Eq 55}$$

The $\Phi(x)$ can be expanded as a Fourier-Floquet expansion and Fourier expand other periodic functions, as shown in Eq 56 and 57:

$$\text{Floquet Theorem: } \Phi(x) = \sum_m \Phi_m\,e^{ik_{xm}x} \qquad \text{Eq 56}$$

$$\text{Fourier Theorem: } \varepsilon(x) = \sum_m \Phi\varepsilon_m\,e^{iKx}$$

$$\eta(x) = \frac{1}{\varepsilon(x)} = \sum_m \eta_m\,e^{iKx} \qquad \text{Eq 57}$$

Where $k_{xm} = k_{x0} + mK$, $K = \frac{2\pi}{\Lambda}$ and $\Phi = \begin{pmatrix} h_{ml} \\ -\zeta_{ml} \end{pmatrix}$ with $m = -N,\ldots,N$ and $-\zeta_{ml} = \chi_{ml}$

For the first row of matrix $L_{TM}(x)$, the Eq 58 is written as below

$$i \tau_l \sum_m h_{ml} \ e^{ik_{xm}x} = i k_0 \sum_{p+n} \varepsilon_n \chi_{pl} \ e^{ik_{x,p+n}x} \qquad \text{Eq 58}$$

and we can simplify it to $\tau_l \ h_{ml} = k_0 \sum_p \varepsilon_{m-p} \ \chi_{pl}$, which leads to $\tau_l \ h_l = k_0 \ [\![\varepsilon]\!] \ \chi_l$ where $[\![\varepsilon]\!]_{mp} = \varepsilon_{m-p}$ is Toeplitz matrix and $h_l$ and $\chi_l$ are the column matrices. Then, for the second row of matrix $L_{TM}(x)$ is shown as Eq 59:

$$i \tau_l \sum_m \chi_{ml} \ e^{ik_{xm}x} = \frac{i}{k_0} \sum_n \frac{\delta}{\delta x} \eta_n e^{iKn} \frac{\delta}{\delta x} \sum_s h_{sl} e^{ik_{xs}x} + i k_0 \sum_p h_{pl} e^{ik_{xp}x} =$$

$$= \frac{-i}{k_0} \sum_{ns} \eta_n \ k_{xs} \ k_{x,n+s} \ h_{sl} \ e^{ik_{x,n+s}x} + i k_0 \sum_p h_{pl} e^{ik_{xp}x} \qquad \text{Eq 59}$$

Eq 59 can be written in the form of Eq 60 by using the Fourier theorem in, follows as:

$$\tau_l \ \chi_{ml} = (\frac{-1}{k_0} k_x \ [\![\eta]\!] \ k_x + k_0) \ h_l \qquad \text{Eq 60}$$

Where $k_x$ is diagonal matrices. Thus, we can construct Eq 58 and 59 in the form of Eq 61.

$$\tau_l \phi_l = \begin{pmatrix} 0 & k_0 \ [\![\varepsilon]\!] \\ k_0 - \frac{1}{k_0} k_x [\![\eta]\!] k_x & 0 \end{pmatrix} \phi_l \qquad \text{Eq 61}$$

Considering the matrix as $M_{TM}$, we have $\tau_l \phi_l = M_{TM} \phi_l$

In the patterned layer, the fields are calculated based on Eq 62 and 63, with A as the expansion coefficient.

$$H_z = \sum_{lm} A_l \ h_{ml} e^{ik_{xm}x} \ e^{i\tau_l y} \qquad \text{Eq 62}$$

$$-E_x = \sum_l A_l \ \chi_{ml} e^{ik_{xm}x} \ e^{i\tau_l y} \qquad \text{Eq 63}$$

## Fourier components of ε and 1/ε

The Fourier theorem for the alternative layers of dielectric 1 and 2 is shown in Figure 22.

| $\varepsilon_{21}$ | $\varepsilon_{22}$ | $\varepsilon_{21}$ | ... |

Figure 22. Schematic of alternative layers of dielectric 1 and 2

Base on Eq 57, the dielectric constant is expanded as Eq 64.

$$\varepsilon(x) = \sum_m \varepsilon_m\, e^{imKx} = \frac{1}{\Lambda}\int_0^\Lambda \varepsilon(x) e^{-isKx} dx = \frac{1}{\Lambda}\int_0^\Lambda (\sum_m \varepsilon_m\, e^{imKx})\, e^{-isKx} dx$$

$$\varepsilon(x) = \frac{1}{\Lambda}\int_0^w \varepsilon_{21}\, e^{-isKx} dx + \frac{1}{\Lambda}\int_w^\Lambda \varepsilon_{22}\, e^{-isKx} dx = \frac{\varepsilon_m}{\Lambda}\sum_m \int_0^\Lambda \sum_m e^{i(m-s)Kx} dx = \varepsilon_s$$

Eq 64

For $s \neq 0$, Eq 64 is written in the form of Eq 65.

$$\varepsilon_s = \varepsilon_{21}\frac{(e^{-isKw}-1)}{-isK\Lambda} + \varepsilon_{22}\frac{(e^{-isK\Lambda} - e^{-isKw})}{-isK\Lambda} = \frac{i\varepsilon_{21}}{2\pi s}(e^{-isKw} - 1) + \frac{i\varepsilon_{22}}{2\pi s}(1 - e^{-isKw})$$

$$\eta_s = \frac{i\frac{1}{\varepsilon_{21}}}{2\pi s}(e^{-isKw} - 1) + \frac{i\frac{1}{\varepsilon_{22}}}{2\pi s}(1 - e^{-isKw}) \qquad \text{Eq 65}$$

For $s \neq 0$, Eq 65 is written in the form of Eq 66.

$$\varepsilon_0 = \frac{w}{\Lambda}\varepsilon_{21} + \frac{(\Lambda-w)}{\Lambda}\varepsilon_{22}$$

$$\eta_0 = \frac{w}{\Lambda}\frac{1}{\varepsilon_{21}} + \frac{(\Lambda-w)}{\Lambda}\frac{1}{\varepsilon_{22}} \qquad \text{Eq 66}$$

## Toeplitz Matrix

Based on Eq 62 and 63, the expanded fields in x-dependent Fourier components of $e^{ik_{xm}x}$ with $m = -N, -N+1, ..., N$ can be used to form a Toeplitz matrices $[\![\varepsilon]\!]$ and $[\![\eta]\!]$ with 2N+1 rows and 2N+1 columns, in the form of Eq 67.

$$[\![\varepsilon]\!] = \begin{pmatrix} [\![\varepsilon]\!]_{-N,-N} & [\![\varepsilon]\!]_{-N,-N+1} & [\![\varepsilon]\!]_{-N,-N+2} & \cdots & [\![\varepsilon]\!]_{-N,N} \\ [\![\varepsilon]\!]_{-N+1,-N} & [\![\varepsilon]\!]_{-N+1,-N+1} & [\![\varepsilon]\!]_{-N+1,-N+2} & \cdots & [\![\varepsilon]\!]_{-N+1,N} \\ [\![\varepsilon]\!]_{-N+2,-N} & [\![\varepsilon]\!]_{-N+2,-N+1} & [\![\varepsilon]\!]_{-N+2,-N+2} & \cdots & [\![\varepsilon]\!]_{-N+2,N} \\ [\![\varepsilon]\!]_{-N+3,-N} & [\![\varepsilon]\!]_{-N+3,-N+1} & [\![\varepsilon]\!]_{-N+3,-N+2} & \cdots & [\![\varepsilon]\!]_{-N+3,N} \\ [\![\varepsilon]\!]_{-N+4,-N} & [\![\varepsilon]\!]_{-N+4,-N+1} & [\![\varepsilon]\!]_{-N+4,-N+2} & \cdots & [\![\varepsilon]\!]_{-N+4,N} \\ [\![\varepsilon]\!]_{-N+5,-N} & [\![\varepsilon]\!]_{-N+5,-N+1} & [\![\varepsilon]\!]_{-N+5,-N+2} & \cdots & [\![\varepsilon]\!]_{-N+5,N} \\ \cdots & \cdots & \cdots & \cdots & \cdots \\ [\![\varepsilon]\!]_{N-1,-N} & [\![\varepsilon]\!]_{N-1,-N+1} & [\![\varepsilon]\!]_{N-1,-N+2} & \cdots & [\![\varepsilon]\!]_{N-1,N} \\ [\![\varepsilon]\!]_{N,-N} & [\![\varepsilon]\!]_{N,-N+1} & [\![\varepsilon]\!]_{N,-N+2} & \cdots & [\![\varepsilon]\!]_{N,N} \end{pmatrix} \quad \text{Eq 67}$$

Since $[\![\varepsilon]\!]_{mp} = \varepsilon_{m-p}$, then Eq 67 is written as Eq 68.

$$[\![\varepsilon]\!] = \begin{pmatrix} \varepsilon_0 & \varepsilon_{-1} & \varepsilon_{-2} & \cdots & \varepsilon_{-2N} \\ \varepsilon_1 & \varepsilon_0 & \varepsilon_{-1} & \cdots & \varepsilon_{-2N+1} \\ \varepsilon_2 & \varepsilon_1 & \varepsilon_0 & \cdots & \varepsilon_{2N+2} \\ \cdots & \cdots & \cdots & \cdots & \cdots \\ \varepsilon_{2N} & \varepsilon_{2N-1} & \varepsilon_{2N-2} & \cdots & \varepsilon_0 \end{pmatrix}$$

Eq 68

The 4N+1 Fourier components of $\varepsilon$ and $\eta$ is required including $e^{-i2NK} \ldots\ldots e^{i2NK}$ components. For any unpatterned layer we have $\varepsilon_m = \varepsilon\, \delta_{m0}$ and $\eta_m = \eta\, \delta_{m0}$.

For the TM polarization, we have $\tau \phi_{TM} = \begin{pmatrix} 0 & k_0\, \varepsilon \\ k_0 - \frac{kx^2}{\varepsilon\, k_0} & 0 \end{pmatrix} \phi_{TM}$.

Considering $\tau H_{zm} = -k_0\, \varepsilon\, \zeta_{xm}$ and $-\tau \zeta_{xm} = (k_0 - \frac{kx^2}{\varepsilon\, k_0})\, H_{zm}$, the Eq 69 is acquired.

$$\tau = \pm\sqrt{\varepsilon\, k_0^2 - k_m^2}$$

$$-\zeta_{xm} = \frac{\tau}{\varepsilon\, k_0}\, H_{zm} \qquad \text{Eq 69}$$

By applying the boundary conditions at y=0, the fields for TM polarization are shown in Eq 70 and 71.

$$H_z = e^{ik_{x0}x} + \sum_n r_n\, e^{ik_{xm}x} = \sum_{lm} h_{ml}\, A_l\, e^{ik_{xm}x}$$

$$r_n - \sum_l h_{nl}\, A_l = -\delta_{n0} \qquad \text{Eq 70}$$

$$-E_x = \frac{\beta_0}{\varepsilon_1\, k0}\, e^{ik_{x0}x} - \frac{1}{\varepsilon_1\, k0} \sum_n \beta_n r_n\, e^{ik_{xm}x} = \sum_{lm} \chi_{ml}\, A_l\, e^{ik_{xm}x}$$

$$-\frac{\beta_n}{\varepsilon_1 \, k0} r_n - \sum_l \chi_{nl} A_l = -\frac{\beta_n}{\varepsilon_1 \, k0} \delta_{n0} \qquad \text{Eq 71}$$

By applying the boundary conditions at y = -t, the fields for TM polarization are shown in Eq 72 and 73.

$$H_z = \sum_{lm} h_{ml} \, e^{-i\tau_l t} A_l e^{ik_{xm}x} = \sum_n t_n \, e^{ik_{xn}x}$$

$$\sum_l h_{nl} \, \gamma_l \, A_l - t_n = 0 \quad \text{where} \quad \gamma_l = e^{-i\tau_l t} \qquad \text{Eq 72}$$

$$-E_x = \sum_{lm} \chi_{ml} \gamma_l \, A_l e^{ik_{xm}x} = \frac{1}{\varepsilon_3 \, k0} \sum_n \tilde{\beta}_n \, t_n \, e^{ik_{xn}x}$$

$$\sum_l \chi_{nl} \gamma_l \, A_l - \frac{\tilde{\beta}_n}{\varepsilon_3 \, k0} t_n = 0 \qquad \text{Eq 73}$$

The unknown expansion coefficients are shown in Eq 74, where l = 1,2,…,4N+2.

$$\Phi = \begin{pmatrix} r_{-N} \\ .. \\ .. \\ r_N \\ A_l \\ t_{-N} \\ .. \\ .. \\ t_N \end{pmatrix} \qquad \text{Eq 74}$$

The initial beam matrix is shown as Eq 75

$$L = \begin{pmatrix} 0 \\ .. \\ .. \\ -1 \\ .. \\ .. \\ 0 \\ 0 \\ .. \\ .. \\ \frac{-\beta_0}{\varepsilon_1 k_0} \\ .. \\ .. \\ 0 \\ 0 \\ .. \\ .. \\ 0 \\ 0 \\ .. \\ .. \\ 0 \end{pmatrix} \begin{array}{l} \text{Row 1} \\ \\ \\ \text{The } (N+1)^{th} \text{row} \\ \\ \\ 2N+1 \\ 2N+2 \\ \\ \\ 3N+2 \\ \\ \\ 4N+2 \\ 4N+3 \\ \\ 6N+3 \\ 6N+4 \\ \\ 8N+4 \end{array}$$

Eq 75

In addition, the coupling matrix is shown in Eq 76.

$$M = \begin{pmatrix} 1 & -[h] & 0 \\ \frac{-\beta_0}{\varepsilon_1 k_0} & -[\chi] & 0 \\ 0 & -[h\gamma] & -1 \\ 0 & -[\chi\gamma] & \frac{\tilde{\beta}_n}{\varepsilon_3 k0} \end{pmatrix}$$

Eq 76

With $[h]_{nl} = h_{nl}$; $[h\gamma]_{nl} = h_{nl}\gamma_l$; $[\chi]_{nl} = \chi_{nl}$; $[\chi\gamma]_{nl} = \chi_{nl}\gamma_l$.

And finally, the $\Phi_{TM}$ is calculated as Eq 77.

$$\Phi_{TM} = M_{TM}^{-1} L_{TM} \quad \text{Eq 77}$$

## Resonant Subwavelength Grating Filters

After discussing the physics of Bragg stacks and dielectric grating in last sections, the resonant subwavelength grating filters is explained in this part. The resonant subwavelength grating filters combine gratings with resonant Bragg Stacks. For TM polarization, let us look at 7-layer structure:

Superstrate $\quad\quad\quad \varepsilon_1$

Layer 2 $\quad\quad\quad \varepsilon_2$

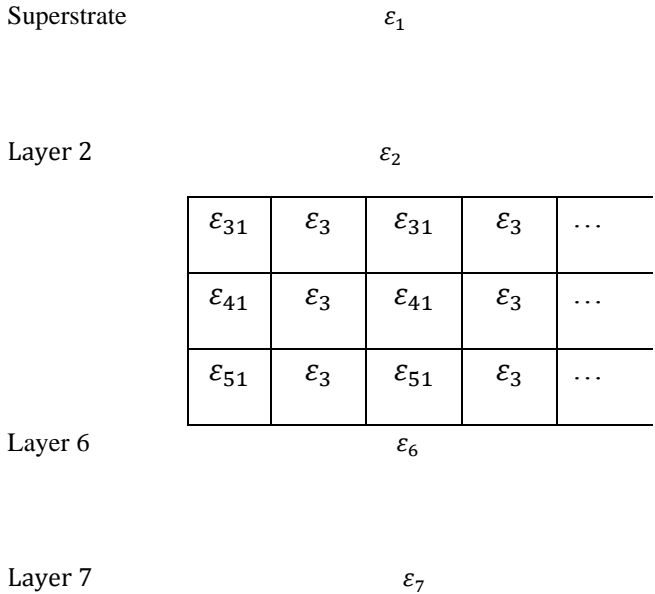

Layer 6 $\quad\quad\quad \varepsilon_6$

Layer 7 $\quad\quad\quad \varepsilon_7$

**Electric and Magnetic Fields in Bragg Stack for TM Polarization**

Assuming $e^{i(k_{x0}x+\beta_0 y)}$ for downward propagating, the electric and magnetic fields in superstrate layer is shown in Eq 78 and 79.

$$H_z = e^{i(k_{x0}x+\beta_0 y)} + \sum_n r_n e^{i(k_{xn}x-\beta_n y)} \qquad \text{Eq 78}$$

$$E_x = \frac{i}{k_0 \varepsilon_1} \frac{\delta H_z}{\delta y} = \frac{-\beta_0}{k_0 \varepsilon} e^{i(k_{x0}x+\beta_0 y)} + \frac{1}{k_0 \varepsilon_1} \sum_n r_n \beta_n e^{i(k_{xn}x-\beta_n y)} \qquad \text{Eq 79}$$

Where $\beta_n = \sqrt{\varepsilon_1 k_0^2 - k_{xn}^2}$ and $k_{xn} = \sqrt{\varepsilon_1} k_0 \sin(\theta) + nK$, $K = \frac{2\pi}{\lambda}$, n = -N,…,0,…,N.

The electric and magnetic fields in unpatented layers (s=2,6) are as Eq 80 and 81.

$$H_{z,s} = \sum_n C_{s,n} e^{i(k_{xn}x+\beta_{s,n}(y+t_s))} + \sum_n D_{s,n} e^{i(k_{xn}x-\beta_{s,n}(y+t_s))} \qquad \text{Eq 80}$$

$$E_{x,s} = -\frac{1}{k_0 \varepsilon_s} \sum_n \beta_{s,n} C_{s,n} e^{i(k_{xn}x+\beta_{s,n}(y+t_s))} + \frac{1}{k_0 \varepsilon_s} \sum_n \beta_{s,n} D_{s,n} e^{i(k_{xn}x-\beta_{s,n}(y+t_s))} \qquad \text{Eq 81}$$

The electric and magnetic fields in substrate layer (s=7) are as Eq 82 and 83.

$$H_{z,s} = \sum_n t_n e^{i(k_{xn}x+\beta_{s,n}(y+t_s))} \qquad \text{Eq 82}$$

$$E_{x,s} = \frac{i}{k_0 \varepsilon_s} \frac{\delta H_z}{\delta y} = -\frac{1}{k_0 \varepsilon_s} \sum_n \beta_{s,n} t_n e^{i(k_{xn}x+\tilde{\beta}_n(y+t_s))} \qquad \text{Eq 83}$$

The electric and magnetic fields in patented layers ($l$=3,4,5) are as Eq 84 and 85.

$$H_z = \sum_{lm} A_{s,l} h_{ml} e^{ik_{xm}x} e^{i\tau_l y} \qquad \text{Eq 84}$$

$$-E_x = \sum_{lm} A_{s,l} \chi_{ml} e^{ik_{xm}x} e^{i\tau_l y} \qquad \text{Eq 85}$$

**Boundary Conditions**

By applying the boundary conditions at y = 0, the electric and magnetic fields are as Eq 86 and 87.

$$H_z: r_n - \gamma_{2,n} C_{2,n} - \gamma_{2,n}^{-1} D_{2,n} = -\delta_{n0} \qquad \text{Eq 86}$$

$$E_x: \frac{\beta_{1,n}}{\varepsilon_1 k_0} r_n + \frac{\beta_{2,n}}{\varepsilon_2 k_0} \gamma_{2,n} C_{2,n} - \frac{\beta_{2,n}}{\varepsilon_2 k_0} \gamma_{2,n}^{-1} D_{2,n} = \frac{\beta_n}{\varepsilon_1 k_0} \delta_{n0} \qquad \text{Eq 87}$$

By applying the boundary conditions at y = $-t_2$, the electric and magnetic fields are as Eq 88 and 89.

$$H_z: C_{2,n} + D_{2,n} - \sum_l h_{nl} \gamma_{3,n} A_{3,l} = 0 \qquad \text{Eq 88}$$

$$E_x: -\frac{\beta_{2,n}}{\varepsilon_2 k_0} C_{2,n} + \frac{\beta_{2,n}}{\varepsilon_2 k_0} D_{2,n} + \sum_l \chi_{nl} \gamma_{3,n} A_{3,l} = 0 \qquad \text{Eq 89}$$

By applying the boundary conditions at y = $-t_3$ and the electric and magnetic fields are as Eq 90.

$$H_z: \sum_l h_{nl} A_{3,l} - \sum_l h_{nl} \gamma_{4,n} A_{4,l} = 0$$

$$E_x: -\sum_l \chi_{nl} A_{3l} + \sum_l \chi_{nl} \gamma_{4,n} A_{4,l} = 0 \qquad \text{Eq 90}$$

By applying the boundary conditions at y = $-t_4$, the electric and magnetic fields are as Eq 91.

$$H_z: \sum_l h_{nl} A_{4,l} - \sum_l h_{nl} \gamma_{5,n} A_{5,l} = 0$$

$$E_x: -\sum_l \chi_{nl} A_{4l} + \sum_l \chi_{nl} \gamma_{5,n} A_{5,l} = 0 \qquad \text{Eq 91}$$

By applying the boundary conditions at y = $-t_5$, the electric and magnetic fields are as Eq 92.

$$H_z: \sum_l h_{nl} A_{5,l} - \gamma_{6,n} C_{6,n} - \gamma_{6,n}^{-1} D_{6,n} = 0$$

$$E_x: -\sum_l \chi_{nl} A_{5,l} + \frac{\beta_{6,n}}{\varepsilon_6 k_0} \gamma_{6,n} C_{6,n} - \frac{\beta_6}{\varepsilon_6 k_0} \gamma_{6,n}^{-1} D_{6,n} = 0 \qquad \text{Eq 92}$$

By applying the boundary conditions at y = $-t_6$, the electric and magnetic fields are as Eq 93.

$$H_z: C_{6,n} + D_{6,n} - t_n = 0$$

$$E_x: -\frac{\beta_{6,n}}{\varepsilon_6 k_0} C_{6,n} + \frac{\beta_{6,n}}{\varepsilon_6 k_0} D_{6,n} + \frac{\beta_{7,n}}{\varepsilon_7 k_0} t_n = 0 \qquad \text{Eq 93}$$

The unknown expansion coefficients are shown as Eq 94.

$$\Phi = \begin{pmatrix} r_{-N} \\ .. \\ r_N \\ C_{2,-N} \\ .. \\ C_{2,N} \\ D_{2,-N} \\ .. \\ D_{2,N} \\ A_{3l} \\ A_{4l} \\ A_{5l} \\ C_{6,-N} \\ .. \\ C_{6,N} \\ D_{6,-N} \\ .. \\ D_{6,N} \\ t_{-N} \\ .. \\ t_N \end{pmatrix} \quad \text{Eq 94}$$

Furthermore, the initial beam column matrix (L) is shown as Eq 95.

$$\begin{pmatrix} 0 \\ .. \\ -1 \\ .. \\ 0 \\ 0 \\ .. \\ \dfrac{-\beta_0}{\varepsilon_1 k_0} \\ .. \\ 0 \\ 0 \\ .. \\ 0 \\ 0 \\ .. \\ 0 \\ 0 \\ .. \\ 0 \\ 0 \\ .. \\ 0 \\ 0 \\ .. \\ 0 \\ 0 \\ .. \\ 0 \\ 0 \\ .. \\ 0 \\ 0 \\ .. \\ 0 \\ 0 \\ .. \\ 0 \\ 0 \\ .. \\ 0 \end{pmatrix}$$

Eq 95

And the coupling matrix is shown in Eq 96.

$M =$

$$\begin{pmatrix}
1 & -\gamma_2 & -\gamma_2^{-1} & 0 & 0 & 0 & 0 & 0 & 0 \\
\frac{\beta_{1,n}}{\varepsilon_1 k_0} & \frac{\beta_{2,n}}{\varepsilon_2 k_0}\gamma_2 & -\frac{\beta_{2,n}}{\varepsilon_2 k_0}\gamma_2^{-1} & 0 & 0 & 0 & 0 & 0 & 0 \\
0 & 1 & 1 & -[h_3\gamma_3] & 0 & 0 & 0 & 0 & 0 \\
0 & -\frac{\beta_{2,n}}{\varepsilon_2 k_0} & \frac{\beta_{2,n}}{\varepsilon_2 k_0} & [\chi_3\gamma_3] & 0 & 0 & 0 & 0 & 0 \\
0 & 0 & 0 & [h_3] & -[h_4\gamma_4] & 0 & 0 & 0 & 0 \\
0 & 0 & 0 & -[\chi_3] & [\chi_4\gamma_4] & 0 & 0 & 0 & 0 \\
0 & 0 & 0 & 0 & [h_4] & -[h_5\gamma_5] & 0 & 0 & 0 \\
0 & 0 & 0 & 0 & -[\chi_4] & [\chi_5\gamma_5] & -\gamma_6 & -\gamma_6^{-1} & 0 \\
0 & 0 & 0 & 0 & 0 & [h_5] & \frac{\beta_{6,n}}{\varepsilon_6 k_0}\gamma_{6,n} & -\frac{\beta_6}{\varepsilon_6 k_0}\gamma_{6,n}^{-1} & 0 \\
0 & 0 & 0 & 0 & 0 & -[\chi_5] & 1 & 1 & -1 \\
0 & 0 & 0 & 0 & 0 & 0 & -\frac{\beta_{6,n}}{\varepsilon_6 k_0} & \frac{\beta_{6,n}}{\varepsilon_6 k_0} & \frac{\beta_{7,n}}{\varepsilon_7 k_0}
\end{pmatrix}$$

<div align="center">Eq 96</div>

Thus, $\Phi_{TM}$ is written as Eq 97

$$\Phi_{TM} = M_{TM}^{-1} L_{TM} \qquad \text{Eq 97}$$

Some parts of the MATLAB code are written below to show the calculation of optical and radiative design of dielectric grating nanostructure based on implementing RCWA.

**Transfer Matrix Method**

The other mathematical modeling of anti-reflective subwavelength structures method for the periodic differential Equations can be expanded using transfer matrix method. In this section a general way to calculate any EM field in dielectric layers of the structures, as shown in Figure 23, is described [22].

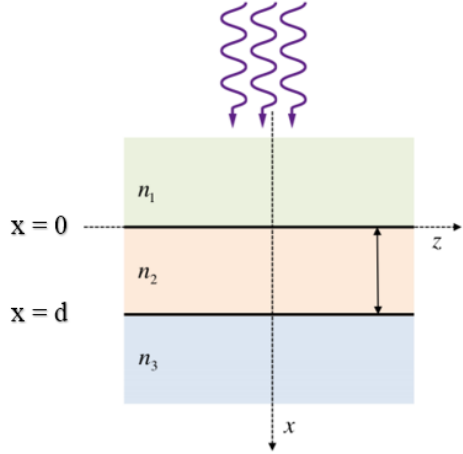

Figure 23 Three-layer material with index of refraction layers [22]

Considering $k_z \approx \beta$ (phase continuity), the electric field for each layer can be written as $\vec{E} = \vec{E}(x)e^{(\omega t - \beta z)}$. Therefore, the boundary conditions for the forward and backward propagation in each three layers of the structure for TM polarization is written as Eq 98.

$$x<0 \quad E = E_1 e^{i\beta_{1x}x} + E_1' e^{-i\beta_{1x}x}$$

$$0<x<d \quad E = E_2 e^{i\beta_{2x}x} + E_2' e^{-i\beta_{2x}x}$$

$$x>d \quad E = E_3 e^{i\beta_{3x}(x-d)} + E_3' e^{-i\beta_{3x}(x-d)} \quad \text{Eq 98 [22]}$$

where d is the thickness of middle layer. Then, for writing the amplitude electric field from layer 1 to 3, the propagation from interface 1 to 2 and interface 2 to 3 in addition to propagation in layer 2 should be calculated. The propagation in homogenous layer 2 is shows in Eq 99.

$$P_2 = \begin{pmatrix} e^{-i\beta_{2x}x} & 0 \\ 0 & e^{i\beta_{2x}x} \end{pmatrix} \quad \text{Eq 99 [22]}$$

Then the equations should be re-written to form the matrix of propagation from interface 1 to 2.

$$H_{1x} + H_{1x}' = H_{2x} + H_{2x}'$$

$$E_{1z} + E_{1z}' = E_{2z} + E_{2z}' \quad \text{changed to} \quad -\frac{\beta_1}{\varepsilon_1} H_{1x} + \frac{\beta_1}{\varepsilon_1} H_{1x}' = -\frac{\beta_2}{\varepsilon_2} H_{2x} + \frac{\beta_2}{\varepsilon_2} H_{2x}'$$

Therefore, the matrixes are written as Eq 100.

$$\begin{pmatrix} 1 & 1 \\ -\frac{\beta_1}{\varepsilon_1} & \frac{\beta_1}{\varepsilon_1} \end{pmatrix} \begin{pmatrix} H_{1x} \\ H'_{1x} \end{pmatrix} = \begin{pmatrix} 1 & 1 \\ -\frac{\beta_2}{\varepsilon_2} & \frac{\beta_2}{\varepsilon_2} \end{pmatrix} \begin{pmatrix} H_{2x} \\ H'_{2x} \end{pmatrix} \quad \text{»»»} \quad D_1 \begin{pmatrix} H_{1x} \\ H'_{1x} \end{pmatrix} = D_2 \begin{pmatrix} H_{2x} \\ H'_{2x} \end{pmatrix} \quad \text{»»»}$$

$$D_{12} \begin{pmatrix} H_{1x} \\ H'_{1x} \end{pmatrix} = \begin{pmatrix} H_{2x} \\ H'_{2x} \end{pmatrix} \quad \text{Eq 100 [22]}$$

Where $D_{12}$ is $D_2^{-1} D_1$. Then the equations should be re-written to form the matrix of propagation from interface 2 to 3.

$$D_{23} \begin{pmatrix} H_{2x} \\ H'_{2x} \end{pmatrix} = \begin{pmatrix} H_{3x} \\ H'_{3x} \end{pmatrix} \quad \text{Eq 101 [22]}$$

For make the connection between layer 1 and layer 3, the Eq 102 should be followed.

$$D_{23} \begin{pmatrix} H_{2x} \\ H'_{2x} \end{pmatrix} = D_{23} P_2 D_{12} \begin{pmatrix} H_{1x} \\ H'_{1x} \end{pmatrix} = \begin{pmatrix} H_{3x} \\ H'_{3x} \end{pmatrix} \quad \text{Eq 102 [22]}$$

The TM transmission of proposed Bragg stack is illustrated in Figure 24. The complete code is referred in Appendix A.

```
%Constructing the TM Transfer Matrices
M11 = 1;
M12 = 1;
M21 = -betaTM(cntlayers)/Epsilon_xy(cntlayers);
M22 = betaTM(cntlayers)/Epsilon_xy(cntlayers);
M_TM(cntlayers).matrix = [M11,M12;M21,M22];

inverseM11 = 1/2;
inverseM12 = -Epsilon_xy(cntlayers)/2/betaTM(cntlayers);
inverseM21 = 1/2;
inverseM22 = Epsilon_xy(cntlayers)/2/betaTM(cntlayers);
inverseM_TM(cntlayers).matrix = [inverseM11,inverseM12;inverseM21,inverseM22];

phiTM_component = exp(1i*betaTM(cntlayers)*c(cntlayers));
inversephiTM_component = exp(-1i*betaTM(cntlayers)*c(cntlayers));
phiTM(cntlayers).matrix = [inversephiTM_component,0;0,phiTM_component];

tempTM = eye(2,2);

for cntlayers = 2:Layers-1
    TtildaTM(cntlayers).matrix = 
inverseM_TM(cntlayers).matrix*M_TM(cntlayers+1).matrix*phiTM(cntlayers+1).matrix;
    tempTM = tempTM*TtildaTM(cntlayers).matrix;
end
```

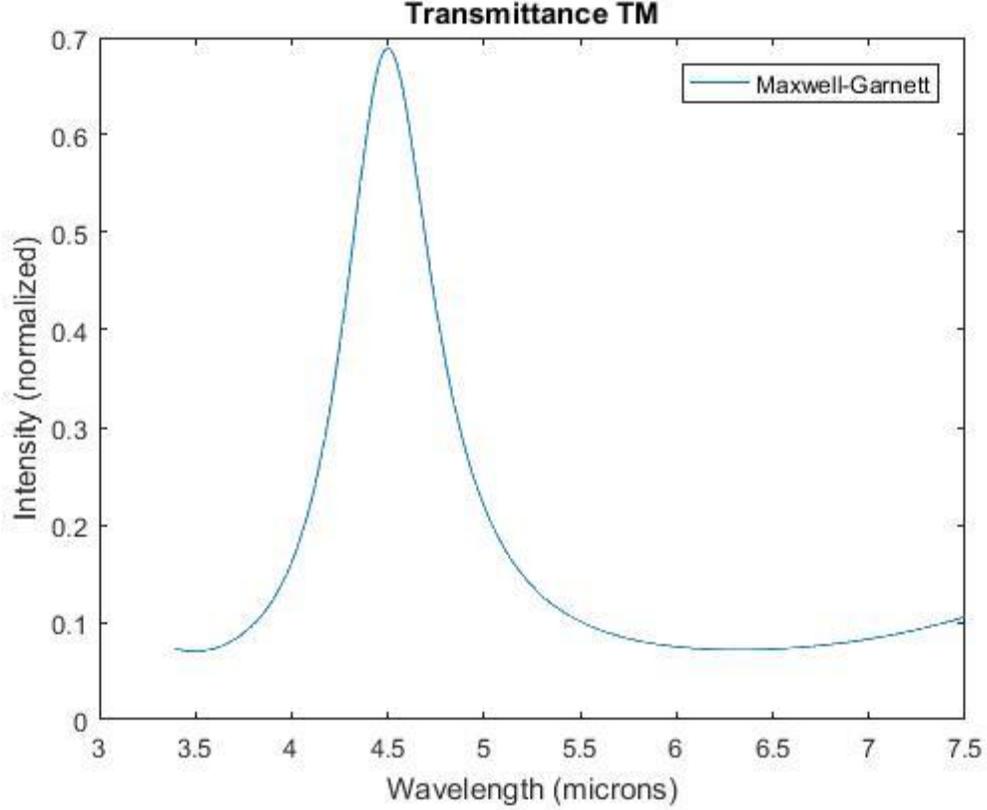
Figure 24. TM Transmittance of Proposed HMBS simulated by MATLAB

## Angle Independency for RCWA and Transfer Matrix Methods

In Fourier model/RCWA theory, it is obvious that the eta used in electric and magnetic fields makes low shift wavelength for very big permittivity of the grating layer on TM polarized angle of MWIR regime. The permittivity of copper in the grating layer of the HMBS at center wavelength 4μm is shown in Figure 25. The high dielectric constant copper in middle resonant layer of HMBS keeps the transmission properties of the proposed filter constant for TM oblique and normal incidence, as shown in Eq 103.

$$\tau^{TM} = \begin{pmatrix} 0 & k_0\,\varepsilon \\ k_0 - \frac{kx^2}{\varepsilon\,k_0} & 0 \end{pmatrix} \quad \text{Eq 103}$$

The angle independency satisfies the Maxwell-Garnett/transfer matrix method; the ratio $\frac{\varepsilon_{xy}}{\varepsilon_{zz}}$ minimize the impact of angle in Beta (~$k_z$) used in TM coupling matrix, as shown in Eq 104.

$$\beta_{TM} = \sqrt{\varepsilon_{xy} k_0^2 - \frac{\varepsilon_{xy}}{\varepsilon_{zz}} k_{xn}^2} \quad \text{Eq 104}$$

However, this condition is not applicable for TE polarization.

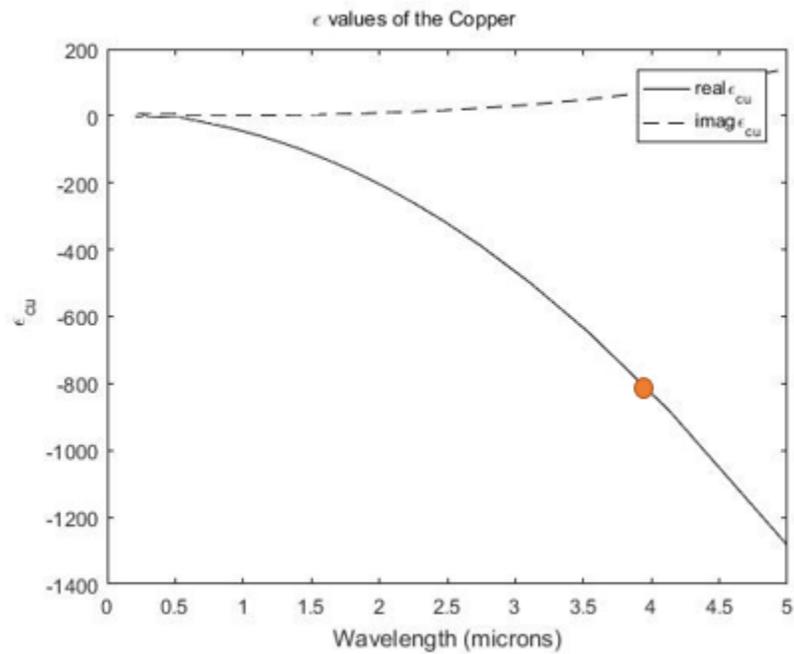

Figure 25. Permittivity of copper at center-wavelength 4 Micrometer

**Surface Plasmon**

SP is a well-known method to measure the absorption of material onto the surface of metal nanoparticles or planar metal [12]. The electromagnetic fields associated to the SP are limited to the metal-dielectric interface, evanescently decrease from the interface as Figure 26; however the field at interface is so strong.

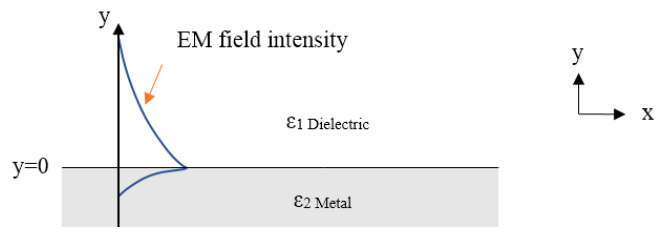

Figure 26. EM Fields of Surface Plasmons

**Solving for the EM Fields of Surface Plasmons**

The Maxwell Equations should be used to solve the filed for SP, as Eq 105.

$$\vec{\Delta} \times \vec{E} = -\frac{\delta \vec{B}}{\delta t} = i\omega\mu\vec{H}$$

$$\vec{\Delta} \times \vec{H} = \frac{\delta \vec{D}}{\delta t} = -i\omega\varepsilon\vec{E}$$

$$\vec{\Delta} \cdot \vec{D} = 0$$

$$\vec{\Delta} \cdot \vec{B} = 0$$

Eq 105

Where $\vec{B} = \mu\vec{H}$ and $\vec{D} = \varepsilon\vec{E}$.

**Region 1: Dielectric with $\varepsilon_1$**

The $E_y$ field component is necessary in order to have the surface plasmon yields the TM polarized field in Eq 106.

$$E_x^1 = A_1\, e^{ik_x x}\, e^{-\alpha_1 y} \qquad E_y^1 = B_1\, e^{ik_x x}\, e^{-\alpha_1 y} \qquad H_z^1 = C_1\, e^{ik_x x}\, e^{-\alpha_1 y} \qquad \text{Eq 106}$$

where $k^2 = \varepsilon_1 \frac{\omega^2}{c^2} = k_x^2 - \alpha_1^2$ and then: $\alpha_1^2 = k_x^2 - \varepsilon_1 \frac{\omega^2}{c^2}$.

The Maxwell's equations are reminded in Eq 107.

$$\vec{\Delta} \times \vec{H} = -i\omega\varepsilon\vec{E}$$

$$\begin{vmatrix} i & j & k \\ \frac{\delta}{\delta x} & \frac{\delta}{\delta y} & 0 \\ 0 & 0 & Hz \end{vmatrix} = -i\omega\varepsilon\vec{E}$$

$$E_x = \frac{i}{\omega\varepsilon}\frac{\delta H_z}{\delta y} \quad \text{and} \quad E_y = -\frac{i}{\omega\varepsilon}\frac{\delta H_z}{\delta x} \qquad \text{Eq 107}$$

The coefficients in Eq 107 can be easily found based on Eq 106, as written in Eq 108.

$$A_1 = -\frac{i\alpha_1 C_1}{\omega\varepsilon_1} \quad \text{and} \quad B_1 = \frac{k_x C_1}{\omega\varepsilon_1} \qquad \text{Eq 108}$$

**Region 2: Metal with $\varepsilon_2$**

The formula in region 1 can be rewritten as Eq 109

$$E_x^2 = A_2\, e^{ik_x x}\, e^{\alpha_2 y}$$

$$E_y^2 = B_2\, e^{ik_x x}\, e^{\alpha_2 y}$$

$$H_z^2 = C_2\, e^{ik_x x}\, e^{\alpha_2 y} \qquad \text{Eq 109}$$

Where $\alpha_2^2 = k_x^2 - \varepsilon_2$ and $\frac{\omega^2}{c^2} = k_x^2 - \varepsilon_2\, k_0^2$.

Also, the coefficients can be easily found, as written in $A1 = \frac{i\,\alpha 2\, C2}{\omega\,\varepsilon 2}$ and $B1 = \frac{k_x C2}{\omega\,\varepsilon 2}$ q 110.

$$A_1 = \frac{i\,\alpha 2\, C2}{\omega\,\varepsilon 2} \quad \text{and} \quad B_1 = \frac{k_x C2}{\omega\,\varepsilon 2} \qquad \text{Eq 110}$$

**Dispersion Relation**

In order to find the dispersion relation, first we start with boundary condition.

At $y = 0$ for $H_z$, the Eq 111 shows $C_1 = C_2 = 1$.

$$C_1\, e^{ik_x x} = C_2\, e^{ik_x x} \qquad \text{Eq 111}$$

For $E_x$ at $y = 0$:

$$A_1\, e^{ik_x x} = A_2\, e^{ik_x x}$$

$$\frac{\alpha_1^{\,2}}{\varepsilon 1^2} = -\frac{\alpha_2^{\,2}}{\varepsilon 2^2}$$

$$\frac{1}{\varepsilon 1^2}\left(k_x^2 - \varepsilon_1 \frac{\omega^2}{c^2}\right) = -\frac{1}{\varepsilon 2^2}\left(k_x^2 - \varepsilon_2 \frac{\omega^2}{c^2}\right) \text{ and therefore: } k_x = \frac{\omega}{c}\left(\frac{\varepsilon 1 \varepsilon 2}{\varepsilon 1 + \varepsilon 2}\right)^{\frac{1}{2}}$$

Using Drude model, the permittivity of metal is mentioned as Eq 112.

$$\varepsilon_m(\omega) = \varepsilon_\infty - \frac{\omega_p^{\,2}}{\omega^2} \qquad \text{Eq 112}$$

The decay constant is calculated as Eq 113:

$$\alpha_1^2 = -\frac{\varepsilon_1^2}{\varepsilon_1 + \varepsilon_m}\, k_0^2 \qquad \text{Eq 113}$$

The permittivity of metal at frequencies less than $\omega_p$ is negative, then for $|\varepsilon_m| \gg |\varepsilon_1|$ and $\text{Re}(\varepsilon_m) < 0$, decay constants are written as Eq 114:

$$\alpha_1 = \frac{\varepsilon_1}{\sqrt{|\varepsilon_1 + \varepsilon_m|}} \frac{\omega}{c} \quad \text{and} \quad \alpha_2 = \frac{|\varepsilon_m|}{\sqrt{|\varepsilon_1 + \varepsilon_m|}} \frac{\omega}{c} \qquad \text{Eq 114}$$

Considering $|\varepsilon_m| \gg |\varepsilon_1|$, it is obvious that $\alpha_2 \gg \alpha_1$. For higher frequency where $|\varepsilon_m|$ is no longer much larger than $|\varepsilon_1|$ we have $|\varepsilon_m| \approx |\varepsilon_1|$. Therefore, $k_x$ goes to $\infty$, as Eq 115.

$$k_x = \frac{\omega}{c} \left( \frac{\varepsilon_1 \varepsilon_m}{\varepsilon_1 + \varepsilon_m} \right)^{\frac{1}{2}} \qquad \text{Eq 115}$$

In $\varepsilon_1 \frac{\omega^2}{c^2} = k_x^2 + k_y^2$, when $k_y = 0$, the $k_x$ is maximum, as Eq 116.

$$k_{x,max} = \frac{\omega}{c} \sqrt{\varepsilon_1} \qquad \text{Eq 116}$$

In lower frequencies for $|\varepsilon_m| \gg |\varepsilon_1|$, as Eq 117, the SP mode is bigger than the light mode propagation. Therefore, the incident or emitted propagating light, can't bound with SP modes.

$$k_{x,SP} = \frac{\omega}{c} \left( \frac{\varepsilon_1 \varepsilon_m}{\varepsilon_1 + \varepsilon_m} \right)^{\frac{1}{2}} > k_x = \frac{\omega}{c} \sqrt{\varepsilon_1} = k_{x,max} \qquad \text{Eq 117}$$

**Light Cone**

In Eq 117, $k_{x,max}$ determines the light cone. Only the modes within the light cone can couples with propagating light. However, SP modes of flat interfaces don't bound with the light cone, since they reside outside of the light cone. On the other hand, SP never intersect the light cone, as Figure 27.

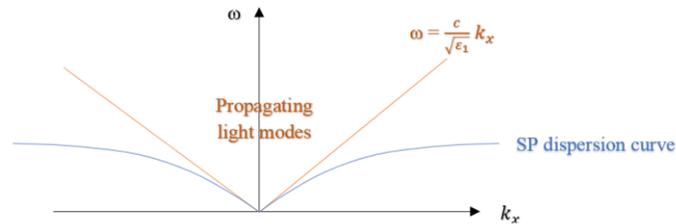

Figure 27. Light Cone and Surface Plasmon Dispersion Curve

Figure 28 shows how flow of SP modes works. In this section the power flow of SP in two regions dielectric and metal are discussed. Then, we explain how SP can couple to radiating light, let us look at power flow of SP modes.

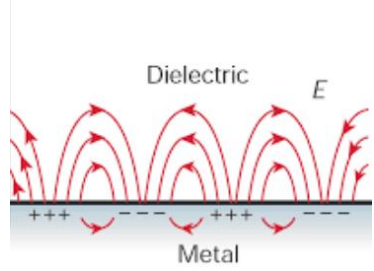

Figure 28. Surface Plasmon [23]

The field in Dielectric region is calculated as Eq 118:

$$E^1_X = -\frac{i}{\omega \varepsilon_1} \alpha_1 e^{ik_x x} e^{-\alpha_1 y}, \quad E^1_y = \frac{k_x}{\omega \varepsilon_1} e^{ik_x x} e^{-\alpha_1 y}, \quad H^1_z = e^{ik_x x} e^{-\alpha_1 y}$$

$$\vec{S^1} = \vec{E^1} \times \vec{H^{1*}} = -i\omega\varepsilon \vec{E} = \begin{vmatrix} i & j & k \\ E^1_X & E^1_y & 0 \\ 0 & 0 & H^{1*}_z \end{vmatrix} = E^1_y H^{1*}_z \hat{x} - E^1_X H^{1*}_z \hat{y}$$

$$= \frac{k_x}{\omega \varepsilon_1} e^{-2\alpha_1 y} \hat{x} + \frac{i}{\omega \varepsilon_1} \alpha_1 e^{-2\alpha_1 y} \hat{y} \qquad \text{Eq 118}$$

Since $\frac{k_x}{\omega \varepsilon_1}$ is real and $\frac{i}{\omega \varepsilon_1} \alpha_1$ is imaginary part of $\vec{S^1}$, the $\vec{S^1}$ is written as Eq 119.

$$S^1_x = \frac{k_x}{\omega \varepsilon_1} e^{-2\alpha_1 y} \qquad \text{Eq 119}$$

The field in Metal region is calculated as follow Eq 120.

$$\vec{S^2} = \frac{k_x}{\omega \varepsilon_2} e^{-2\alpha_2 y} \hat{x} - \frac{i}{\omega \varepsilon_1} \alpha_2 e^{-2\alpha_1 y} \hat{y} = \frac{k_x}{\omega \varepsilon_2} e^{-2\alpha_2 y} \quad \text{Eq 120}$$

The total energy flow in x direction is the sum of the energy flow in the dielectric and metal layer and calculated as Eq 121:

$$P_{\text{total}} = P_1 + P_2 = \int_0^\infty S_{1x}\, dy + \int_{-\infty}^0 S_{2x}\, dy = \frac{k_x}{\omega \varepsilon_1} \int_0^\infty e^{-2\alpha_1 y}\, dy + \frac{k_x}{\omega \varepsilon_2} \int_{-\infty}^0 e^{2\alpha_2 y}\, dy =$$

$$= \frac{k_x}{2\omega} \left( \frac{1}{\alpha_1 \varepsilon_1} + \frac{1}{\alpha_2 \varepsilon_2} \right) \qquad \text{Eq 121}$$

The energy of metal layer in $-\hat{y}$ direction makes the sign of P$_2$ change to positive. The power $\frac{1}{\alpha_1 \varepsilon_1}$ flows in $+\hat{x}$ direction and power $\frac{1}{\alpha_2 \varepsilon_2}$ flow in $-\hat{x}$ direction. Based on the ratio of the power in Eq 122, as k$_x$ goes to $\infty$ for $\omega = \frac{\omega_p}{\sqrt{\varepsilon_1 + \varepsilon_\infty}}$, we see that $\varepsilon_1 = -\varepsilon_m$, and therefore, P$_2$ = - P$_1$. This shows they are canceling each other and there is no power flow in the x direction.

$$R = \frac{P_1}{P_2} = \frac{\alpha_2 \varepsilon_2}{\alpha_1 \varepsilon_1} = \frac{|\varepsilon_m| \varepsilon_2}{\varepsilon_1^2} = \frac{-\varepsilon_m^2}{\varepsilon_1^2} \qquad \text{Eq 122}$$

SPs never intersect the light cone and therefore, there is no coupling between them. Because of the conservation of energy (~ ω) and momentum (~ k$_x$), we can't have that with SP and radiated light modes for any flat metal-dielectric interface.

SP can be residing, but never interact with outside. However, using methods like "pattern the surface" or "use double layer interface", we can excite those SPs or radiating light. Two methods are discussed to interact SPs with propagating light: Kretschemann configuration and patterned metal surface.

**Method 1 Kretschemann Configuration:**

In this method, a thin film layer with high dielectric material (like gold) is surrounded by two dielectrics (like air and glass) and therefore, we have two light cones. Part of SP is bonded and excited by a beam incident upon the metal film from glass part. The light is being incident from glass and if the field can penetrate through the gold, it can excite the SP modes on top of the interface between air and gold. As shown in Figure 29, the SP area, surrounded by light cone 1 (air) and 2 (glass), can be excited by a beam incident upon the metal film from the glass.

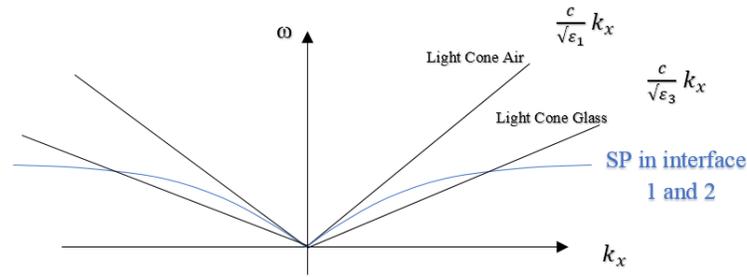

Figure 29. Kretschemann Configuration

Surface plasmon resonance is a helpful setup to recognize and characterize chemical species, that is illustrated in Figure 30.

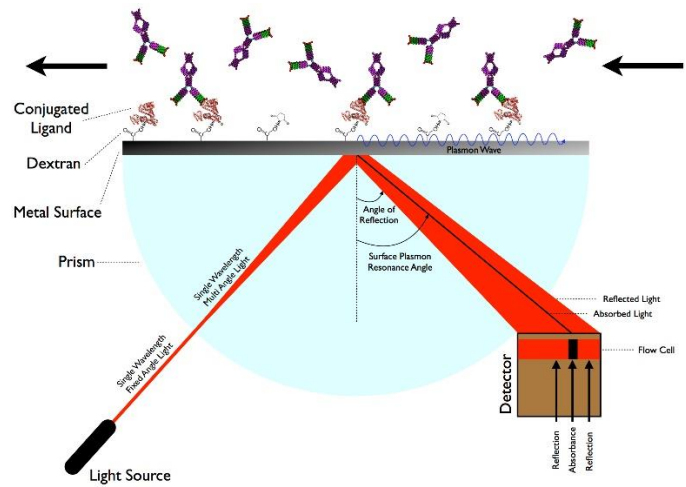

Figure 30. Biacore's SP Resonance [24]

**Method 2 Patterned Metal Surfaces:**

The other way to get SPs to interact with light cone is using the patterned materials. Instead of a planar metal surface, a periodically undulating film is used to form a "band-folded" dispersion curve into the first Brillouin zone [25]. On the other hand, in the periodic grating structure with period Λ, the plasmon of wave numbers $\frac{2n\pi}{\Lambda}$ are bounded with each other. Because of the Bragg diffraction, multiple band gap open and make these bonds shape the standing waves on the Brillouin zone [26].The SP modes in the light cone can be excited by or decay into propagating light modes, as illustrated in Figure 31. These periodic patterns can be a simple curved surface, or a whole array.

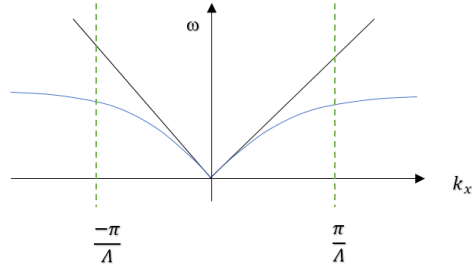

Figure 31. Band Diagram

At first the band is folded (fold-banded) and intersect the Brillouin zone boundaries with zero slope. Then the band repel each other, but don't cross each other, as shown in Figure 32. Every SP mode can be excited by the incident light in the light cone or decay propagating light modes.

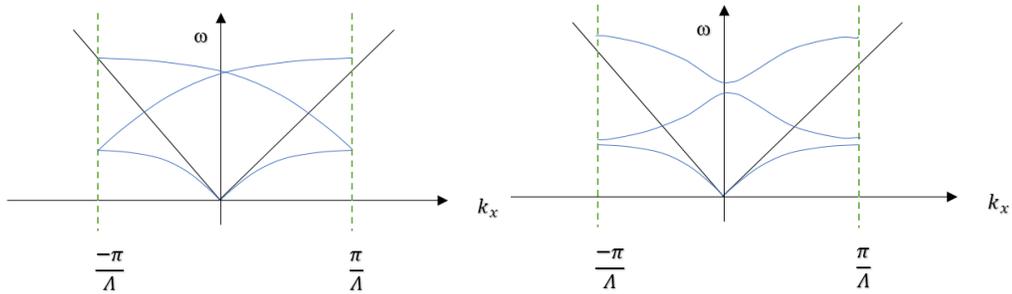

Figure 32. Band -folded (left). Bands repel each other (right)

**Extraordinary Optical Transmission (EOT)**

Incident light can excite SPs in hole arrays and create very strong EM fields at the holes and interface. These fields are strong enough to go through the holes and excite SPs on the other surface, decay into radiating transmitted beam, as shown in Figure 33.

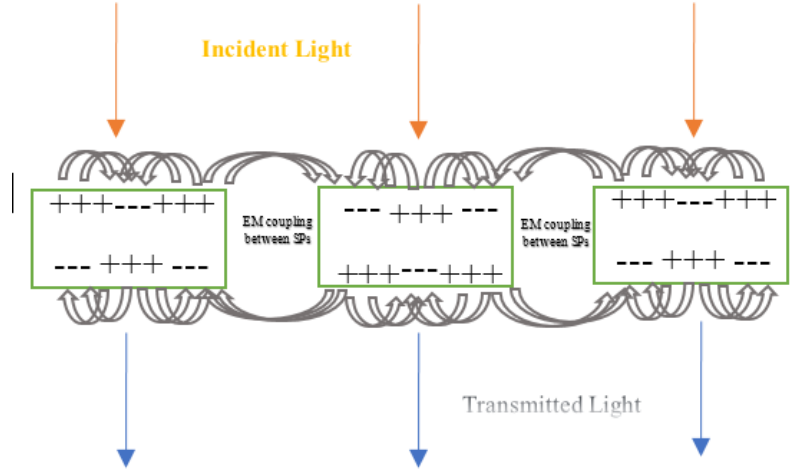

Figure 33. Light flow in Extraordinary Optical Transmission

## SP on Cylinders:

The electromagnetic behavior of cylinder helps us to understand the physics behind the metallic cylinders within a dielectric matrix. Thus, the approach for metallic cylinders are discussed in this section. The transverse component of E and H are determined in term of longitudinal $E_z$ and $H_z$, as shown in Eq 123.

$$\vec{E_t} = \frac{i}{\gamma^2}(k_2 \Delta_t E_z - \omega\mu_0 \hat{z}\, \Delta_t H_z) \qquad \vec{H_t} = \frac{i}{\gamma^2}(k_2 \Delta_t H_z + \omega\mu\varepsilon_0\varepsilon \hat{z}\, \Delta_t E_z) \qquad \text{Eq 123 [27]}$$

With $\Delta_t = \hat{\rho}\frac{\delta}{\delta\rho} + \hat{\Phi}\frac{1}{\rho}\frac{\delta}{\delta\Phi}$, $\gamma^2 = n^2 k_0^2 - k_z^2$, $k_0 = \frac{\omega}{c}$, $\hat{z} \times \hat{\rho} = \hat{\Phi}$ and $\hat{z} \times \hat{\Phi} = -\hat{\rho}$

Maxwell's Equations for inside of a cylinder with radius a ($\rho < a$) is written as Eq 124.

$$E_z = A_e\, J_m(\gamma\rho)\, e^{im\Phi} \qquad H_z = A_h\, J_m(\gamma\rho)\, e^{im\Phi} \qquad \text{Eq 124}$$

where $\gamma^2 = n_1^2 k_0^2 - k_z^2$

Maxwell's Equations for outside of a cylinder with radius a ($\rho > a$) is written as Eq 125.

$$E_z = B_e\, K_m(\beta\rho)\, e^{im\Phi} \qquad H_z = B_h\, K_m(\beta\rho)\, e^{im\Phi} \qquad \text{Eq 125}$$

where $\beta^2 = k_z^2 - n_2^2 k_0^2$

**Inside the Cylinder**

The electric and magnetic fields inside the cylinder are shown as Eq 126 and 127.

$$\vec{E_t} = \frac{i}{\gamma^2} e^{im\Phi}[(k_z \gamma A_e J'_m + \frac{im\omega\mu_0}{\rho} A_h J_m)\hat{\rho} + (\frac{imk_z A_e}{\rho} J_m - \gamma A_h \omega \mu_0 J'_m)\hat{\Phi}] \quad \text{Eq 126}$$

$$\vec{H_t} = \frac{i}{\gamma^2} e^{im\Phi}[(k_z \gamma A_h J'_m - \frac{i\omega\varepsilon_0 n_1^2 m}{\rho} A_e J_m)\hat{\rho} + (\frac{imk_z}{\rho} A_h J_m + \omega\varepsilon_0 n_1^2 \gamma A_e J'_m)\hat{\Phi}] \quad \text{Eq 127}$$

With $J'_m = \frac{\delta J_m}{\delta(\gamma\rho)}$.

**Outside the Cylinder**

The electric and magnetic fields outside the cylinder are shown as Eq 128 and 129.

$$\vec{E_t} = -\frac{i}{\beta^2} e^{im\Phi}[(\beta k_z B_e K'_m + \frac{im\omega\mu_0}{\rho} B_h K_m)\hat{\rho} + (\frac{imk_z}{\rho} B_e k_m - \beta B_h \omega\mu_0 K'_m)\hat{\Phi}] \quad \text{Eq 128}$$

$$\vec{H_t} = -\frac{i}{\beta^2} e^{im\Phi}[(k_z \beta B_h K'_m - \frac{i\omega\varepsilon_0 n_2^2 m}{\rho} B_e K_m)\hat{\rho} + (\frac{im\, k_z}{\rho} B_h K_m + \omega\varepsilon_0 n_2^2 \beta B_e K'_m)\hat{\Phi}] \quad \text{Eq 129}$$

From Eq 117 and 118, the $\hat{z}$ component condition are written as Eq 130 and 131.

$$A_e J_m = B_e K_m \text{ where } J_m = J_m(\gamma a) \quad \text{Eq 130}$$

$$A_h J_m = B_h K_m \text{ where } K_m = K_m(\gamma a) \quad \text{Eq 131}$$

From the $\hat{\Phi}$ component condition of Eq 121 and 122, we can conclude Eq 132 and 133.

$$\frac{-mk_z}{\gamma^2 a} A_e J_m - \frac{i\omega\mu_0}{\gamma} A_h J'_m = \frac{mk_z}{a\beta^2} B_e k_m - \frac{i\omega\mu_0}{\beta} B_h K'_m \quad \text{Eq 132}$$

$$\frac{-mk_z}{a\gamma^2} A_h J_m + \frac{i\omega\varepsilon_0 n_1^2}{\gamma} A_e J'_m = \frac{m\, k_z}{a\beta^2} B_h K_m + \frac{-i\omega\varepsilon_0 n_2^2}{\beta} B_e K'_m \quad \text{Eq 133}$$

In order to write the Eq 132 and 133 in the form of matrix, we consider $\zeta = \begin{pmatrix} A_e \\ A_h \\ B_e \\ B_h \end{pmatrix}$ and therefore, $M\zeta = 0$, as shown in Eq 134.

$$M = \begin{pmatrix} J_m & 0 & -K_m & 0 \\ 0 & J_m & 0 & -K_m \\ \frac{-mk_z}{\gamma^2 a} J_m & \frac{-i\omega\mu_0}{\gamma} J'_m & \frac{-mk_z}{a\beta^2} k_m & \frac{-i\omega\mu_0}{\beta} K'_m \\ \frac{i\omega\varepsilon_0 n_1^2}{\gamma} J'_m & \frac{-mk_z}{\gamma^2 a} J_m & \frac{i\omega\varepsilon_0 n_2^2}{\beta} K'_m & \frac{-m\, k_z}{a\beta^2} K_m \end{pmatrix} \quad \text{Eq 134}$$

Considering det (M) = 0, the Eq 135 is calculated.

$$k_0^2(\frac{n_2^2}{\beta K_m}K'_m + \frac{n_1^2}{\gamma J_m}J'_m)(\frac{1}{\beta K_m}K'_m + \frac{1}{\gamma J_m}J'_m) = \frac{m^2 k_z^2}{a^2}(\frac{n_2^2 k_0^2 - \beta^2}{\beta^2} + \frac{n_1^2 k_0^2 - \gamma^2}{\gamma^2})(\frac{1}{\beta^2} + \frac{1}{\gamma^2}) \quad \text{Eq 135}$$

Assuming $k_z^2 = n_1^2 k_0^2 - \gamma^2 = n_1^2 k_0^2 - \beta^2$, the Eq 136 is written.

$$(\frac{n_2^2}{\beta K_m}K'_m + \frac{n_1^2}{\gamma J_m}J'_m)(\frac{1}{\beta K_m}K'_m + \frac{1}{\gamma J_m}J'_m) = \frac{m^2}{a^2}(\frac{n_1^2}{\gamma^2} + \frac{n_2^2}{\beta^2})(\frac{1}{\beta^2} + \frac{1}{\gamma^2}) \quad \text{Eq 136}$$

In terms of permittivity, the Eq 136 is written in form of Eq 137. If m=0, the dispersion relation is satisfied.

$$(\frac{\varepsilon_m}{\beta K_m}K'_m + \frac{\varepsilon_d}{\gamma J_m}J'_m)(\frac{K'_m}{\beta K_m} + \frac{J'_m}{\gamma J_m}) = \frac{m^2}{a^2}(\frac{\varepsilon_m}{\gamma^2} + \frac{\varepsilon_d}{\beta^2})(\frac{1}{\beta^2} + \frac{1}{\gamma^2}) \quad \text{Eq 137}$$

The Bessel function and Hankel function in Eq 137, are shown in Eq 138.

$$J'_{n-1}(x) = \frac{1}{2}(J_{n-1}(x) - J_{n+1}(x))$$

$$J'_0(x) = -J_1(x) \text{ since } J_{-n}(x) = (-1)^n J_n(x)$$

$$K'_n(x) = -\frac{1}{2}(K_{n-1}(x) + K_{n+1}(x))$$

$$K_n(x) = \frac{\pi}{2}(i)^{n+1} H_n^{(1)}(ix) \qquad \text{Eq 138}$$

where $H_n^{(1)}$ is Hankel function of the first kind.

$$K'_0(x) = -K_1(x) \quad \text{since } K_{-n}(x) = K_n(x) \qquad \text{Eq 139}$$

Thus, for m=0, the Eq 140 is written.

$$(\frac{\varepsilon_m}{\gamma}\frac{J_1}{J_0} + \frac{\varepsilon_d}{\beta}\frac{K_1}{K_0})(\frac{1}{\gamma}\frac{J_1}{J_0} + \frac{1}{\beta}\frac{K_1}{K_0}) = 0 \qquad \text{Eq 140}$$

**Surface Plasmon Dispersion Curve Implemented by MATLAB**

Some parts of the MATLAB code are written below to show the calculation of surface plasmon dispersion curve of metallic cylinder, the result is illustrated in Figure 34. The complete code is referred in Appendix B.

```
tempc=0.5*(besselj(cntm-1,gamma*radius) - besselj(cntm+1,gamma*radius));
tempd=-0.5*(besselk(cntm-1,gamma*radius)+ besselk(cntm+1,gamma*radius));
```

```matlab
        T1 = epsilonrod./gamma.*tempc./besselj(cntm,gamma*radius);
        T2 = epsilonmatrix./beta.*tempd./besselk(cntm,beta*radius);
        T3 = 1./gamma.*tempc./besselj(cntm,gamma*radius);
        T4 = 1./beta.*tempd./besselk(cntm,beta*radius);
        LHS = (T1 + T2).*(T3 + T4);

        tempe = gamma.^2;
        tempf = beta.^2;
        T1 = epsilonrod./tempe;
        T2 = epsilonmatrix./tempf;
        T3 = 1./tempe;
        T4 = 1./tempf;
        RHS = cntm^2/radius^2.*(T1 + T2).*(T3 + T4);

for cntomega = 1:wavelengthpnts-1

   if real(LHS(cntomega))>real(RHS(cntomega)) &&
     real(LHS(cntomega+1)) < real(RHS(cntomega+1))

         wavelength_mode_HO = [wavelength_mode_HO,wavelength(cntomega)];
         kz_mode_HO = [kz_mode_HO,kz(cntkz)];
         m_mode_HO = [m_mode_HO,cntm];
         epsilonrod_mode_HO = [epsilonrod_mode_HO,epsilonrod(cntomega)];
         epsilonmatrix_mode_HO = [epsilonmatrix_mode_HO,epsilonmatrix];

       elseif real(LHS(cntomega)) < real(RHS(cntomega)) &&        real(LHS(cntomega+1)) >
       real(RHS(cntomega+1))

         wavelength_mode_HO = [wavelength_mode_HO,wavelength(cntomega)];
         kz_mode_HO = [kz_mode_HO,kz(cntkz)];
         m_mode_HO=[m_mode_HO,cntm];
         epsilonrod_mode_HO=[epsilonrod_mode_HO,epsilonrod(cntomega)];
         epsilonmatrix_mode_HO = [epsilonmatrix_mode_HO,epsilonmatrix];
   end

end

tempg = 2*pi./wavelength_mode_HO*light;
energy_mode_HO = (hbar/joule)*tempg;
```

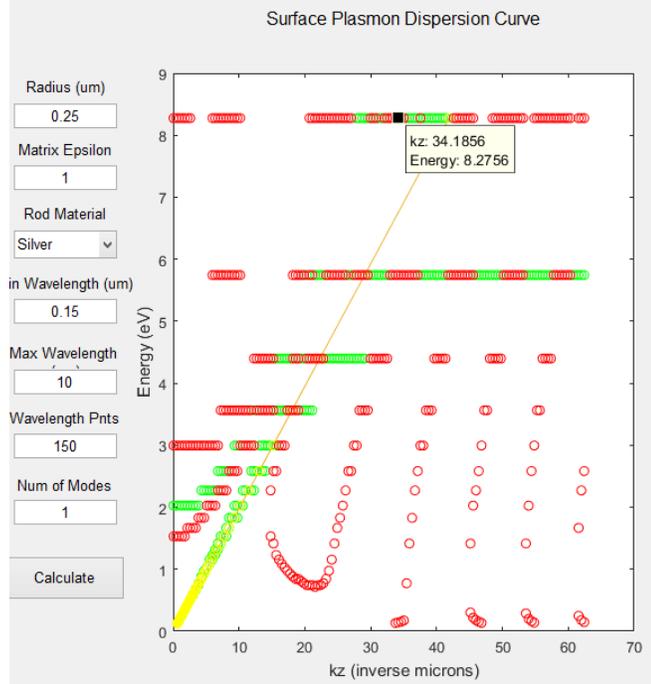

Figure 34. Surface Plasmon Dispersion Curve

**Electric Filed of Metallic Cylinder Implemented by MATLAB**

Considering Ae = 1, the fields are easily calculated as Eq 141. Since the $B_e = \frac{J_m(\gamma a)}{K_m(\beta a)}$ and

$B_h = \frac{J_m(\gamma a)}{K_m(\beta a)} A_h$ and $-i\omega\mu_0 \left(\frac{J_m K'_m}{\beta K_m} + \frac{J'_m}{\gamma}\right) A_h = \frac{m k_z}{a} J_m \left(\frac{1}{\beta^2} + \frac{1}{\gamma^2}\right)$ and therefore;

$$A_h = \frac{i m k_z}{\omega\mu_0 a} \frac{\left(\frac{1}{\beta^2}+\frac{1}{\gamma^2}\right)}{\left(\frac{J_m K'_m}{\beta K_m}+\frac{J'_m}{\gamma}\right)} \text{ and } B_h = \frac{J_m}{K_m} A_h \qquad \text{Eq 141}$$

Some parts of the MATLAB code are written below to show the calculation of $E_\rho$ filed of metallic cylinder, the result is illustrated in Figure 35. The complete code is referred in Appendix B.

```
tempd = 0.5*(besselj(m-1,gamma*radius) - besselj(m+1,gamma*radius));
tempe = -0.5*(besselk(m-1,gamma*radius) + besselk(m+1,gamma*radius));
tempf = tempd/gamma + bjm/bkm*tempe/beta;
```

```
Ah = 1i*m*kz/omega/mu_o/radius*tempc/tempf;
Bh = bjm/bkm*Ah;

delta_r = radius/250;
[r_wire,t_wire] = meshgrid(0:delta_r :radius,0:pi/30:(2*pi));
x_wire = r_wire.*cos(t_wire);
y_wire = r_wire.*sin(t_wire);
z_wire = zeros(size(x_wire));

tempg = 0.5*(besselj(m-1,gamma*r_wire) - besselj(m+1,gamma*r_wire));
tempi = besselj(m,gamma*r_wire);
E_rho_wire=1i/gamma^2.*exp(1i*m*t_wire).*(gamma*kz*tempg+
1i*omega*mu_o*m./r_wire*Ah.*tempi);
```

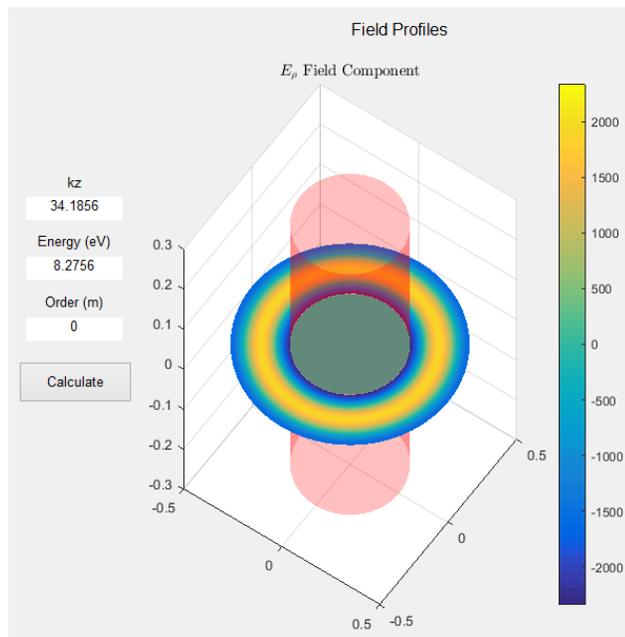

Figure 35. $E_\rho$ filed of Metallic Cylinder

# CHAPTER III: METHODOLOGY

## STRUCTURE DESIGN

This section is written based on my article "Metamaterials based hyperspectral and multi-wavelength filters for imaging applications" published at SPIE Defense + Commercial Sensing, 2019, Baltimore, Maryland, United States [1].

### Conventional Bragg Stack Filters

A traditional Bragg stack is composed of alternative vertical dielectric layers. Each pair of Bragg stack consists of a layer with higher dielectric constant than the other layer. The thickness of each layer is calculated by $t_j = \lambda_c/4n_j$, where $n_j$ is the index of refraction of the $j^{th}$ layer; for the normal incidence of light, the center wavelength $\lambda_c$ is centered in a wide spectral stopband. Based on electromagnetic wave theory, the center wavelength has the destructive interference experience while going through the whole layers of the Bragg stack, leads to holding back the transmission of incidence light. However, by increasing the thickness of a layer of Bragg stack (namely middle layer), twice to the original size with the thickness of $t_j = \lambda_c/2n_j$, a resonant condition happens for the incident center wavelength of light in this layer, makes the narrow passband $\lambda_c$ in the wider stopband [1, 28]. The resonance happens if the round-trip phase of the wavelength of light would be an integer of $2\pi$ in $k\,d + \varDelta = 2\pi m$, where k is wave vector, d is length of roundtrip propagation and $\varDelta$ is the phase shift related to the reflection. The mode order is defined by the integer m, for resonant the simplest mode 2 is usually considered.

One of the limitations of notch narrow passband filters is that the center wavelength $\lambda_c$ is dependent on the angle of the incident light. The resonant condition can be written in terms of wave vector and off-normal incidence, as shown in Eq 142 [1].

$$2k_{j,z}t_j = 2n_j k \cos(\theta_j) t_j = 2\pi \qquad \text{Eq 142 [1]}$$

with $k_{j,z}$ is the wave vector in the z direction (parallel to the normal vector of layers). $k = 2\pi/\lambda_o$ and $n_j = \sqrt{\varepsilon_j}$, $\lambda_0$ is the free-space wavelength of light. The angle of light relative to the normal vector in resonant layer is shown by $\theta_j = arcsin(n_1/n_j sin(\theta_1))$ where $\theta_1$ is the angle of incidence of the beam for the top layer of Bragg stack. Based on this angle, we can rewrite the resonant condition for the $\lambda_c$, as shown in Eq 143 [1].

$$\lambda_c = 2n_j cos(\theta_j) t_j = 2n_j cos(sin^{-1}\left(\frac{n_1}{n_j}sin(\theta_1)\right)) t_j \qquad \text{Eq 143 [1]}$$

The light reaching to the filter should be collimated to overcome the problem of angle of incidence dependency of the center wavelength of light (dispersion). Therefore, there is need to some extra optic setups to collimate the light that add cost and weigh to the hyperspectral imaging system. The hyperspectral imaging systems characterize the objects by collecting a continuum of wavelengths to get lots of images from material in infrared or ultraviolet regime. Furthermore, complex optical train reduces the tolerance of vibrations related to optical systems. To achieve a small DBR filters that be able to collimate the light, we need to remove the dependency of angle of light ($cos(\theta_j)$) by using hyperbolic metamaterial in DBR design [1].

**Hyperbolic Metamaterials Bragg Stack**

Hyperbolic metamaterial can be defined based on the optical properties of the crystals. In such materials, the relation between electric and Magnetic field E and H with electric displacement D and magnetic flux density B can be defined as Eq 144 and 145:

$$\epsilon \epsilon_0 E = D \quad \text{Eq 144}$$

$$\mu \mu_0 H = B \quad \text{Eq 145}$$

where the $\varepsilon_0$ and $\mu_0$ are vacuum permittivity and permeability. Also, $\varepsilon$ and $\mu$ are the relative permittivity and permeability tensor. The electric permittivity tensor in Eq 144 gives the directional dependence of $\varepsilon$. One type of the hyperbolic metamaterials is the material with wire mesh (Fakir's bed) which is composed of a vertical array of metal wires.

Eq 146 shows the Maxwell-Garnett theory, providing $\varepsilon_{zz}$ and $\varepsilon_{xy}$ which are calculated based on from permittivity of wire $\varepsilon_w$, permittivity of dielectric $\varepsilon_d$ and N is the ratio of the area of wire to the total area (fill factor). For anisotropic uniaxial crystal, the three permittivity values of the tensor are not the same: $\epsilon_{xx} = \epsilon_{yy} \neq \epsilon_{zz}$. The electric permittivity tensor is shown for the non-magnetic material in Eq 146 [1].

$$\epsilon = \begin{pmatrix} \epsilon_{xy} & 0 & 0 \\ 0 & \epsilon_{xy} & 0 \\ 0 & 0 & \epsilon_{zz} \end{pmatrix}$$

$$\epsilon_{xy} = \frac{(1+N)\epsilon_d \epsilon_w + (1-N)\epsilon_d^2}{(1-N)\epsilon_w + (1+N)\epsilon_d} \qquad \epsilon_{zz} = N\epsilon_w + (1-N)\epsilon_d \qquad \text{Eq 146 [1]}$$

To solve the optical properties of subwavelength periodic nanostructures, Maxwell equations are considered to calculate the vector and polarization components of electromagnetic waves [12]. To find the dispersion relation, we mention two important Maxwell's equations, given by Eq 147 [29]:

$$\nabla \times E = -\frac{\delta B}{\delta t} \quad \nabla \times H = -\frac{\delta D}{\delta t} \qquad \text{Eq 147 [29]}$$

Where H is the magnetic fields and B is the magnetic induction. The general format for the plane wave expressions for electric and magnetic fields with angular frequency ω and wave vector k are given by Eq 148:

$$E = E_0\, e^{\,i(\omega t - k.r)} \quad H = H_0\, e^{\,i(\omega t - k.r)} \qquad \text{Eq 148 [29]}$$

Therefore, the eigenvalue problem for electric field E is given by Eq 149:

$$K \times (K \times E) + \omega^2 \mu_0 \epsilon_0 \epsilon E = 0 \qquad \text{Eq 149 [29]}$$

Solving the Eq 149 into the matrix from, two solutions are accomplished. One solution is for extra ordinary waves (TM) shows the wave polarization in plane containing the optical axis z, while the other solution is for ordinary waves (TE) shows the wave polarization in the XY plane.

Since one of the $\varepsilon_{zz}$ and $\varepsilon_{xy}$ is negative, here $\varepsilon_{zz}<0$, the anisotropy structure is called indefinite, in the quadric form. The hyperboloidal iso-frequency surface is shaped when the permittivity components of a media have different sign, and therefore, mathematically result in a hyperbolic medium (dielectric) with large wavevectors and propagating TM polarization [29]. This is in contrast with isotropic structure that have evanescent polarization related to the iso-frequency contour. The Figure 36 shows the iso-frequency surfaces for an isotropic material ($\varepsilon_{zz}>0$ and $\varepsilon_{xy}>0$), hyperbolic medium with dielectric (type I) ($\varepsilon_{zz}>0$ and $\varepsilon_{xy}<0$) and one-fold hyperboloid called metallic medium (type II) ($\varepsilon_{zz}<0$ and $\varepsilon_{xy}>0$) [29, 30].

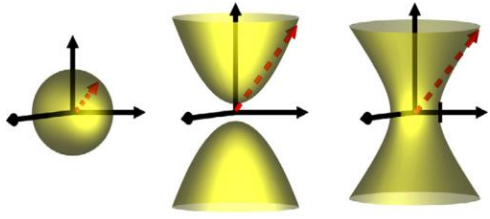

Figure 36. Iso-frequency surfaces for an isotropic material (left), hyperbolic medium with dielectric medium-type I(center) and metallic medium type II(right) [29].

The dispersion relation for TM polarized light is given in Eq 150, where ε and k are the permittivity and wave number, respectively [31, 7, 32, 33]. In hyperbolic metamaterial, waves have the following dispersion relation [7]:

$$\frac{k_x^2 + k_y^2}{\epsilon_{zz}} + \frac{k_z^2}{\epsilon_{xy}} = k_o^2$$

Eq 150 [1]

To calculate the wavenumber in z direction ($k_z$), Eq 150 can be transformed to Eq 151, letting $k_y$=0. If $|\epsilon_{xy}| \ll |\epsilon_{zz}|$, $k_z$ is not dependent to the angle of incidence largely [7, 34, 35, 3, 6, 1].

$$k_z(\theta) = \sqrt{\epsilon_{xy} k_o^2 - \frac{\epsilon_{xy}}{\epsilon_{zz}} k_x^2(\theta)} \approx \sqrt{\epsilon_{xy}} k_o = k_z(\theta = 0)$$

Eq 151 [1]

As long as $|\epsilon_{xy}| \ll |\epsilon_{zz}|$, the dispersion of the narrow passband is decreased [7, 36, 37, 11, 38, 39]. As shown in Figure 37, this condition is accountable for the shortwave and midwave infrared regions for a material composed of silver wires of radius 0.2μm in the arrays of square shaped lattice with period (lattice constant) of 1μm, electric permittivity database and a dielectric medium with a $\epsilon_d = 2.1025$. Although the material obviously shows optical loss, the condition of $|\epsilon_{zz}| \gg |\epsilon_{xy}|$ holds.

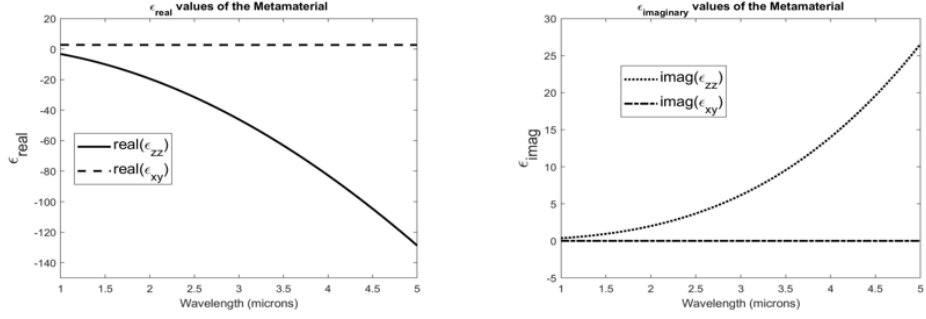

Figure 37. The real (left) and imaginary (right) parts of $\varepsilon_{xy}$ and $\varepsilon_{zz}$ of a wire mesh grid [1].

The values of $\varepsilon_{xy}$ and $\varepsilon_{zz}$ in Maxwell-Garnett theory can be approximated Eq 152.

$$\epsilon_{xy} \approx \frac{(1+N)}{(1-N)} \epsilon_d \qquad \epsilon_{zz} \approx N\epsilon_w \qquad \text{Eq 152}$$

In our case, the metal wires occupied half of each unit cell, therefore we can say the ratio of occupied wire area to the unit cell area is 0.5. Considering the very big permittivity of copper wires at center wavelength 4$\mu m$, shown in Figure 37, the ratio of $\varepsilon_{xy}$ / $\varepsilon_{zz}$ is calculated by Eq 153.

$$|\frac{\epsilon_{xy}}{\epsilon_{zz}}| \approx \frac{(1+N)}{(1-N)} \frac{\epsilon_d}{N(\epsilon_w)} \approx \frac{(1.5)}{(0.5)} \frac{2.1025}{0.5(-800)} \approx |-0.01576| \qquad \text{Eq 153}$$

In this work, the proposed structure is modeled and simulated, shown in Figure 38. The Bragg stack is composed of seven-layers of dielectrics $SiO_2$ as the low index material (LoK) and a-Si as the high index material (HiK), with copper wires perpendicular in three middle layers, resonant layer and the adjacent two HiK a-Si layers. The thick middle layer, known as resonant layer, is made of $SiO_2$. The fabrication the structure, such that metal wires go through the whole five layers, was not possible, and the alternative designed was designed. The other two $SiO_2$ layers have a thickness of $t_{SiO2} = 880nm$ and the middle resonant layer thickness is twice of this amount. Also, thickness of two a-Si layers are $t_{Si}$

= 330*nm*. The radius of copper wires is 0.25*μm* and the square period (lattice constant) is 1*μm*.

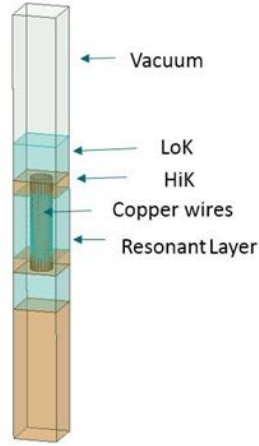

Figure 38. The proposed HMBS simulated by HFSS[1]

The proposed HMBS transmits different wavelengths at different points on the filter by changing the lattice constant (unit cell of crystal lattice) and splits the light to different spectral wavelength. By sweeping the lattice constant from 1 to 2.4*μm*, separate wavelengths are transmitted at different points on the hyperspectral filter to each pixel of the array, as shown in Figure 39. Changing the lattice constant modifies the *N*. Consequently, $\varepsilon_{xy} / \varepsilon_{zz}$ changes according to Eq 153. Finally, $k_z$ in Eq 151 varies by tuning $\varepsilon_{xy} / \varepsilon_{zz}$ that leads to changing the center wavelength and resonant condition in Eq 143.

In conclusion, changing the lattice constant leads to create the resonant condition in hyperspectral Bragg stack [1, 40, 41, 42, 43]. The light propagates through the several boundaries of BS, forming a region of graded refractive index and makes the continuous change of index of refraction smoother. This small change makes the light transmits as much as possible and leads to less reflection within the subwavelength structure [44, 45, 46, 47, 48, 49, 50]

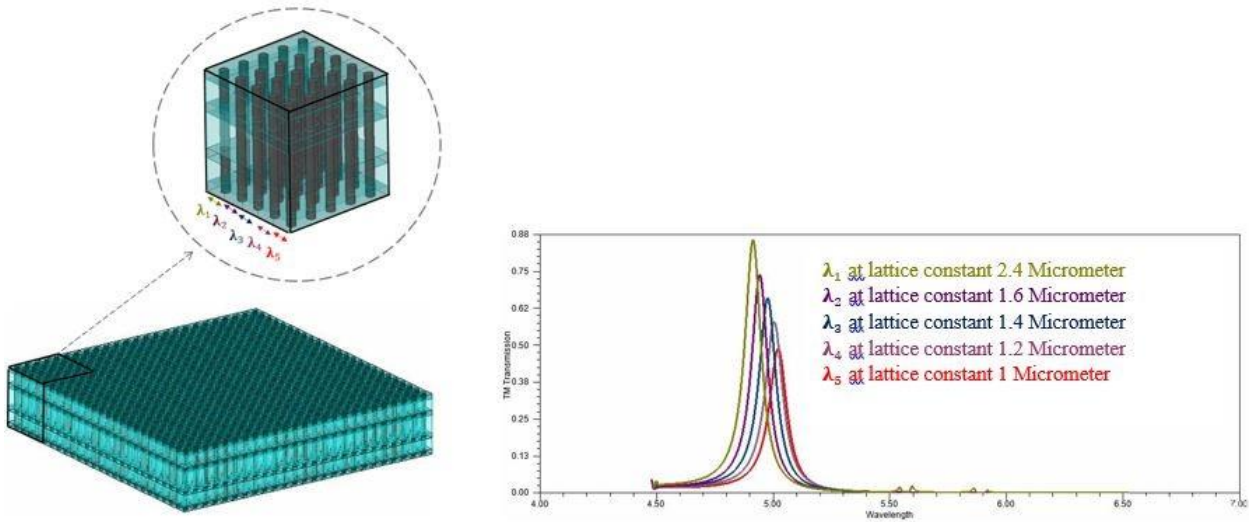

Figure 39. Each pixel of the filter has different geometry, is for different wavelength(left). Center wavelength changes with changing the lattice constant(right)

## FEM Simulation

### Simulation with High Frequency Structural Simulator

Since in the Maxwell-Garnett consider the composition of the material is considered, full-wave finite element modeling (FEM) was also implemented using ANSYS' high frequency structural simulator (HFSS). In contrast with Maxwell-Garnett theory, the HFSS doesn't use effective media approximation technique to approximate permittivity but uses the frequency dependence permittivity of the materials in full-wave electromagnetic modeling.

In finite element method, the floquet ports were used as excitation port for our periodic structure that is defined by Master-Slave boundaries. The plane waves of these ports define the propagation direction by the geometry of the periodic structure, phasing and frequency. The Master and slave boundaries were used to model periodic plane waves, where every point of the slave boundry should be match with the electric field of corresponding point on the master boundary. The unit cell geometry of the square lattice has the period of 1μm with the radius of wire about 0.25μm.

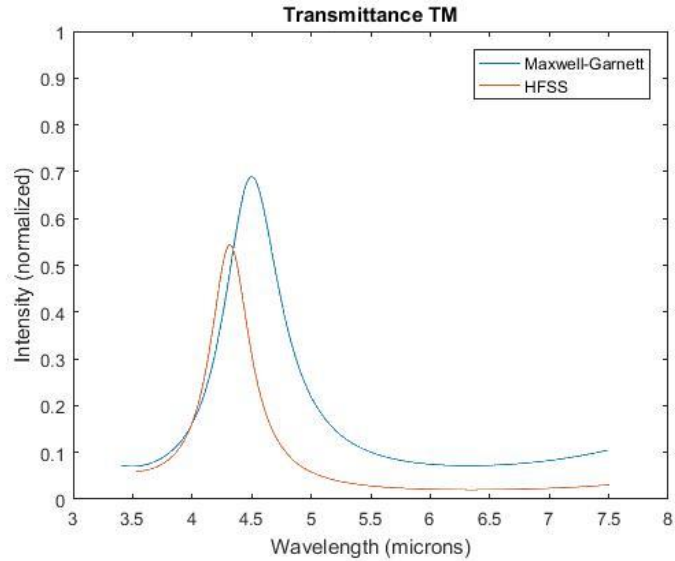

Figure 40. The transmittance of 7 layers HMBS simulated by Maxwell-Garnett theory and FEM modeling

Therefore, there is a big difference in results simulated by FEM modeling and Maxwell-Garnett theory. The results of both simulations show different transmissions, as seen in Figure 40. Although the wire radius used in Maxwell-Garnett theory is not too large, but we saw weaker result in comparison with HFSS. In addition, the field is uniformly distributed around the wire in the resonance layer in FEM modeling; however, Maxwell-Garnett theory makes the field close to the wires. A similar example to show the field is close to wire in the middle layer of HMBS, simulated by Maxwell-Garnett theory, is shown in Figure 41.

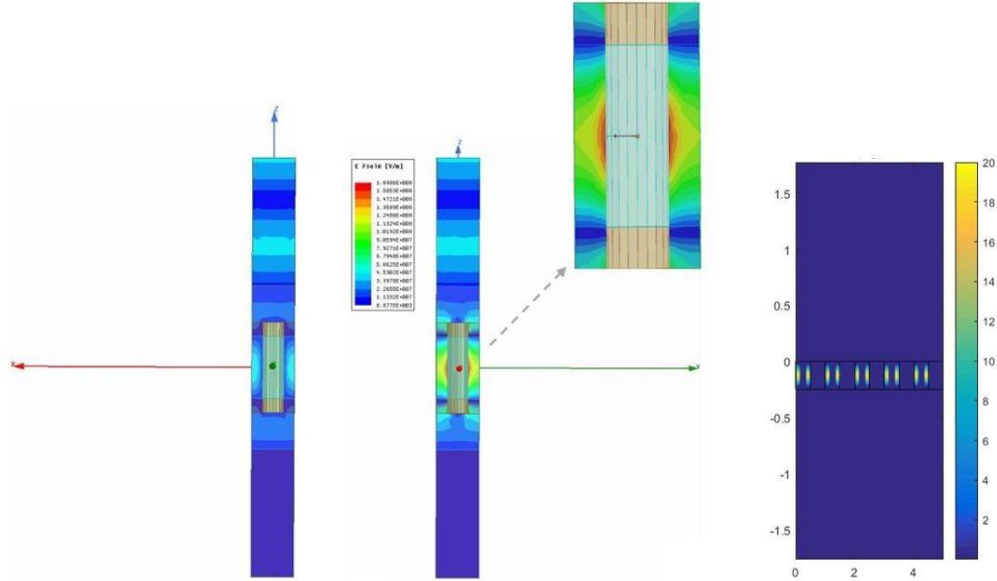

Figure 41. E-field in XZ plane (left) and YZ plane (middle) of HMBS simulated by FEM, compared to E-field simulated by Maxwell-Garnett theory (right).

Figure 41 shows the electric field of HMBS at resonant frequency 65.789THz (center wavelength 4.556μm) simulated by HFSS comparing with simulated by Maxwell-Garnett theory. The electric field simulated by HFSS is in the $\hat{y}$ direction at the resonant frequency, which is visible in YZ plane. The excitation from SP (SP) is strongest at the middle dielectric layer, seen in YZ plane, showing that the majority of the energy of the EM fields are in the middle $SiO_2$ layer, where the resonant condition happened. The collective electron oscillations in the SP creates electromagnetic field around the copper wires. Figure 42 shows the electric field direction in X axis ($E_x$) of HMBS at resonant frequency 65.789THz (4.556μm) simulated by HFSS.

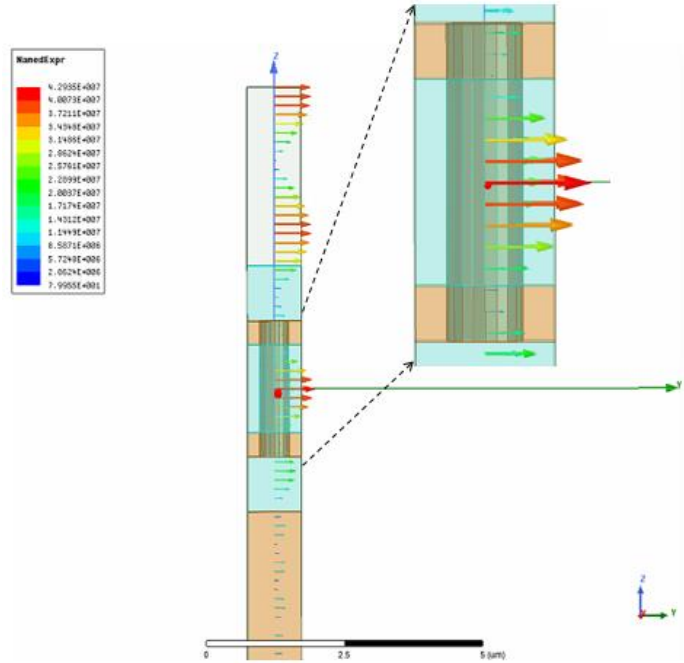

Figure 42. Electric field in the resonant layer

Figure 43 shows the large negative permittivity the of the copper wire in the infra-red wavelength range, makes no field inside of the wire. However, the field simulated by Maxwell-Garnett theory is closer to the cavities (the BS with cavities is used to show the field similar to the HMBS with copper wires).

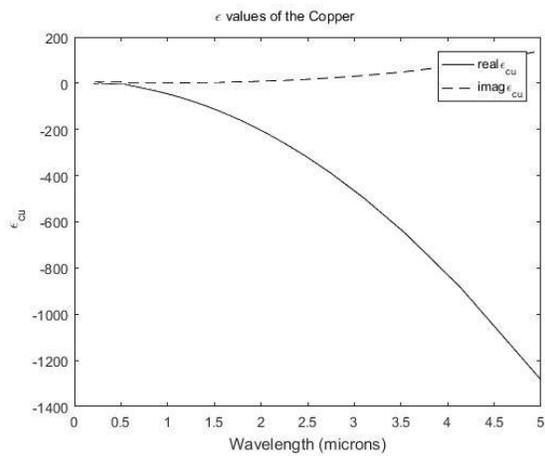

Figure 43. Real and Imaginary parts of permittivity values of the Copper

The transmittance of the DBR stack without the copper wires was modeled in HFSS and shown in Figure 44 for TE and Figure 45 for TM. The TE and TM polarized incident beams were used for the excitation with angles of incidence varying from 0 to 25 degrees. As the angle of incidence light changed from 0 to 25 degrees, shown in Figure 44 and Figure 45, the center wavelength of the transmission band for the traditional BS shifted by ~300nm. This is in contrast with the trivial wavelength shift of the transmission peaks for the hyperbolic metamaterial Bragg stack for 0 and 25 degrees, which practically center wavelengths of different angles of light are locked on top of each other. The TM transmittance of the DBR stack with the copper wires was simulated to show that center wavelength changes by ~15nm as the angle of incidence changed from 0 to 25 degrees, as shown in Figure 46.

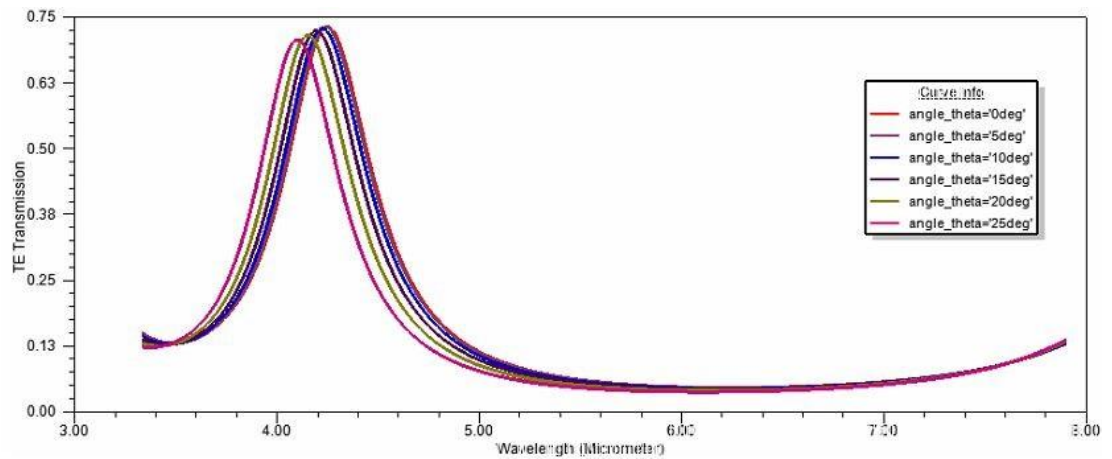

Figure 44. HFSS simulations that show a large wavelength shift for the TE transmission peak for the traditional Bragg stack filter.

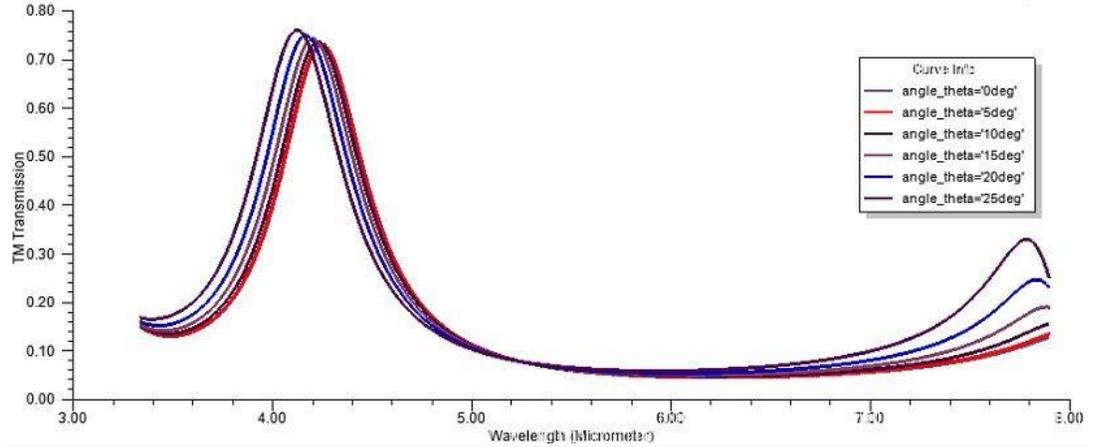

Figure 45. HFSS simulations that show a large wavelength shift for the TM transmission peak for the traditional BS filter.

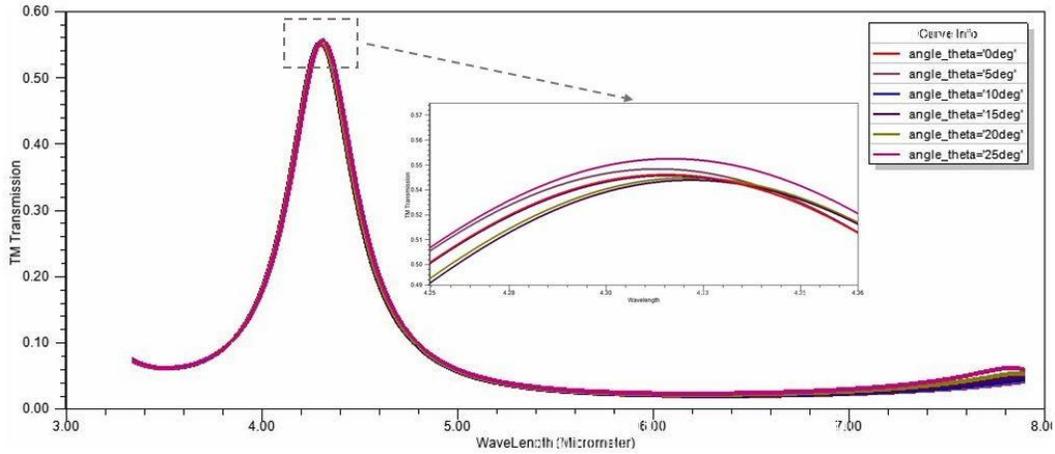

Figure 46. HFSS simulations that show a very small wavelength shift for angle of incidence 0 to 25 degree in the hyperbolic metamaterial Bragg stack (HMBS) filter.

## FABRICATION

The simulated Bragg stack is composed of alternative dielectric layers of a-Si as the high dielectric (HiK) material and silicon oxide ($SiO_2$) as low dielectric (LoK) material, with the 1μm unit cell, compatible to optical lithography process. One of the big challenges in fabrication was etching the middle three layers and the electrodeposition of the copper in holes. However, all the technical steps of the fabrication process are outlined below.

**Deposit alternative layers**

The proposed Bragg Stack is composed of the seven alternative layers of $SiO_2$ and a-Si. The $SiO_2$ and a-Si layers were deposited onto a double side polished 4-inch silicon wafer with Plasma-enhanced chemical vapor deposition (PECVD) method. PECVD is a chemical vapor deposition procedure which happens when the reacting gas creates plasma, leads to form a vapor state to a solid state and therefore, deposits thin films on the wafer. The plasma is created by AC (or DC) and released between the space of two electrodes charged by reacting gases. The $SiO_2$ layers ware deposited at 1800 mTorr pressure, with the $SiH_4$ flow rate of 20 sccm, the $N_2O$ flow rate of 2500 sccm, and a RF power (AC) of 140 Watts at 300° C. The a-Si layers were deposited at 3 Torr pressure, with the $SiH_4$ flow rate of 25 sccm, the Ar flow rate of 475 sccm at 200° C. The first four layers, including 880nm $SiO_2$, 330nm a-Si, 1500 nm resonant $SiO_2$ layer and the second 330 nm a-Si layer were deposited on a both side polished 525$\mu m$ silicon wafer, as depicted in the Figure 47.

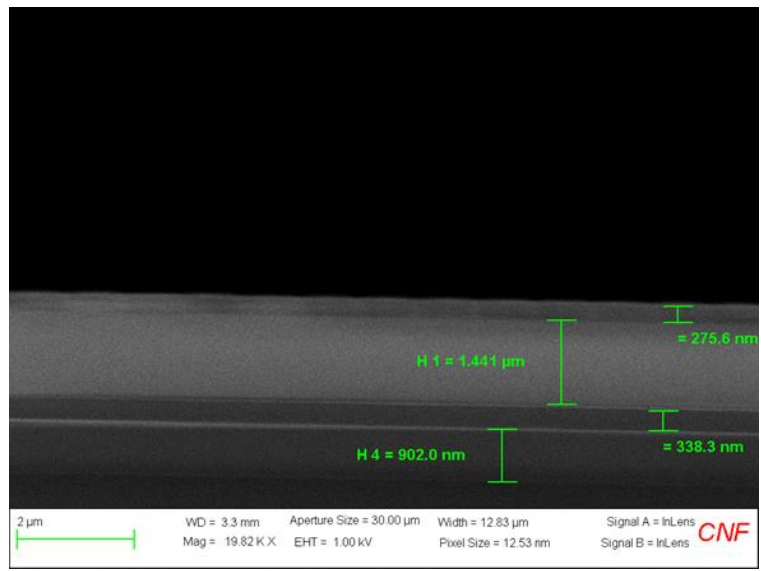

Figure 47. SEM image of deposited alternative layers of $SiO_2$ and a-Si with PECVD

The deposition of alternative layers of the a-Si and SiO$_2$ increase the stress on the wafer, makes the wafer bowing that lead to be rejected by lithography tool, ASML. Using a thicker wafer 625$\mu m$ instead of regular 400 $\mu m$ helped to solve this issue

**Photolithography**

The recent advanced technology in ion-beam lithography and nanolithography created small wavelength-based patterns as small as 100 nm. Semiconductor Lithography (Photolithography) is a method to fabricate the micro electromechanical systems (MEMS) structures. This technique is used to make different feature sizes of some materials in the form of thin layers, such that thin film can be removed in selected parts of the film and other parts remained unchanged on wafer.

The first step in lithography is making a mask. A mask is made of the chrome which is patterned on a glass plate. The wafer first is spin-coated with photoresist, which is a sensitive polymer to the ultraviolet light. Ultraviolet light exposes on the wafer through the mask, makes a chemical change on the photoresist. This change causes some of the photoresists to be disappeared by a specific solution in the next step. The next step is developing, means that the photoresist exposed by the light and therefore, the pattern of the mask appears on the photoresist. The common type of photoresist is positive photoresist that exposed area is soluble in developing process. With negative resist, in contrast with positive, the unexposed area is soluble, as shown in Figure 48 [51, 52].

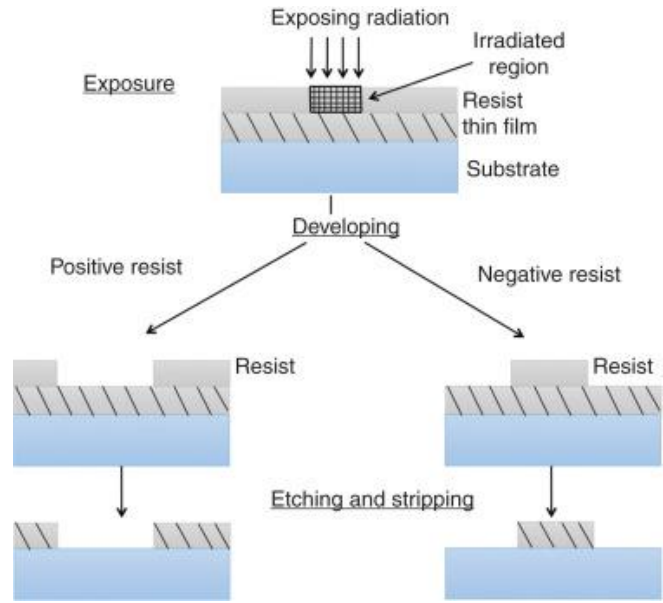

Figure 48. Schematic diagram of the lithography steps for positive and negative resist [52].

In the lithography process, first the anti-reflective coating (ARC) coated on the wafer, which acts as a good adhesion for the UV210 photoresist of 600nm thickness, that was spin coated on the wafer next. Then wafer was pre-baked and prepared for exposure. In order to find the best energy and dose for the simulated Bragg stack, a matrix of different energies and doses was designed with the help of lithography tool (ASML), as depicted in Figure 49.

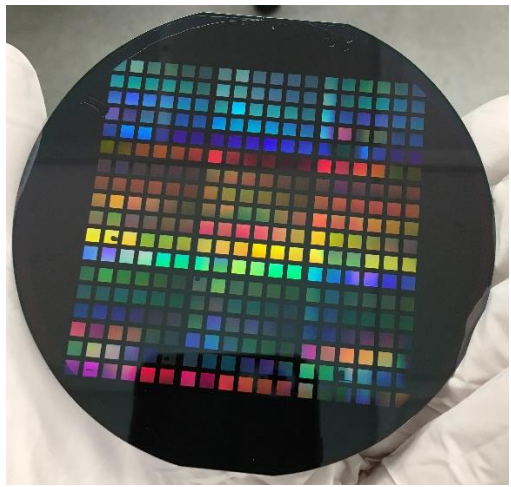

Figure 49. Matrix of different energy and dose in lithography for positive PR

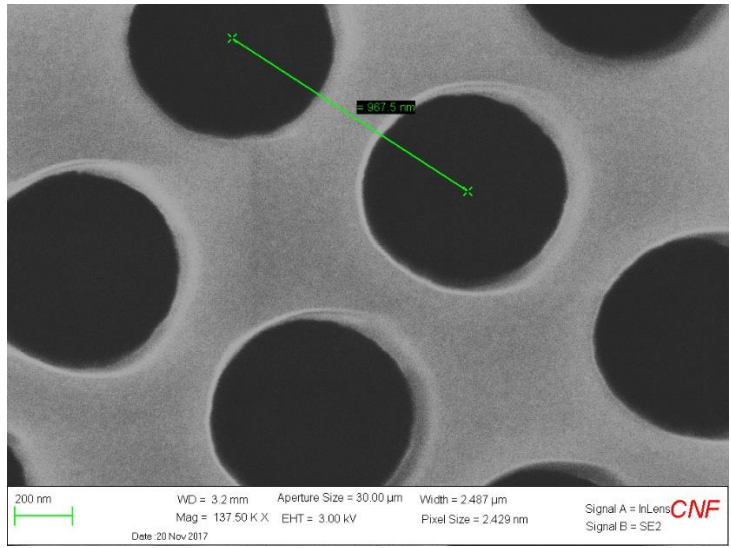

Figure 50. SEM image of lithography

Figure 50 shows the SEM top view of the best feature size after the lithography, holding the period 1 μm. Since the etching process for the seven layers Bragg stack was different than the one-layer Bragg stack, several alternative solutions were tested, including thicker photoresists (600nm and 1100 nm), hard mask or separate lithography. Lithography pattern with negative photoresist and bright field mask after Pt sputtered is shown in Figure 51.

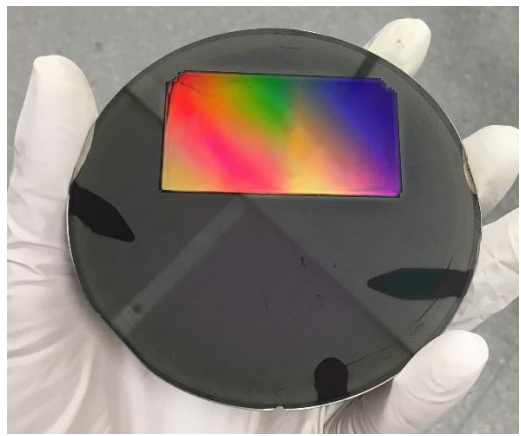

Figure 51. Lithography pattern with negative PR and bright field mask

**Etching**

Etching the layers were performed by the reactive ion etching (RIE). RIE is the dry etching method that uses high energy ions of the plasma to attack and etch the wafer. The plasma is created under low pressure vacuum by the electromagnetic filed. After etching the ARC, we prepared the wafer for the etching of the a-Si and SiO$_2$ layers. First, one-layer SiO2 layer Bragg stack was designed and fabricated by depositing, and then etching the 1.5$\mu m$ thick SiO$_2$ resonance layer with a 20 sccm flow rate of CH$_2$F$_2$ gas (18sccm CH2F2, 72sccm He, ICP 1500W, temperature 10C, RF 35W and pressure 4mTorr). Then 7-layer Bragg stack was fabricated, the top and underlying a-Si layers were etched using a HBr RIE etch process (20sccm HBr, pressure 15mTorr, HF 30W, temperature 16C).

During etching of the middle SiO$_2$ layer with CH$_2$F$_2$, the etching gas interacted with a-Si, made polymer material in the holes and leads to faster photoresist consumption and slowing the etching process, as depicted in Figure 52. Using thicker photoresist (UV1400) with thickness of 1200nm didn't help as well because of the aspect ratio problem.

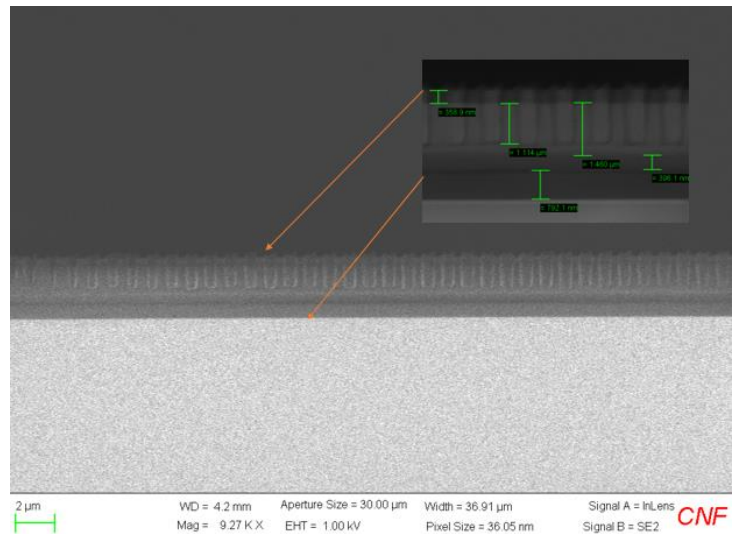

Figure 52. Etching the top a-Si and middle SiO$_2$ consumed all the resists

The other solution for etching the whole three middle layers, is using a hard mask on top of the wafer. The hard mask, which can act as a photoresist, can be aluminum oxide or Chrome. Aluminum oxide can be removed by wet etching. However, Chrome can be removed by the tools, which is easier. Therefore, 110nm of Cr hard mask was deposited on the top a-Si layer, to act as the photoresist, as depicted in Figure 53.

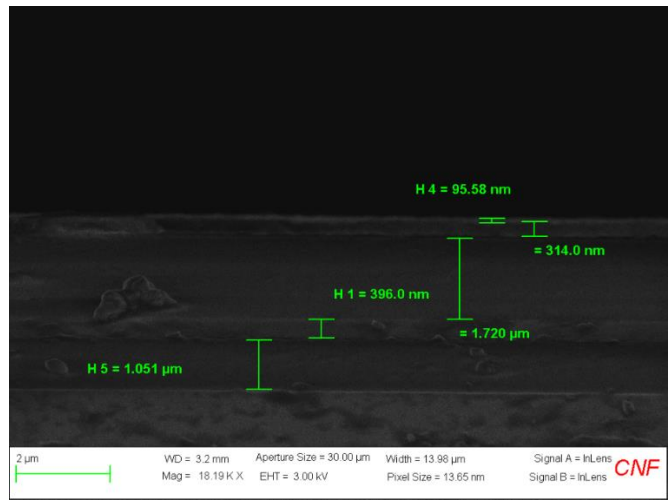

Figure 53. 110nm Cr Hard Mask

After etching the ARC about 110 second with Arc Etch (42.5sccm Ar, 7.5sccm O2, pressure 15mT, RF 40W and rate 48nm/min), the next step is etching the Chrome (27sccm $Cl_2$, 1sccm $O_2$, 2sccm Ar, RIE/ICP: 5W/800W, etch rate 38nm/min) by Trion Etcher tool. Trion Etcher is a tool who purpose is to etch the Chrome. We used a two layers a-Si and SiO2 Bragg Stack to measure the rate of etching for Cr. Therefore, 4minutes and 24seconds is enough to etch down the 110nm of Cr as hard mask, as shown in Figure 54.

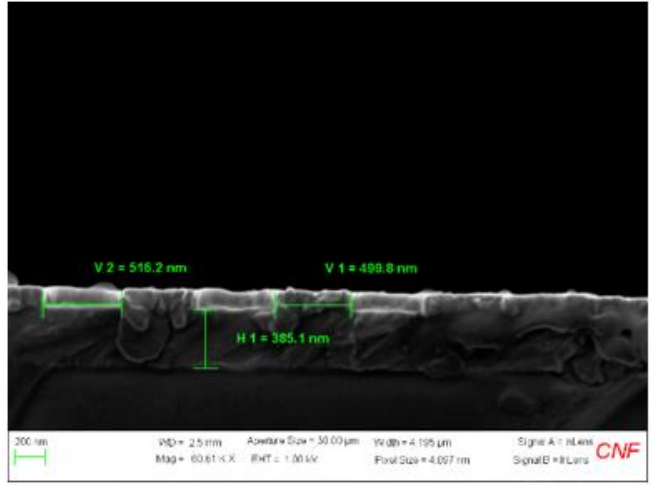
Figure 54. SEM view of the etched Cr

After etching the Cr hard mask, the first a-Si layer was reactive ion etched using HBr gas next. Since the a-Si could oxidize after deposition, we used $BCl_3$ etching gas before the HBr gas, in order to remove the polymer material. The etching rate of HBr gas for etching a-Si is 2 nm/s. Therefore, for etching the top a-Si, 3 minutes and 20 seconds of etching gas was required, as shown in Figure 55.

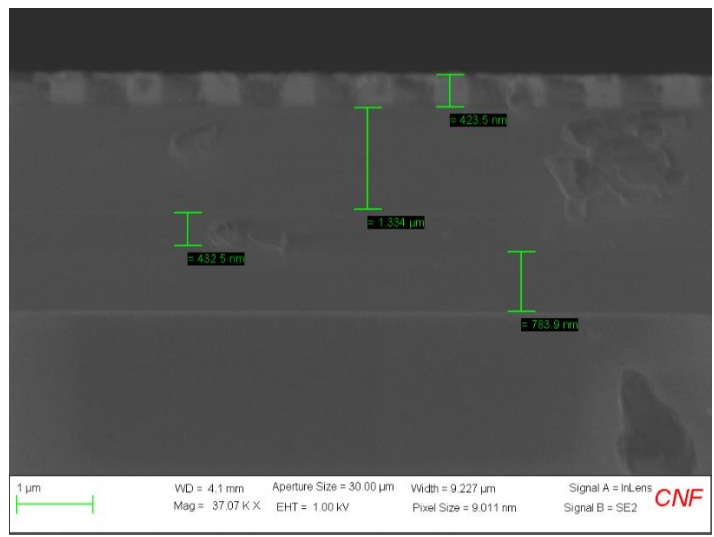
Figure 55. Etching the top a-Si layer with HBr etching gas

The $CH_2F_2$ high helium etching gas was used to etch the middle resonant layer for 15 minutes, made the holes bigger than we expected, depicted in Figure 56.

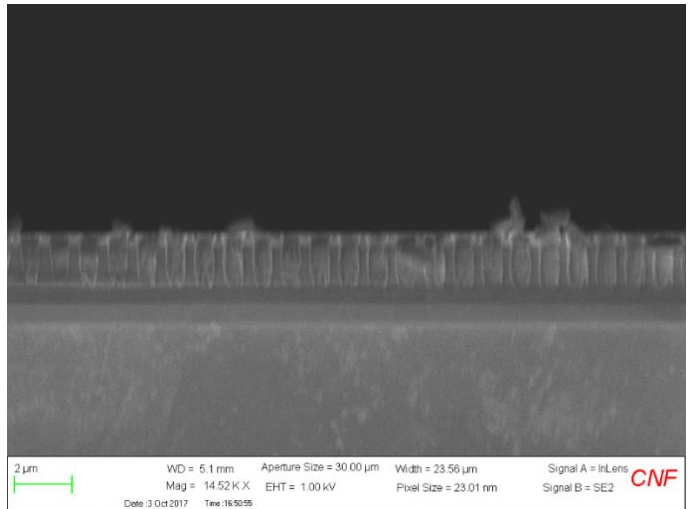
Figure 56. Etching the $SiO_2$ by $CH_2H_2$

The next step was etching the last a-Si layer with HBr and $BCl_3$ gas. Finding the etching time of 14 minutes and half for middle layer $SiO_2$, we started to find the etch time for the second a-Si. Since the aspect ratio changed, the second a-Si should be etched more than 3 minutes and 20 second, (the etch time for the first a-Si). Figure 56 shows the SEM result for etching the second a-Si for 4 minutes and 30 seconds, clarifies the first a-Si feature mouth is bigger than usual, while the second a-Si barely etched. We repeated the last test by etching the middle $SiO_2$ with $CH_2F_2$ gas about 13 minutes and etching the second a-Si with 40 seconds of $BCl_3$ and 5 minutes and 30 seconds of HBr gas, shown in Figure 57.

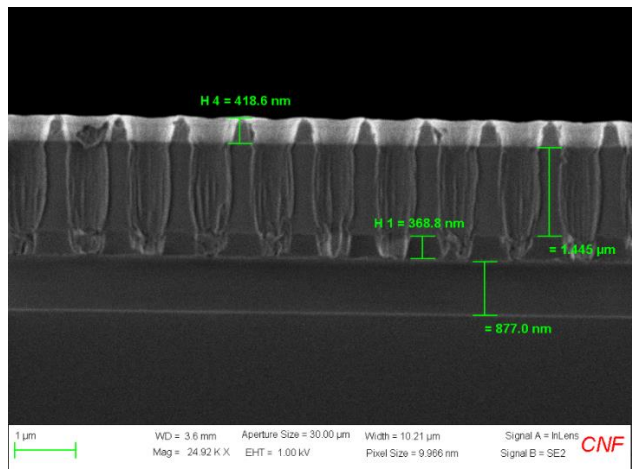
Figure 57. Etching the top and bottom a-Si with HBr and middle SiO2 with $CH_2F_2$

In order to get rid of the polymer created with $CH_2F_2$ gas in middle $SiO_2$ layer, we used another etching gas. The $CHF_3O_2$ created less polymer rather than $CH_2F2$. Increasing the etching time of $SiO_2$ by $CHF_3O_2$, make the walls wider and the hole bigger, as depicted in Figure 58.

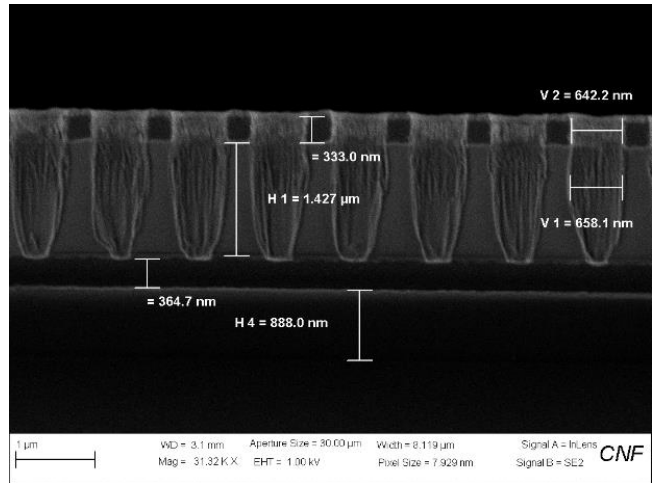
Figure 58. Etching middle SiO2 with $CHF_3O_2$

We repeat the test for etching the first layer of a-Si with $CF_4$ gas for 2 minutes and 30 seconds, then etching the middle sio2 with $CH_2F_2$ gas about 12 minutes, as depicted in Figure 59. However, none of the photoresists were not enough to go through the whole layers.

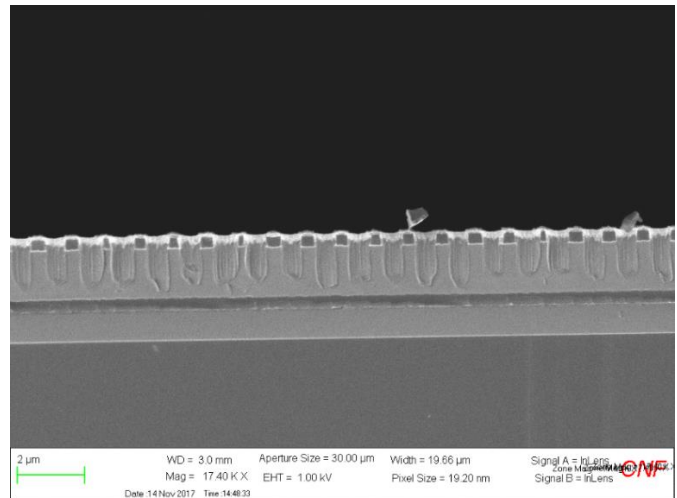
Figure 59. Etching top a-Si with HBr and middle $SiO_2$ for shorter etching time

Therefore, after multiple unsuccessful tests, another trick was devised. The back-side alignment for the purpose of the separate lithography processes was used. We applied the negative resist and, it resulted the best multi etching in the Bragg stack. By putting 4 marks on the wafer during lithography process, either front or back of the wafer, the ASML stepper tool find the marks each time during lithography process. The misalignment after 6 lithography was less than 5 nm and the result after electroplating and polishing is shown in Figure 65.

During the separate lithography process, the holes were filled with so much resist. Therefore, there were two problems related to this thick resist. One problem was focusing over that much dept and distance is very limit for stepper tool. The other problem was the dose must be so much bigger for thick resist, that leads to change the size of the feature. By using the bright field mask instead of dark field mask, the negative resist is exposed flat on the wafer in the same thickness, not in the etched holes. This makes the size of the holes remain the same. Developer solution removes exposed positive photoresist, but for negative photoresist, the unexposed material is removed. Therefore, using the negative resist makes the resist leaves where light exposed and therefore, the photoresist in the holes go away after development, in contrast with positive photoresist which leaves resist where no light exposure, as shown in Figure 60.

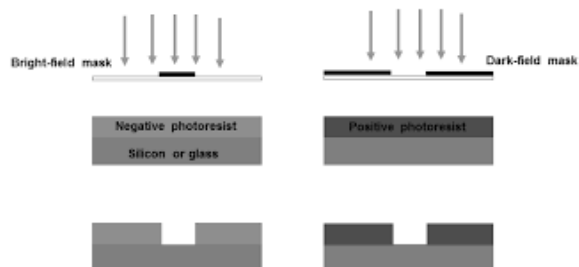

Figure 60: Negative resist with clear field mask (left). Positive resist with dark field mask (right) [53].

**Electroplating**

Electroplating, also known as electrodeposition, is the process of deposit material within electric circuit. In another word, a thin layer of material is bonded on the surface of another substrate based on molecule level. Electroplating is acting like an electrolytic cell, which directs the electricity to transfer metal ions on a conductive surface. The cell includes two separate metallic conductors; metal electrode (anode) and brass electrode (cathode). Two electrodes are inserted in an electrolyte solution consists of the ions and connected to each other through a circuit with a power supply, like a battery. Before electroplating, we did sputter 80nm Pt (pressure 3mTorr, rate 4.69A/sec, DC target) on BS as the adhesion layer for copper. The 5nm Ti (pressure 3mTorr, rate 1.67A/sec, stress 15.2MPa) was sputtered to stick the Pt on BS.

First, we must check the electrode to plate is completely clean. If the plate is not clean, the metal atoms will not form a good connection and rub of the substrate. By turning on the electric current, the electrolyte solution splits up and some of the metal atoms in solution are electroplated in a thin layer on one of the electrodes. On the other hand, positive charged ions in the solutions moving forward to the cathode (negative electrode), which makes they lose their positive charge. Meanwhile, the negative ions of electrolyte solution go to the anode (positive electrode), which makes they lose their negative charge after changing their electrons [54]. For our purpose, when the power is on, the copper sulfate solution breaks into ions. Positive copper ions move to the cathode electrode and depositing a thin layer of copper. In another side, negative sulfate ions attracted to the copper anode electrode and losing electrons and move to the negative electrode. The time of electroplating the copper depends on the current and the concentration of the solution. By increasing this time, ions move faster and copper plates quickly.

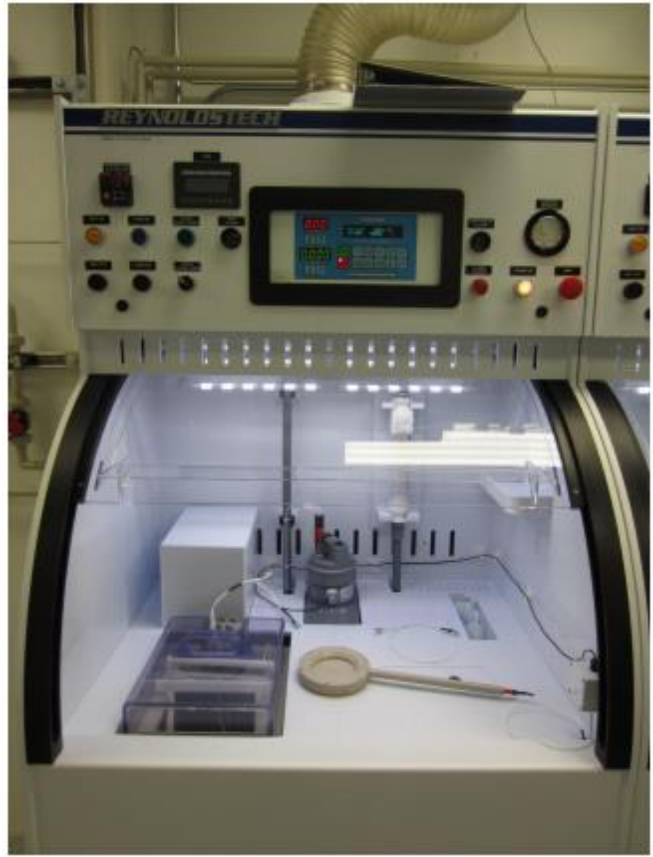

Figure 61. Schematic of CNF Electroplating Station/Hood

The electroplating copper started with the two layers Brag stack. After striping the photoresist from the wafer by $O_2$, first 8nm of titanium sputtered into the holes to cover the edges conformally, as the seed layer for platinum. Then 80nm of platinum sputtered in the holes to cover the edges more thoroughly before electroplating. Pt helps the electroplating process to create a good conductive layer on the insulator walls of the feature. The copper started to be electroplated on the wafer about 4 hours. The electroplating result is seen in Figure 64, for one layer (up) and for 7layer (down) of the BS. Electroplating Cu in the holes was one of the biggest obstacles.

For the seven layers Bragg stack, we tried different current for different forward and reverse pulse time. For the forward, the current is 120 mA, the total time for the forward current is 200 ms, while 60 ms on and 20 ms off. For the reverse, the current is 350 mA, the total time for the reverse current is 180 ms, while 2 ms on and 178 ms off. Figure 61 shows the schematic of CNF Electroplating Station/Hood.

**Chemical Mechanical Polishing**

Chemical mechanical polishing (CMP) is a method that planarize the materials by using the chemical material (slurry) to smooth the surfaces of the variety of materials, including semiconductors, dielectrics or polymers to have the best performance in microprocessors or other industries. CMP has been the second biggest market of nanofabrication industry, a schematic of CMP operation is illustrated in Figure 62.

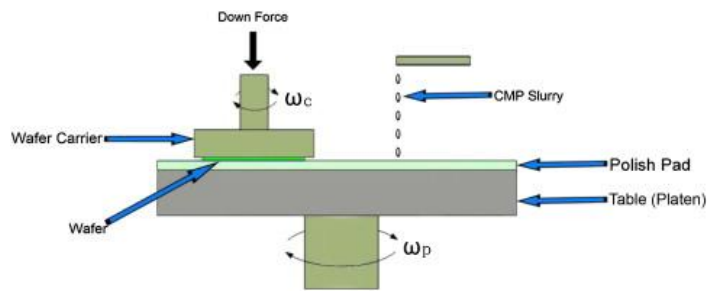

Figure 62. Schematic of CMP operation [55].

The abrasive particles in slurries play important role in CMP. There are different kind of inorganic slurries, including silica, alumina and hydrogen peroxide [55]. The slurry we used for our Bragg stack, M8540-P6, is created for polishing the copper and $SiO_2$ and it is mixed with one-part Hydrogen Peroxide and 14.8 parts M8540 by volume. Figure 63 shows the CMP tool in CNF lab at Cornell University [56].

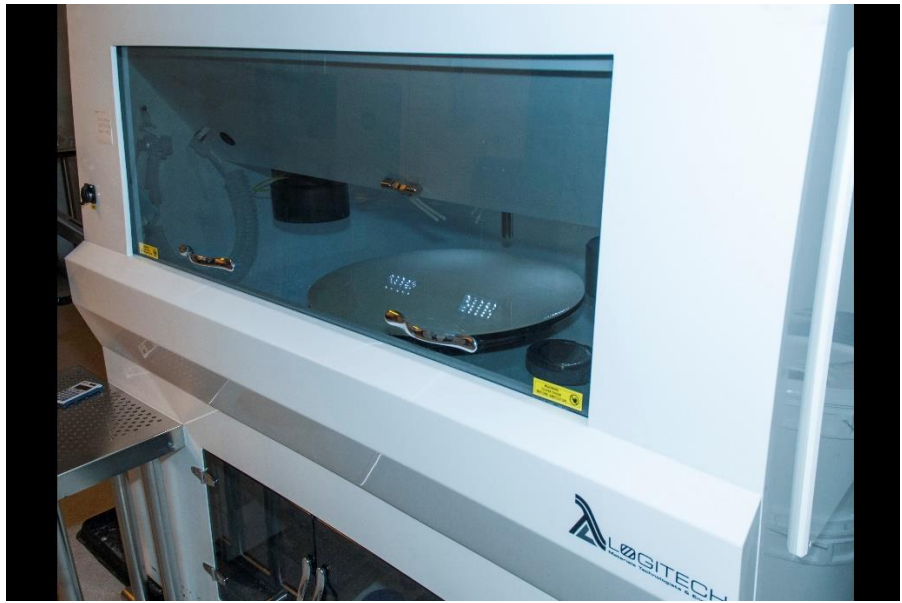

Figure 63. CMP tool in CNF lab at Cornell University [56]

The chemical mechanical polishing (CMP) were performed following the electrodeposition Cu to remove the metal on top of the a-Si layer of 7-layer Bragg stack, leaving only the copper in the holes, as depicted in Figure 64.

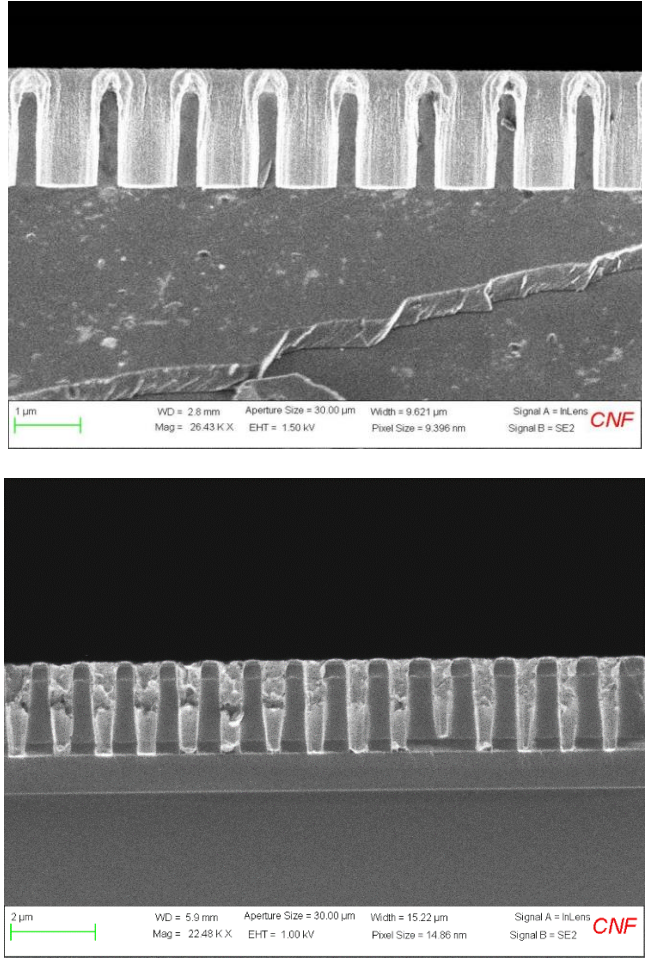

Figure 64. SEM view of electroplating Cu of 1-layer BS (top) and 6-layer HMBS (bottom) after CMP

After polishing the 6-layer BS, the last SiO$_2$ layer was deposited with PECVD method, with the exact rate and time we did for the first SiO$_2$ layer. The result is shown in Figure 65.

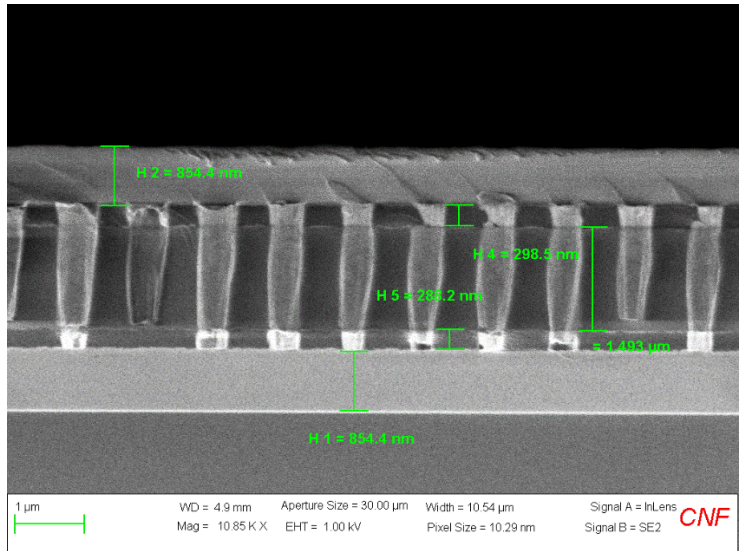
Figure 65. The SEM view of final fabricated 7-layer HMBS

# FTIR Spectrometer

## Fourier Transform Infrared (FTIR)

Fourier transform infrared (FTIR) is the most common type of infrared spectroscopy since the 1988's. When IR radiation goes through the material, some part of the radiation transmits the sample, and some is absorbed by the material. The FTIR detector receives the signals and interprets them to the signals. In another word, the detector recognizes the signals and shows the unique result as a molecular fingerprint In FTIR method, the information of the material placed in the IR is recorded by interferometry. The spectral results by Fourier Transform are used to quantify and analyze the material. The interferograms then decode the spectra into specific recognizable IR fingerprint [57].

## Interferometer

In Fourier transform interferometer, the detected intensity as the function of retardation is the Fourier transform of the intensity as the function of wavelength [58]. In interferometry, the electromagnetic waves are superimposed, which leads to interference and therefore, extract information. The combination of two waves ($E_1$ and $E_2$) with the same frequency ($\omega$) results in a wave ($E_T$) that is defined by the phase difference ($\theta$), as shown in Eq 154.

$$E_T = E_1 \, e^{i\omega t} + E_2 \, e^{i\omega t + \theta} \quad \text{Eq 154}$$

Based on the above equation, when the waves are in-phase the interference is constructive, while out of phase waves make destructive interference. The interference pattern is illustrated in Figure 66. The source emits the IR light that reaches to beamsplitter at C. Then beamsplitter splits the light, transmits it to Point A and reflects it to point B. After that, the two reflected beams rejoint to each other at point C, which is detected by detector.

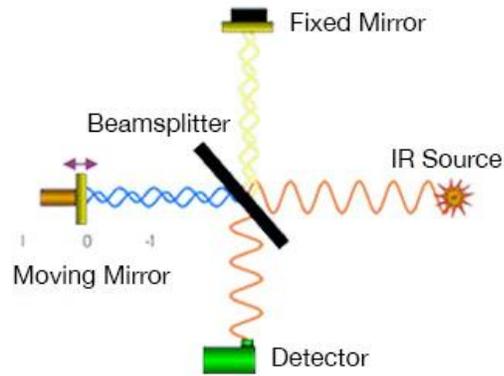

Figure 66. Schematic of Michelson Interferometer [57]

Therefore, the retardation (optical path difference) between two beams is calculated based on Eq 155.

$$\Delta = 2(AC - BC) \quad \text{Eq 155}$$

Also, the phase difference between two beams after rejoint each other for wavelength ($\lambda$) and wavenumber ($\sigma$) is calculated based on Eq 156.

$$\theta = \frac{2\pi}{\lambda} \Delta = 2\pi \sigma \Delta \quad \text{Eq 156}$$

The intensity of light for a monochromatic source, include $I_0$ (intensity at zero path difference (ZPD) when $\Delta = 0$), is calculated based on the Eq 157.

$$I(\Delta) = \frac{1}{2} I_0 (1+\cos(2\pi\sigma\Delta)) \quad \text{Eq 157}$$

Assuming the two beams reflected from the mirrors have the same amplitude, the only variable in Eq 157 is the retardation ($\Delta$), shows that interference can be adjusted by changing the distance of mirrors in respect to the beamsplitter. The total detected spectral radiance is the integral of all incoherent wavenumbers, as shown in Eq 158.

$$I(\Delta) = \frac{1}{2} \int_0^\infty I(\sigma) (1 + \cos(2\pi\sigma\Delta)) \, d\sigma \quad \text{Eq 158}$$

## Fourier Transform Interferometer

The Fourier transform interferometer indicates that the rejoint beams intensity as the function of retardation (I(Δ)) is the Fourier transform of light source intensity as the form of wavenumber (I(σ)). The modulated AC part detector signal is known as interferogram is shown in Eq 159.

$$I(\Delta) = \frac{1}{2} \int_0^\infty I(\sigma) \cos(2\pi\sigma\Delta) \, d\sigma \quad \text{Eq 159}$$

Considering the spectral signal as symmetrical distribution, the interferogram can be shown as Eq 160.

$$I(\Delta) = \frac{1}{2} \int_{-\infty}^\infty I(\sigma) \, e^{i\pi\sigma\Delta} \, d\sigma \quad \text{Eq 160}$$

Using inverse Fourier transform, we can get the intensity as a function of wavenumber, as written in Eq 161.

$$I(\sigma) = 2 \int_{-\infty}^\infty I(\Delta) \, e^{-i\pi\sigma\Delta} \, d\Delta \quad \text{Eq 161}$$

The detector records the intensity related to the constant velocity of the mirror moves as a function of time. The detected intensity, which is the retardation as the function of constant velocity, is transformed to the intensity as the function of wavenumber with the help of inverse Fourier transform [58].

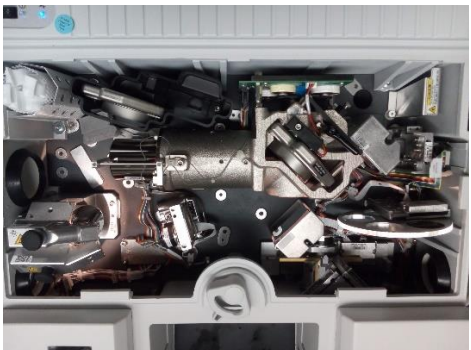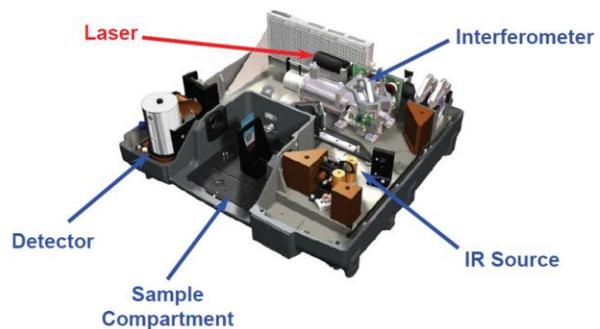

Figure 67. Internal view of the Nicolet 380 FT-IR spectrometer [57]

Figure 67 shows the internal optical view of the Nicolet 380 FT-IR spectrometer. The parts shown in the Figure 67 are basic in all FTIR spectrometers. The broad band visible or IR light are projected from the source to the interferometer. The beamsplitter inside of the interferometer splits the entrance light, and as a partial reflector, transmits the half beam to the moving mirror and reflects the other half to the fixed mirror. Moving mirror makes a path length difference by scanning back and forth. This difference is captured in time with respect to the laser, makes the mirror position specific in each scan. The two reflected beams join back at the beamsplitter and direct to the sample compartment. The detector can easily detect and measure the intensity of the beam, create the interferogram [57].

**Path Difference**

In Michelson Interferometer, the one wavelength of the light coming from the source to the beamsplitter splits to two parts. When the moving mirror has the same distance to the beamsplitter with the fixed mirror, the half part is transmitted to the moving mirror and the other half is reflected to the fixed. Since the distance travel between two mirrors with respect to the beampliter is the same, the constructive interference makes the detected signal big. In this situation, the location of the mirror is named as Zero Path Difference (ZPD), as depicted in Figure 68 [57].

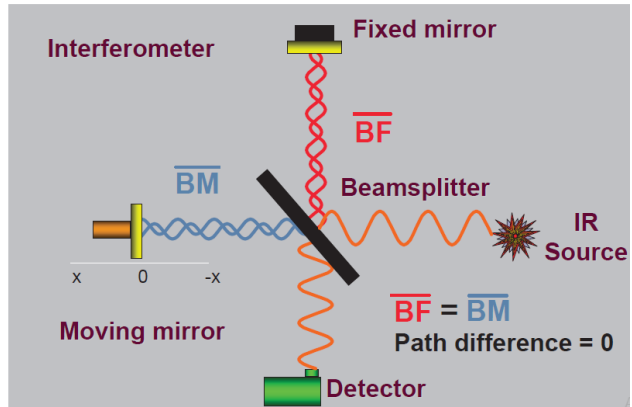
Figure 68. Zero Path Difference [57]

When the distance between the moving mirror and beamsplitter is bigger that distance of the fixed mirror and beamsplitter, the path difference makes an interreference effect. The interference is constructive, when the distance difference between the beams is ¼ of a wavelength and the beams are not considerably out-of-phase. The interference is destructive when the former distance is ½ of a wavelength and there is no signal to be detected, as shown in Figure 69.

When the two interference waves have the same wavelength and phase, and start at the same moment of time, the combination wave has the same wavelength. However, the amplitude is the sum of the amplitude of the both waves, which lead to the constructive interreference. In the same way, when two waves start at the different moment of time (out of phase) make the resulting wave amplitude zero. This interference is called destructive.

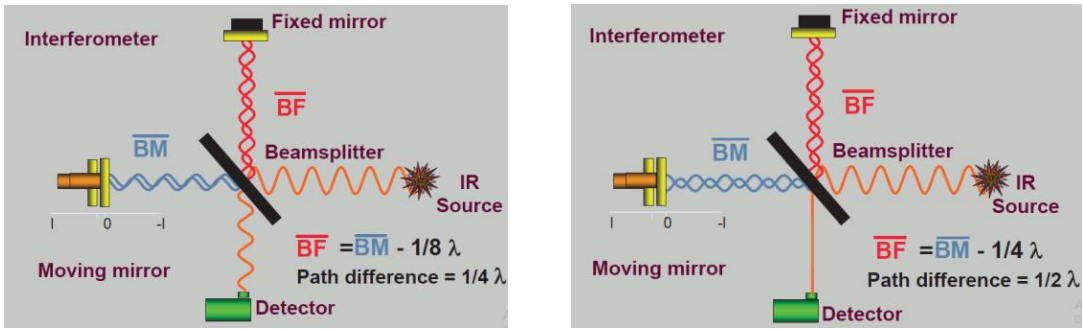

Figure 69. Constructive Interference (left). Destructive Interference (right) [57]

When the interference and path difference are observed in time by keep moving the moving mirror, the wave pattern is measured through time by the detector, which can be seen as picture Figure 70.

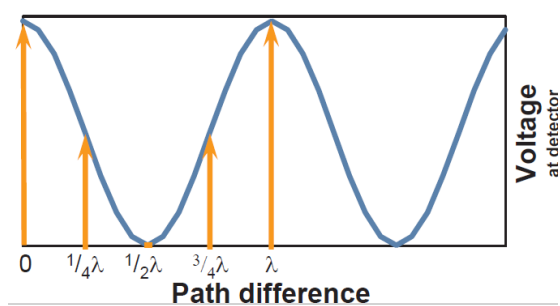

Figure 70. Signal at the Detector [57]

For each frequency, the radiation from the broad band source results in a constructive/destructive cosine interference. The detector detects signals (interferogram) as the sum of the separate interference pattern of each frequency. The energy changes are recorded by detector and shown in terms of data points and voltages. The data point represents the spectral information of interferogram as a function of resolution.

A Fast Fourier transform (FFT) is the method that interprets the raw information and calculates the frequency and intensity from the interferogram and is useful for multiple waves calculation. As depicted in Figure 71, the calculated spectrum in the right is the interferogram for no sample (background), interpreted by FFT. The spectrum shows the amount of signal identified by the detector.

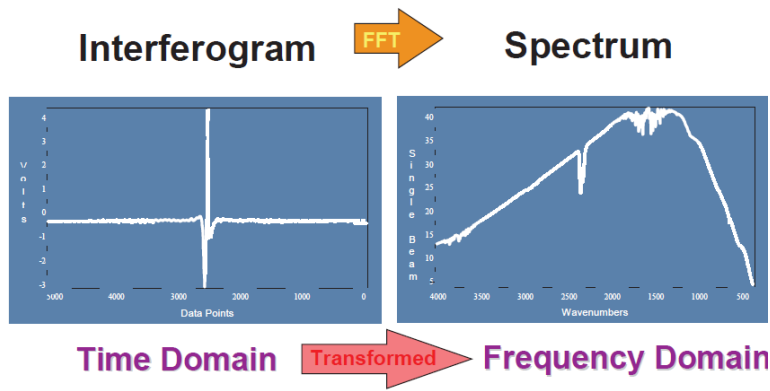

Figure 71. Fast Fourier Transformation [57]

The single beam spectrum shown in Figure 72 is for a sample in sample compartment, which is used to test the tool (signal/noise). If the industrial noise isn't changed enough, the single beam background can be used for subsequent runs and save time.

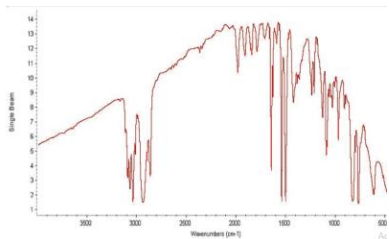

Figure 72. Single Beam Sample [57]

In order to calculate the transmission, all we need is measuring the energy passing it through the sample divided by the background energy, as it is depicted in Figure 73. Light scattering and refraction may cause the shift in the transmission result. The unique data encoded in IR spectrum is very helpful to recognize the material.

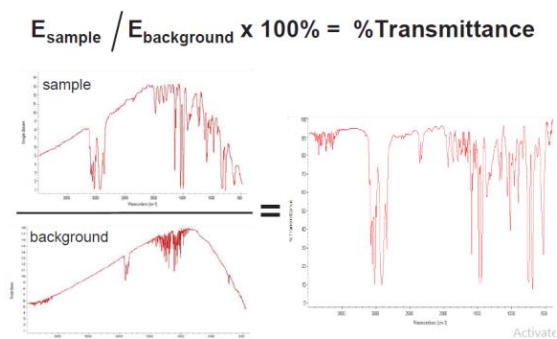

Figure 73. Processing Single Beam [57]

**Resolution, Aperture and Data Spacing in OMNIC software**

In this section, we are going to learn more about the experiment setup in OMNIC software for FTIR Spectrometer. The optical velocity is the speed at which the moving mirror in the interferometer travels. That speed is listed in cm/sec. This speed will affect how quickly the scans can be collected. The slower the speed the longer it will take to collect a given number of scans. For MCT detectors we use of 1.8988 for the optical velocity and for a DTGS detector we recommend 0.4747 or 0.6329, as shown in Figure 74. The Gain setting is an electronic gain that the instrument can apply to the detector signal. This setting can be used if you have a small natural signal at the detector. Increasing the gain will increase the interferogram signal to better fill the analog to digital converter (ADC). The reasoning to do this is typically to help with signal to noise characteristics for a small signal. If we choose to use a higher gain setting, the interferogram maximum should not exceed 9 volts, otherwise we risk oversaturating the ADC, which will cause strange looking backgrounds and sample spectra.

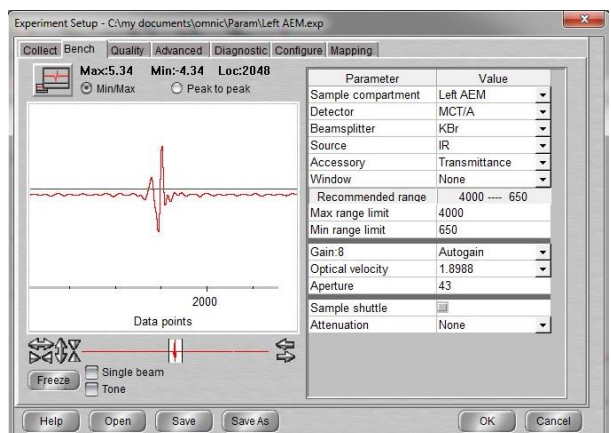
Figure 74. A Schematic of the Bench tab in OMNIC software for FTIR Spectrometer

The distance between two closely peaked in the spectrum is called resolution. The highest resolution in iS50 spectrometer belongs to the 0.125 (cm$^{-1}$) data spacing setting. The data spacing shows how far two data point will be in the detected spectra. The aperture size changes automatically, whenever the resolution changes. On the other hand, the large aperture increases the band distortion at high resolution collections and therefore, it is not effective for collecting the high-resolution spectra.

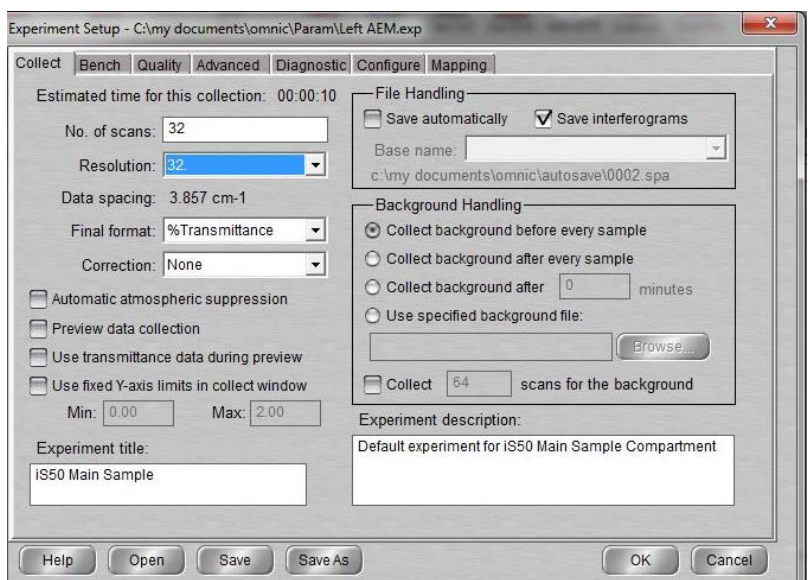
Figure 75. A Schematic of the Experiment Setup in OMNIC software for FTIR

Figure 75 shows a schematic of the experiment setup in OMNIC software for FTIR Spectrometer. The zero filling options in experiment set up makes the software add points between collected data point in order to improve the spectrum shape and doesn't affect the spectrum resolution. Figure 76 shows the resolution of Full width at half maximum (FWHM) 0.3777 $\mu m$ for the simulated spectra of the proposed seven-layer Bragg stack at center wavelength 4.32 $\mu m$. For our center wavelength, the data spacing 3.857 cm$^{-1}$ available in the software, satisfied the target resolution, as calculated in Eq 162.

$$\text{Data Spacing } (\Delta\omega) = \frac{1}{\lambda_1} - \frac{1}{\lambda_2} = \frac{\lambda_2 - \lambda_1}{\lambda_1 \lambda_2} = 3.857 \text{ cm}^{-1} \quad \text{Eq 162}$$

For two close points $\lambda_1 \approx \lambda_2$, $\Delta\lambda = 3.857 \lambda^2$. For 3.857 cm$^{-1}$ * ($\frac{1cm}{10^4 \mu m}$) = 3.857 *$10^{-4}$ $\mu m$ at $\lambda = 4.38$ $\mu m$, we have $\Delta\lambda = 0.0074$ $\mu m$, which satisfies the resolution FWHM 0.3777 $\mu m$ in our proposed HMBS.

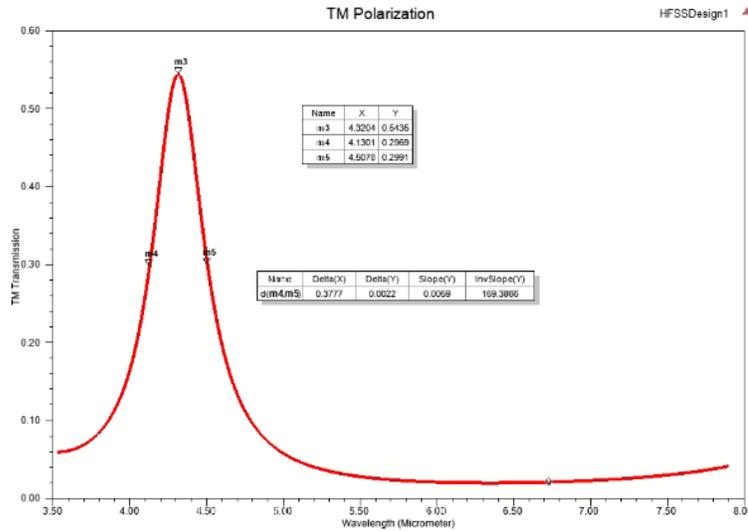

Figure 76. Full Width at Half Maximum (resolution) of proposed HMBS simulated in HFSS

## Test the Fabricated HMBS Filter by FTIR

**Collimating the light**

The FTIR setup consists of an IR light source, KBr beam splitter and nitrogen cooled DTGS detector with a spot size of $100 \times 100 \ \mu m^2$. The output signal in compartment of iS50 FTIR has the diameter of 1 inch, which is so much bigger than the sample pieces used for the testing. In order to have smaller light, using aperture and the light collimation is necessary. These requirements lead us to direct the light out of the compartment, as depicted in Figure 77. The passport mirror inside FTIR is moved to the glass window at the right and hits the reflective beam expander. The beam expander has entrance and exit apertures of 4 mm and 20mm, that the exit part is used to create the narrow bandwidth light. The two mirrors beside expander are the reflectors to direct the light to the detector.

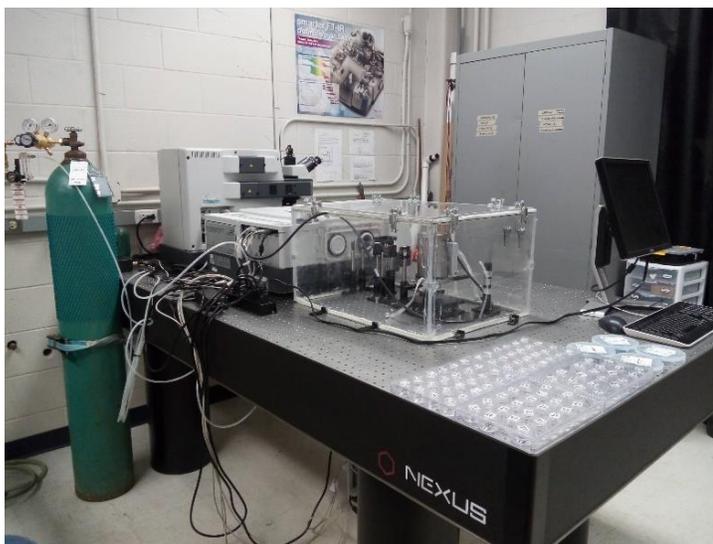
Figure 77. Optical setup to collimate the light and purging the system

**Purge Gas for FTIR Systems**

FTIR is a powerful analytical technique of scanning IR spectrophotometer. However, there are significant disadvantage in FTIR systems; including the additional peaks in the result spectra. These peaks belong to the water vapor or carbon monoxide in the chamber compartment and make difficult to interpret the important peaks of the final spectra. Therefore, the entire FTIR system should be purged and sealed with the $CO_2$ and humidity free gases [59]. For the proposed research the nitrogen gas is used as dry purging gas in a sealed plastic box, while the purge gas regulator is tuned to the pressure of 20 psig[1], as demonstrated in Figure 78. We used two IR convex lenses (120mm and 200mm) to transfer the collimated light to the MCT detector. The nitrogen is lighter than air and combines with air at room temperature.

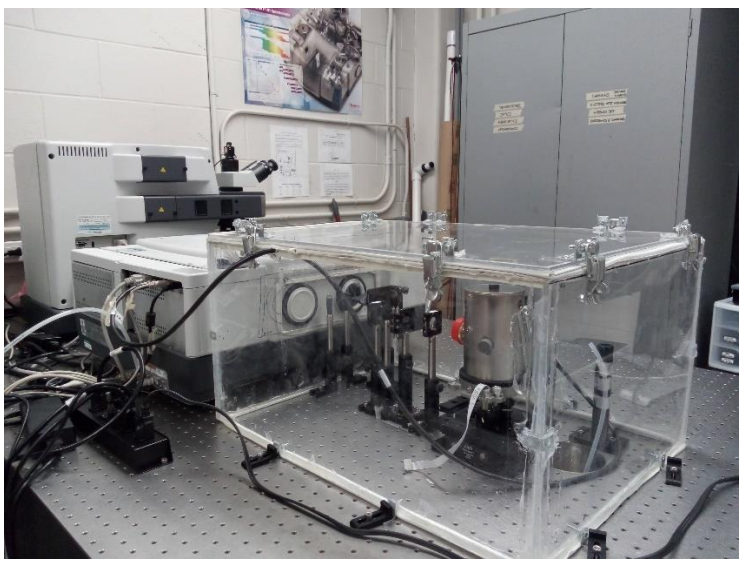

Figure 78. Purging set up at Clarkson University lab

---

[1] Pounds per Square Inch Gauge

**Measuring the Transmission of Fabricate HMBS**

Comparing the HFSS, after setup the tool, the HMBS was tested with the MCT detector in FTIR. The Figure 79 shows the measured TM transmission of the proposed HMBS in FTIR at Clarkson University. The filter shows the high narrowband band peak for the transmission of %33 at the wavelength of $4.33 \mu m$.

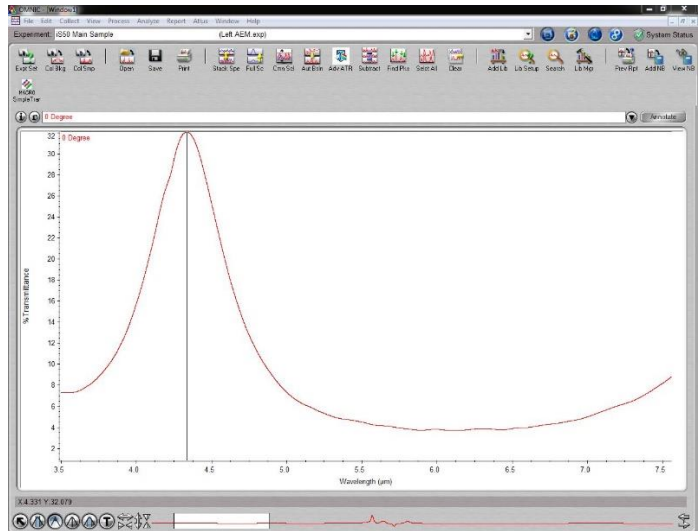

Figure 79. The measured TM transmission of the proposed HMBS in FTIR at Clarkson University

The main purpose of our proposed filter is showing dependence free center-wavelength to the different angles of the incident TM polarized light. This goal is achieved by testing the fabricated 7-layer HMBS for different angles of light incidence from 0 to 25 degree, by step of 5 degree, which is shown in Figure 80.

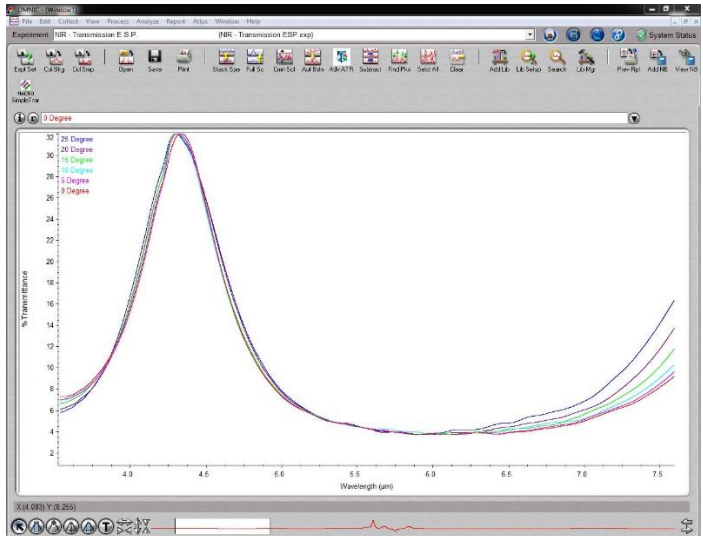
Figure 80. Angle sweep of 0-25 degree for fabricated HMBS in FTIR

As it is clear in Figure 81 , Center wavelength is locked at 4.32 $\mu m$ for different angles of incident light.

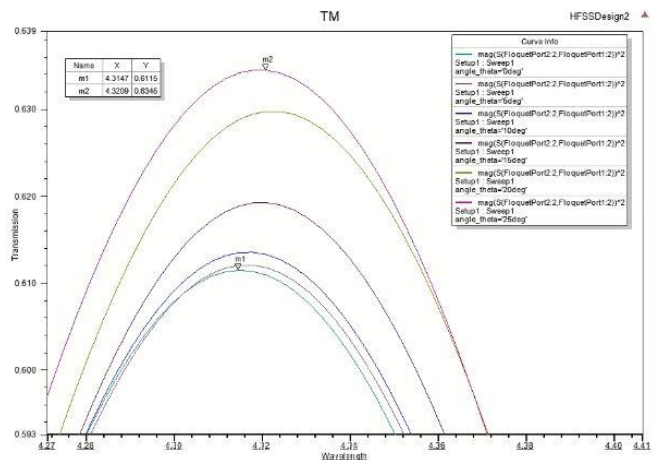
Figure 81. Locked Center wavelength at 4.32 micrometer for different angles

### Comparing the Results

Now that we implemented the Maxwell-Garnett theory and finite element method (HFSS) to simulate the proposed HMBS, and test the fabricated HMBS with FTIR, it's time to compare the three results. The finite element method shows the precise transmission with calculating the frequency dependent permittivities of the materials used in the simulated

HMBS. This makes it different than the result from Maxwell-Garnett theory that the $\varepsilon_{zz}$ and $\varepsilon_{xy}$ are used to calculate the energy and transmission of the filter. Although the $0.5\mu m$ diameter of the wire is not a big number, but it makes it harder to calculate the transmission with Maxwell-Garnett theory, in comparison with HFSS.

The yellow line in Figure 82 shows lower transmission in comparison to the HFSS and Maxwell-Garnett theory. This is because the author had to do the last step of fabrication in a different way. After CMP, the author broke the wafer to see the cross section in SEM. Seeing Cu on top of the last a-Si with SEM, the author had to do more polishing. However, the tool accepted just the whole wafer (not a piece of wafer) and author had to continue polishing it with hand. Therefore, CMP brushing operation was not done for the polished wafer and few "picometer" of the Cu remained on the wafer. However, the main goal of this project, angle-independent center-wavelength of HMBS to the different angle of incident TM light, is achieved as shown in Figure 81.

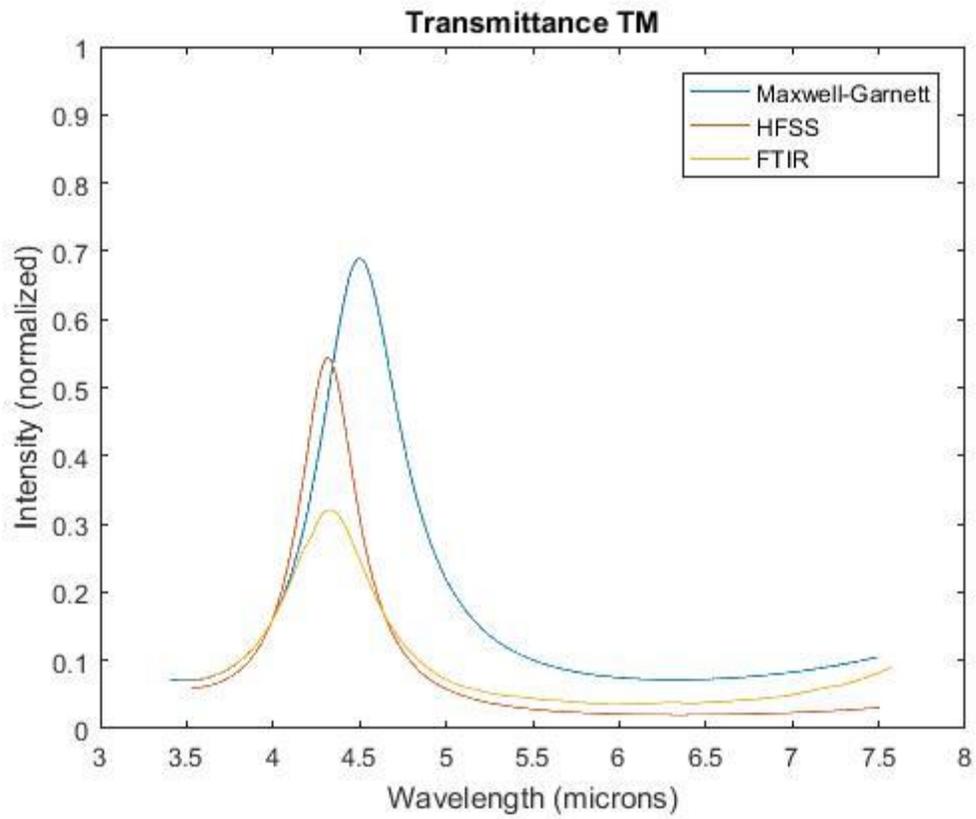

Figure 82. Comparison the measured transmission of the proposed HMBS with Maxwell-Garnet theory (blue), finite element method (red) and FTIR (yellow)

# CHAPTER IV: CONCLUSION

In this work, the angle-independent hyperbolic metamaterial filter was simulated and analyzed with finite element method and the Maxwell-Garnett theory in MWIR regime. The integration of hyperbolic metamaterial with traditional notch filters let us to eliminate the changing of transmission properties on normal and oblique TM polarized incidence. The focus of our thesis was districting the shift to shorter wavelength on angle; this is more desirable by fabricating the metal wires array in the dielectric layers of the SiO2/a-Si filter. Furthermore, by changing the geometry of the unit cell of the filter, resonant condition is achieved for the wide MIIR spectral band. The narrowband TM transmission peak of the proposed HMBS is locked at the special center-wavelength for all the angle of incident light, where the Cu is low-loss at the MWIR range. The novel method in the fabrication of the proposed filter implemented multiple lithography steps for etching the three middle dielectric layers, represents the magnificent alignment in the etched holes. The state-of-the-art technique is published in the NanoMeter newsletter of the Cornell Nanoscale Facility, shown in Figure 83. By modifying the geometric size of the structure and the materials, HMBS filters can operate in SWIR, MWIR, LWIR and visible spectral range in different applicable applications.

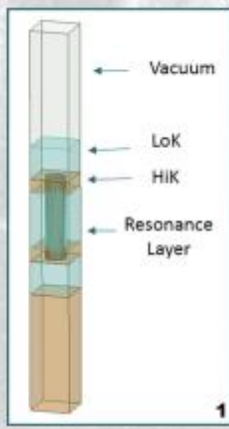
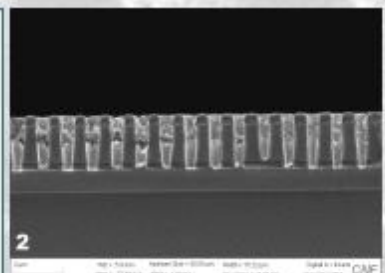
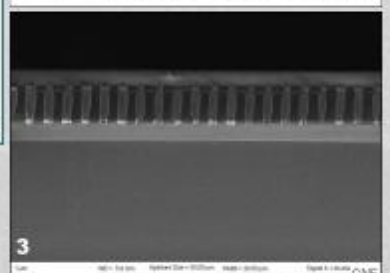
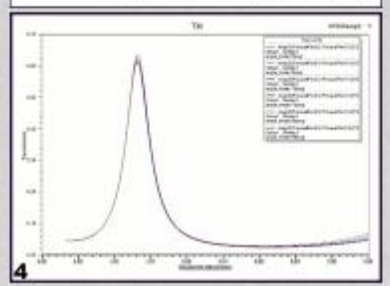

Figure 83. Published work at Nanometer newsletter of CNF at Cornell University [60]

# Appendix A

Here is the MATLAB code to calculate the transmission of the proposed 7-layerd Bragg stack with RCWA method, explained in Chapter II.

## MATLAB Code for TM Transmittance of Proposed Bragg Stack

```
CenterWavelength = 5;           %Units of microns

%Rod Material:  1 for Ag, 2 for Al, 3 for Au, 77 for Cu
Rod_Material = 77;
Rod_Epsilon = -800;
MaterialA = 1.45^2; %LoK SiO2
MaterialB = 3.43^2; %HiK a-Si
radius = .25;                   %Units of microns
period = 1;                     %Units of microns

Filename = 'Test_File_09_21_2016F.mat';

%Layers should be an odd number
Layers = 9;
Interfaces = Layers-1;

Epsilon_Matrix(1) = 1;
Epsilon_Matrix(2) = 1.45^2;
Epsilon_Matrix(Layers-1) = 1.45^2;
Epsilon_Matrix(Layers) = 3.43^2;

Epsilon_Rod = DielectricConstantCorrected(Rod_Material,Rod_Epsilon,1.2408/CenterWavelength);

for cnt = 3:2:Layers-2
    Epsilon_Matrix(cnt) = MaterialB;  %Epsilon for the Bulk dielectric material
    c(cnt) =
wiremesh_thickness_calculator(CenterWavelength,Epsilon_Matrix(cnt),Epsilon_Rod,period,radius);

    Epsilon_Matrix(cnt+1) = MaterialA;  %Epsilon for the Bulk dielectric material
    c(cnt+1) =
wiremesh_thickness_calculator(CenterWavelength,Epsilon_Matrix(cnt+1),Epsilon_Rod,period,radius);

end

c(5) = 0.0; %Thickness of Layers
c(1) = 3;
c(2)=CenterWavelength/4/1.45;
c(Layers) = 3;
c(Layers-1)=CenterWavelength/4/1.45;

theta = 0; %0/180*pi;   %Theta (angle of incidence) will be in radians
wavelengthmin = 3.5;            %Units in microns
```

```matlab
wavelengthmax = 8.5;          %Units in microns
wavelengthpnts = 750;         %Units in nm
%%%%%%%%%%%%%%%%%%%%%%%%%%%%%%%%%%%%%%%%%%%%%

wavelength = linspace(wavelengthmin,wavelengthmax,wavelengthpnts);
reflectionTM = zeros(wavelengthpnts,1);
transmissionTM = zeros(wavelengthpnts,1);
TotalEnergyTM = zeros(wavelengthpnts,1);
betaTM = zeros(Layers,1);
betatest=zeros(wavelengthpnts,1);

for cnt = 1:wavelengthpnts
    knot = 2*pi/wavelength(cnt);
    kxnot = sqrt(Epsilon_Matrix(1))*knot*sin(theta);

    Epsilon_Rod(cnt) = DielectricConstantCorrected(Rod_Material,Rod_Epsilon,1.2408/wavelength(cnt));

    [EpsilonA_xx(cnt),EpsilonA_yz(cnt)] = Epsilon_wiremesh(Epsilon_Matrix(2),Epsilon_Rod(cnt),period,radius);
    [EpsilonB_xx(cnt),EpsilonB_yz(cnt)] = Epsilon_wiremesh(Epsilon_Matrix(3),Epsilon_Rod(cnt),period,radius);

    for cntlayers=1:Layers

        if cntlayers==1
            Epsilon_xy(cntlayers) = Epsilon_Matrix(cntlayers);
            Epsilon_zz(cntlayers) = Epsilon_Matrix(cntlayers);
            betaTM(cntlayers) = sqrt(Epsilon_Matrix(cntlayers)*knot^2 - kxnot^2);

        elseif cntlayers==Layers
            Epsilon_xy(cntlayers) = Epsilon_Matrix(cntlayers);
            Epsilon_zz(cntlayers) = Epsilon_Matrix(cntlayers);
            betaTM(cntlayers) = sqrt(Epsilon_Matrix(cntlayers)*knot^2 - kxnot^2);

        elseif cntlayers==Layers-1
            Epsilon_xy(cntlayers) = Epsilon_Matrix(cntlayers);
            Epsilon_zz(cntlayers) = Epsilon_Matrix(cntlayers);
            betaTM(cntlayers) = sqrt(Epsilon_Matrix(cntlayers)*knot^2 - kxnot^2);

        elseif cntlayers==2
            Epsilon_xy(cntlayers) = Epsilon_Matrix(cntlayers);
            Epsilon_zz(cntlayers) = Epsilon_Matrix(cntlayers);
            betaTM(cntlayers) = sqrt(Epsilon_Matrix(cntlayers)*knot^2 - kxnot^2);

        else

[Epsilon_zz(cntlayers),Epsilon_xy(cntlayers)]=Epsilon_wiremesh(Epsilon_Matrix(cntlayers),Epsilon_Rod(cnt),period,radius);
            betaTM(cntlayers) = sqrt(Epsilon_xy(cntlayers)*knot^2 - Epsilon_xy(cntlayers)/Epsilon_zz(cntlayers)*kxnot^2);
```

```matlab
        end
        
        %Constructing the TM Transfer Matrices
        M11 = 1;
        M12 = 1;
        M21 = -betaTM(cntlayers)/Epsilon_xy(cntlayers);
        M22 = betaTM(cntlayers)/Epsilon_xy(cntlayers);
        M_TM(cntlayers).matrix = [M11,M12;M21,M22];
        
        inverseM11 = 1/2;
        inverseM12 = -Epsilon_xy(cntlayers)/2/betaTM(cntlayers);
        inverseM21 = 1/2;
        inverseM22 = Epsilon_xy(cntlayers)/2/betaTM(cntlayers);
        inverseM_TM(cntlayers).matrix = [inverseM11,inverseM12;inverseM21,inverseM22];
        
        phiTM_component = exp(1i*betaTM(cntlayers)*c(cntlayers));
        inversephiTM_component = exp(-1i*betaTM(cntlayers)*c(cntlayers));
        phiTM(cntlayers).matrix = [inversephiTM_component,0;0,phiTM_component];
    end
    
    
    tempTM = eye(2,2);
    for cntlayers = 2:Layers-1
        TtildaTM(cntlayers).matrix = inverseM_TM(cntlayers).matrix*M_TM(cntlayers+1).matrix*phiTM(cntlayers+1).matrix;
        tempTM = tempTM*TtildaTM(cntlayers).matrix;
    end
    TtildaTM_total = tempTM;
    
    PTM = M_TM(2).matrix*phiTM(2).matrix*TtildaTM_total;
    
    transmissionTM(cnt) = -2*betaTM(1)/Epsilon_Matrix(1)/(PTM(2,1)-betaTM(1)/Epsilon_Matrix(1)*PTM(1,1));
    reflectionTM(cnt) = PTM(1,1)*transmissionTM(cnt) - 1;
    ReflectanceTM(cnt) = abs(reflectionTM(cnt))^2;
    TransmittanceTM(cnt) = Epsilon_Matrix(1)/Epsilon_Matrix(Layers)*betaTM(Layers)/betaTM(1)*abs(transmissionTM(cnt))^2;
    TotalEnergyTM(cnt) = ReflectanceTM(cnt) + TransmittanceTM(cnt);
    
end

Figure;
plot(wavelength,real(Epsilon_Rod),'r',wavelength,imag(Epsilon_Rod),'k');

Figure;
plot(wavelength,TransmittanceTM)
ylim([0 1])
```

# Appendix B

The surface plasmon dispersion curve and electric fields are coded with MTALAB, explain in Chapter II.

## MATLAB Code for Surface Plasmon Dispersion Curve

```
knot = 2*pi./wavelength;

temp_a  = 2*pi/wavelengthmax*sqrt(epsilonmatrix);
temp_b  = 2*pi/wavelengthmin*sqrt(epsilonmatrix);
kz_min = 0.1*temp_a;
kz_max = 1.50*temp_b;
kz = linspace(kz_min,kz_max,wavelengthpnts);

wavelength_mode1_0th = [];
kz_mode1_0th = [];
m_mode1_0th = [];
epsilonrod_mode1_0th = [];
epsilonmatrix_mode1_0th = [];

wavelength_mode2_0th = [];
kz_mode2_0th = [];
m_mode2_0th = [];
epsilonrod_mode2_0th = [];
epsilonmatrix_mode2_0th = [];

wavelength_mode_HO = [];
kz_mode_HO = [];
m_mode_HO = [];
epsilonrod_mode_HO = [];
epsilonmatrix_mode_HO = [];

for cntm = 0:maxm
   for cntkz = 1:wavelengthpnts-1
      beta= sqrt(kz(cntkz)^2 - knot.^2*epsilonmatrix);      %Outside the cylinder
      gamma = sqrt(knot.^2.*epsilonrod - kz(cntkz)^2);       %Inside the cylinder

      if cntm == 0
         LHS1 = epsilonrod./gamma.*besselj(1,gamma*radius)./besselj(0,gamma*radius);
         RHS1 = -epsilonmatrix./beta.*besselk(1,beta*radius)./besselk(0,beta*radius);
         LHS2 = 1./gamma.*besselj(1,gamma*radius)./besselj(0,gamma*radius);
         RHS2 = -1./beta.*besselk(1,beta*radius)./besselk(0,beta*radius);

         for cntomega = 1:wavelengthpnts-1
            if real(LHS1(cntomega)) > real(RHS1(cntomega)) && real(LHS1(cntomega+1)) < real(RHS1(cntomega+1))
```

```
                wavelength_mode1_0th = [wavelength_mode1_0th,wavelength(cntomega)];
                kz_mode1_0th = [kz_mode1_0th,kz(cntkz)];
                m_mode1_0th = [m_mode1_0th,cntm];
                epsilonrod_mode1_0th = [epsilonrod_mode1_0th,epsilonrod(cntomega)];
                epsilonmatrix_mode1_0th = [epsilonmatrix_mode1_0th,epsilonmatrix];
            elseif real(LHS1(cntomega)) < real(RHS1(cntomega)) && real(LHS1(cntomega+1)) > real(RHS1(cntomega+1))
                wavelength_mode1_0th = [wavelength_mode1_0th,wavelength(cntomega)];
                kz_mode1_0th = [kz_mode1_0th,kz(cntkz)];
                m_mode1_0th = [m_mode1_0th,cntm];
                epsilonrod_mode1_0th = [epsilonrod_mode1_0th,epsilonrod(cntomega)];
                epsilonmatrix_mode1_0th = [epsilonmatrix_mode1_0th,epsilonmatrix];
            end

            if real(LHS2(cntomega)) > real(RHS2(cntomega)) && real(LHS2(cntomega+1)) < real(RHS2(cntomega+1))
                wavelength_mode2_0th = [wavelength_mode2_0th,wavelength(cntomega)];
                kz_mode2_0th = [kz_mode2_0th,kz(cntkz)];
                m_mode2_0th = [m_mode2_0th,cntm];
                epsilonrod_mode2_0th = [epsilonrod_mode2_0th,epsilonrod(cntomega)];
                epsilonmatrix_mode2_0th = [epsilonmatrix_mode2_0th,epsilonmatrix];
            elseif real(LHS2(cntomega)) < real(RHS2(cntomega)) && real(LHS2(cntomega+1)) > real(RHS2(cntomega+1))
                wavelength_mode2_0th = [wavelength_mode2_0th,wavelength(cntomega)];
                kz_mode2_0th = [kz_mode2_0th,kz(cntkz)];
                m_mode2_0th = [m_mode2_0th,cntm];
                epsilonrod_mode2_0th = [epsilonrod_mode2_0th,epsilonrod(cntomega)];
                epsilonmatrix_mode2_0th = [epsilonmatrix_mode2_0th,epsilonmatrix];
            end
        end

        temph = 2*pi./wavelength_mode1_0th*light;
        energy_mode1_0th = (hbar/joule)*temph;
        tempi = 2*pi./wavelength_mode2_0th*light;
        energy_mode2_0th = (hbar/joule)*tempi;

    else
        tempc = 0.5*(besselj(cntm-1,gamma*radius) - besselj(cntm+1,gamma*radius));
        tempd = -0.5*(besselk(cntm-1,beta*radius) + besselk(cntm+1,beta*radius));

        T1 = epsilonrod./gamma.*tempc./besselj(cntm,gamma*radius);
        T2 = epsilonmatrix./beta.*tempd./besselk(cntm,beta*radius);
        T3 = 1./gamma.*tempc./besselj(cntm,gamma*radius);
        T4 = 1./beta.*tempd./besselk(cntm,beta*radius);
        LHS = (T1 + T2).*(T3 + T4);

        tempe = gamma.^2;
        tempf = beta.^2;
        T1 = epsilonrod./tempe;
        T2 = epsilonmatrix./tempf;
        T3 = 1./tempe;
        T4 = 1./tempf;
        RHS = cntm^2/radius^2.*(T1 + T2).*(T3 + T4);

        for cntomega = 1:wavelengthpnts-1
```

```matlab
                if real(LHS(cntomega)) > real(RHS(cntomega)) && real(LHS(cntomega+1)) < real(RHS(cntomega+1))
                    wavelength_mode_HO = [wavelength_mode_HO,wavelength(cntomega)];
                    kz_mode_HO = [kz_mode_HO,kz(cntkz)];
                    m_mode_HO = [m_mode_HO,cntm];
                    epsilonrod_mode_HO = [epsilonrod_mode_HO,epsilonrod(cntomega)];
                    epsilonmatrix_mode_HO = [epsilonmatrix_mode_HO,epsilonmatrix];
                elseif real(LHS(cntomega)) < real(RHS(cntomega)) && real(LHS(cntomega+1)) > real(RHS(cntomega+1))
                    wavelength_mode_HO = [wavelength_mode_HO,wavelength(cntomega)];
                    kz_mode_HO = [kz_mode_HO,kz(cntkz)];
                    m_mode_HO = [m_mode_HO,cntm];
                    epsilonrod_mode_HO = [epsilonrod_mode_HO,epsilonrod(cntomega)];
                    epsilonmatrix_mode_HO = [epsilonmatrix_mode_HO,epsilonmatrix];
                end
            end
            tempg = 2*pi./wavelength_mode_HO*light;
            energy_mode_HO = (hbar/joule)*tempg;

        end

        %disp(['m=',num2str(cntm),'  cntkz=',num2str(cntkz)]);

    end

    if cntm == 0
        Mode.kz = kz_mode1_0th;
        Mode.energy = energy_mode1_0th;
        Mode.m = m_mode1_0th;
        Mode.epsilonrod = epsilonrod_mode1_0th;
        Mode.epsilonmatrix = epsilonmatrix_mode1_0th;

        Mode.kz = [Mode.kz,kz_mode2_0th];
        Mode.energy = [Mode.energy,energy_mode2_0th];
        Mode.m = [Mode.m,m_mode2_0th];
        Mode.epsilonrod = [Mode.epsilonrod,epsilonrod_mode2_0th];
        Mode.epsilonmatrix = [Mode.epsilonmatrix,epsilonmatrix_mode2_0th];
    else
        Mode.kz = [Mode.kz,kz_mode_HO];
        Mode.energy = [Mode.energy,energy_mode_HO];
        Mode.m = [Mode.m,m_mode_HO];
        Mode.epsilonrod = [Mode.epsilonrod,epsilonrod_mode_HO];
        Mode.epsilonmatrix = [Mode.epsilonmatrix,epsilonmatrix_mode_HO];
    end

end

hold off;
light_line = (hbar/joule)*knot*light*sqrt(epsilonmatrix);

SPHandles = guihandles;
axes(SPHandles.Dispersion)
```

```matlab
plot(kz_mode1_0th,energy_mode1_0th,'Marker','o','MarkerEdgeColor','green','LineStyle','none');    %y values are in eV
ylabel('Energy (eV)');
xlabel('kz (inverse microns)');
hold on;
plot(kz_mode2_0th,energy_mode2_0th,'Marker','o','MarkerEdgeColor','green','LineStyle','none')
plot(knot,light_line,'Marker','o','MarkerEdgeColor','Yellow');

frequency_mode_HO = 2*pi./wavelength_mode_HO*light;
energy_mode_HO = (hbar/joule)*frequency_mode_HO;
plot(kz_mode_HO,energy_mode_HO,'Marker','o','MarkerEdgeColor','red','LineStyle','none');    %y values are in eV

Mode1_0th.kz = kz_mode1_0th;
Mode1_0th.energy = energy_mode2_0th;
Mode2_0th.kz = kz_mode1_0th;
Mode2_0th.energy = energy_mode2_0th;
ModeHO.kz = kz_mode_HO;
ModeHO.energy = energy_mode_HO;

Structure.epsilonmatrix = epsilonmatrix;
Structure.epsilonrod = epsilonrod;
Structure.radius = radius;
Structure.DispersionInfo = Mode;

set(gca,'Tag','Dispersion');
set(SPHandles.MainFigure,'UserData',Structure);
```

## MATLAB Code for Eρ filed of Metallic Cylinder

```matlab
wavelength = 1.2408/energy;
omega = 2*pi*light/wavelength;
knot = 2*pi/wavelength;
beta= sqrt(kz^2 - knot^2*epsilonmatrix);        %Outside the cylinder
gamma = sqrt(knot^2.*epsilonrod - kz^2);         %Inside the cylinder

bjm = besselj(m,gamma*radius);
bkm = besselk(m,gamma*radius);

Be = bjm/bkm;
tempa = gamma^2;
tempb = beta^2;
tempc = 1/tempa + 1/tempb;

tempd = 0.5*(besselj(m-1,gamma*radius) - besselj(m+1,gamma*radius));
tempe = -0.5*(besselk(m-1,gamma*radius) + besselk(m+1,gamma*radius));
tempf = tempd/gamma + bjm/bkm*tempe/beta;
Ah = 1i*m*kz/omega/mu_o/radius*tempc/tempf;

Bh = bjm/bkm*Ah;

h6 = findobj('Tag','Fields');
axes(h6);
```

```matlab
delta_r = radius/250;
[r_wire,t_wire] = meshgrid(0:delta_r :radius,0:pi/30:(2*pi));
x_wire = r_wire.*cos(t_wire);
y_wire = r_wire.*sin(t_wire);
z_wire = zeros(size(x_wire));

tempg = 0.5*(besselj(m-1,gamma*r_wire) - besselj(m+1,gamma*r_wire));
tempi = besselj(m,gamma*r_wire);
E_rho_wire = 1i/gamma^2.*exp(1i*m*t_wire).*(gamma*kz*tempg + 1i*omega*mu_o*m./r_wire*Ah.*tempi);
surf(x_wire,y_wire,z_wire,real(E_rho_wire),'LineStyle','none');

hold on;
[r_matrix,t_matrix] = meshgrid(radius+delta_r:delta_r :2*radius,0:pi/30:(2*pi));
x_matrix = r_matrix.*cos(t_matrix);
y_matrix = r_matrix.*sin(t_matrix);
z_matrix = zeros(size(x_matrix));

tempj = -0.5*(besselk(m-1,beta*r_matrix) + besselk(m+1,beta*r_matrix));
tempk = besselk(m,beta*r_matrix);
E_rho_matrix = -1i/beta^2.*exp(1i*m*t_matrix).*(beta*kz*Be*tempj + 1i*omega*mu_o*m./r_matrix*Bh.*tempk);
surf(x_matrix,y_matrix,z_matrix,real(E_rho_matrix),'LineStyle','none');

[Xc,Yc,z_temp] = cylinder(radius,100);
Zc = 3*wavelength*z_temp;
surf(Xc,Yc,Zc,'FaceAlpha',0.25,'EdgeColor','none','FaceColor','red');
view([35,40]);
title('$E_{\rho}$ Field Component','interpreter','latex');
cb = colorbar;
templma = max(max(abs(E_rho_matrix)));
templmb = max(max(abs(E_rho_wire)));
if templma>templmb
    set(cb,'Limits',[-templma,templma]);
else
    set(cb,'Limits',[-templmb,templmb]);
end

Field.x_wire = x_wire;
Field.y_wire = y_wire;
Field.E_rho_wire = E_rho_wire;
Field.x_matrix = x_matrix;
Field.y_matrix = y_matrix;
Field.E_rho_matrix = E_rho_matrix;
Field.Xc = Xc;
Field.Yc = Yc;
Field.Zc = Zc;

hold off;
set(gca,'Tag','Fields');
set(gca,'UserData',Field);

h6 = findobj('Tag','Zslider');
set(h6,'Max',2*wavelength);
```

```matlab
set(h6,'Min',-2*wavelength);
set(h6,'MajorTicks',[-2*wavelength -wavelength 0 wavelength 2*wavelength]);
set(h6,'Enable','on');
```

# BIBLIOGRAPHY

Golsa Mirbagheri

Candidate for the Degree of Doctor of Philosophy

Thesis:

HYPERBOLIC METAMATERIAL FILTER FOR ANGLE-INDEPENDENT TM-TRANSMISSION IN IMAGING APPLICATIONS

Major Field: Computer and Electrical Engineering

Biographical: -

Personal Data: -

Education: Master of Science – Information Technology Management and Technology

Completed the requirements for the Doctor of Philosophy in Computer and Electrical Engineering at Clarkson University, Potsdam, New York in November 2020.

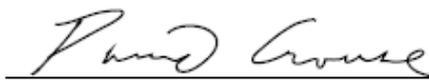

ADVISER'S APPROVAL: Dr. David Crouse